%% file: main.tex
\documentclass[12pt, review]{elsarticle}

\input{Inputs/preamble.tex}

\input{./Inputs/acronyms.tex}

\journal{Computer Methods in Applied Mechanics and Engineering}

\begin{document}

\begin{frontmatter}

\title{A modular massively parallel computing environment for three-dimensional multiresolution simulations of compressible flows}

\author[1]{Nils Hoppe\corref{cor1}}
\cortext[cor1]{Corresponding author}
\ead{nils.hoppe@tum.de}

\author[2]{Stefan Adami}
\author[3]{Nikolaus A. Adams}

\address{Chair of Aerodynamics and Fluid Mechanics, Boltzmannstr. 15, D-85748 Garching bei München}

\begin{abstract}
   \input{./Inputs/abstract.tex}
\end{abstract}



\end{frontmatter}


\input{./Inputs/introduction.tex}
\input{./Inputs/preliminaries.tex}
\input{./Inputs/modular_block_mr.tex}
\input{./Inputs/algorithm_implementation.tex}
\input{./Inputs/Benchmarking/benchmarking.tex}

\input{./Inputs/conclusion.tex}
\input{./Inputs/acknowledgment.tex}

\appendix

\input{./Inputs/appendix.tex}

\bibliographystyle{elsarticle-num}
\bibliography{main}
\end{document}

%% file: Inputs/preamble.tex
\usepackage{graphicx}
\usepackage{amsmath}
\usepackage{amssymb}
\usepackage{mathtools}
\usepackage{subcaption}
\usepackage{algorithm}
\usepackage{algpseudocode}
\usepackage{adjustbox}
\usepackage{cleveref}
\usepackage[T1]{fontenc}

\usepackage{xcolor}
\definecolor{TUMBlau}{RGB}{0,100,189}
\definecolor{TUMBlauDunkel}{RGB}{0,82,147}
\definecolor{TUMBlauSehrDunkel}{RGB}{0,51,81}
\definecolor{TUMBlauHell}{RGB}{152,198,234}
\definecolor{TUMBlauMittel}{RGB}{100,160,200}
\definecolor{TUMElfenbein}{RGB}{218,215,203}
\definecolor{TUMGruen}{RGB}{162,173,0}
\definecolor{TUMOrange}{RGB}{227,114,34}
\definecolor{TUMGrauZwanzig}{RGB}{217,218,219}

\usepackage{pgfplots}
\pgfplotsset{compat=newest}

\pgfplotscreateplotcyclelist{TUMcyclelist}{%
TUMBlau,every mark/.append style={solid,fill=TUMBlau},mark=*\\%
TUMOrange,every mark/.append style={solid,fill=TUMOrange},mark=diamond*\\%
TUMGruen,every mark/.append style={solid,TUMGruen},mark=x\\%
TUMBlauDunkel,every mark/.append style={solid,fill=TUMBlauDunkel},mark=triangle*\\%
TUMBlauMittel,every mark/.append style={solid,fill=TUMBlauMittel},mark=diamond*\\%
TUMBlau,densely dashed,every mark/.append style={solid,fill=TUMBlau},mark=triangle*\\%
TUMOrange,densely dashed,every mark/.append style={solid,fill=TUMOrange},mark=square*\\%
TUMBlauDunkel,densely dashed,mark=star,every mark/.append style={solid,fill=TUMBlauDunkel},mark=x\\%
TUMBlauMittel,densely dashed,every mark/.append style={solid,fill=TUMBlauMittel},mark=square*\\%
TUMGruen,densely dashed,every mark/.append style={solid,TUMGruen},mark=*\\%
}

\pgfplotscreateplotcyclelist{TUMcyclelistNomark}{%
TUMBlau,solid,mark=none\\
TUMOrange,dashed,mark=none\\
TUMGruen,dotted,mark=none\\
TUMBlauDunkel,dashdotted,mark=none\\
TUMBlauHell,densely dashed,mark=none\\
TUMBlau,densely dotted,mark=none\\
TUMOrange,loosely dashed,mark=none\\
TUMGruen,loosely dotted,mark=none\\
}

\usetikzlibrary{patterns}

\usetikzlibrary{spy,calc}

\newif\ifblackandwhitecycle
\gdef\patternnumber{0}

\pgfkeys{/tikz/.cd,
    zoombox paths/.style={
        draw=orange,
        very thick
    },
    black and white/.is choice,
    black and white/.default=static,
    black and white/static/.style={
        draw=white,
        zoombox paths/.append style={
            draw=white,
            postaction={
                draw=black,
                loosely dashed
            }
        }
    },
    black and white/static/.code={
        \gdef\patternnumber{1}
    },
    black and white/cycle/.code={
        \blackandwhitecycletrue
        \gdef\patternnumber{1}
    },
    black and white pattern/.is choice,
    black and white pattern/0/.style={},
    black and white pattern/1/.style={
            draw=white,
            postaction={
                draw=black,
                dash pattern=on 2pt off 2pt
            }
    },
    black and white pattern/2/.style={
            draw=white,
            postaction={
                draw=black,
                dash pattern=on 4pt off 4pt
            }
    },
    black and white pattern/3/.style={
            draw=white,
            postaction={
                draw=black,
                dash pattern=on 4pt off 4pt on 1pt off 4pt
            }
    },
    black and white pattern/4/.style={
            draw=white,
            postaction={
                draw=black,
                dash pattern=on 4pt off 2pt on 2 pt off 2pt on 2 pt off 2pt
            }
    },
    zoomboxarray inner gap/.initial=5pt,
    zoomboxarray columns/.initial=2,
    zoomboxarray rows/.initial=2,
    subfigurename/.initial={},
    figurename/.initial={zoombox},
    zoomboxarray/.style={
        execute at begin picture={
            \begin{scope}[
                spy using outlines={%
                    zoombox paths,
                    width=\imagewidth / \pgfkeysvalueof{/tikz/zoomboxarray columns} - (\pgfkeysvalueof{/tikz/zoomboxarray columns} - 1) / \pgfkeysvalueof{/tikz/zoomboxarray columns} * \pgfkeysvalueof{/tikz/zoomboxarray inner gap} -\pgflinewidth,
                    height=\imageheight / \pgfkeysvalueof{/tikz/zoomboxarray rows} - (\pgfkeysvalueof{/tikz/zoomboxarray rows} - 1) / \pgfkeysvalueof{/tikz/zoomboxarray rows} * \pgfkeysvalueof{/tikz/zoomboxarray inner gap}-\pgflinewidth,
                    magnification=3,
                    every spy on node/.style={
                        zoombox paths
                    },
                    every spy in node/.style={
                        zoombox paths
                    }
                }
            ]
        },
        execute at end picture={
            \end{scope}
            \node at (image.north) [anchor=north,inner sep=0pt] {\subcaptionbox{\label{\pgfkeysvalueof{/tikz/figurename}-image}}{\phantomimage}};
            \node at (zoomboxes container.north) [anchor=north,inner sep=0pt] {\subcaptionbox{\label{\pgfkeysvalueof{/tikz/figurename}-zoom}}{\phantomimage}};
     \gdef\patternnumber{0}
        },
        spymargin/.initial=0.5em,
        zoomboxes xshift/.initial=1,
        zoomboxes right/.code=\pgfkeys{/tikz/zoomboxes xshift=1},
        zoomboxes left/.code=\pgfkeys{/tikz/zoomboxes xshift=-1},
        zoomboxes yshift/.initial=0,
        zoomboxes above/.code={
            \pgfkeys{/tikz/zoomboxes yshift=1},
            \pgfkeys{/tikz/zoomboxes xshift=0}
        },
        zoomboxes below/.code={
            \pgfkeys{/tikz/zoomboxes yshift=-1},
            \pgfkeys{/tikz/zoomboxes xshift=0}
        },
        caption margin/.initial=4ex,
    },
    adjust caption spacing/.code={},
    image container/.style={
        inner sep=0pt,
        at=(image.north),
        anchor=north,
        adjust caption spacing
    },
    zoomboxes container/.style={
        inner sep=0pt,
        at=(image.north),
        anchor=north,
        name=zoomboxes container,
        xshift=\pgfkeysvalueof{/tikz/zoomboxes xshift}*(\imagewidth+\pgfkeysvalueof{/tikz/spymargin}),
        yshift=\pgfkeysvalueof{/tikz/zoomboxes yshift}*(\imageheight+\pgfkeysvalueof{/tikz/spymargin}+\pgfkeysvalueof{/tikz/caption margin}),
        adjust caption spacing
    },
    calculate dimensions/.code={
        \pgfpointdiff{\pgfpointanchor{image}{south west} }{\pgfpointanchor{image}{north east} }
        \pgfgetlastxy{\imagewidth}{\imageheight}
        \global\let\imagewidth=\imagewidth
        \global\let\imageheight=\imageheight
        \gdef\columncount{1}
        \gdef\rowcount{1}
        
    },
    image node/.style={
        inner sep=0pt,
        name=image,
        anchor=south west,
        append after command={
            [calculate dimensions]
            node [image container,subfigurename=\pgfkeysvalueof{/tikz/figurename}-image] {\phantomimage}
            node [zoomboxes container,subfigurename=\pgfkeysvalueof{/tikz/figurename}-zoom] {\phantomimage}
        }
    },
    color code/.style={
        zoombox paths/.append style={draw=#1}
    },
    connect zoomboxes/.style={
    spy connection path={\draw[draw=none,zoombox paths] (tikzspyonnode) -- (tikzspyinnode);}
    },
    help grid code/.code={
        \begin{scope}[
                x={(image.south east)},
                y={(image.north west)},
                font=\footnotesize,
                help lines,
                overlay
            ]
            \foreach \x in {0,1,...,9} {
                \draw(\x/10,0) -- (\x/10,1);
                \node [anchor=north] at (\x/10,0) {0.\x};
            }
            \foreach \y in {0,1,...,9} {
                \draw(0,\y/10) -- (1,\y/10);                        \node [anchor=east] at (0,\y/10) {0.\y};
            }
        \end{scope}
    },
    help grid/.style={
        append after command={
            [help grid code]
        }
    },
}

\newcommand\phantomimage{%
    \phantom{%
        \rule{\imagewidth}{\imageheight}%
    }%
}
\newcommand\zoombox[2][]{
    \begin{scope}[zoombox paths]
        \pgfmathsetmacro\xpos{
            (\columncount-1)*(\imagewidth / \pgfkeysvalueof{/tikz/zoomboxarray columns} + \pgfkeysvalueof{/tikz/zoomboxarray inner gap} / \pgfkeysvalueof{/tikz/zoomboxarray columns} ) + \pgflinewidth
        }
        \pgfmathsetmacro\ypos{
            (\rowcount-1)*( \imageheight / \pgfkeysvalueof{/tikz/zoomboxarray rows} + \pgfkeysvalueof{/tikz/zoomboxarray inner gap} / \pgfkeysvalueof{/tikz/zoomboxarray rows} ) + 0.5*\pgflinewidth
        }
        \edef\dospy{\noexpand\spy [
            #1,
            zoombox paths/.append style={
                black and white pattern=\patternnumber
            },
            every spy on node/.append style={#1},
            x=\imagewidth,
            y=\imageheight
        ] on (#2) in node [anchor=north west] at ($(zoomboxes container.north west)+(\xpos pt,-\ypos pt)$);}
        \dospy
        \pgfmathtruncatemacro\pgfmathresult{ifthenelse(\columncount==\pgfkeysvalueof{/tikz/zoomboxarray columns},\rowcount+1,\rowcount)}
        \global\let\rowcount=\pgfmathresult
        \pgfmathtruncatemacro\pgfmathresult{ifthenelse(\columncount==\pgfkeysvalueof{/tikz/zoomboxarray columns},1,\columncount+1)}
        \global\let\columncount=\pgfmathresult
        \ifblackandwhitecycle
            \pgfmathtruncatemacro{\newpatternnumber}{\patternnumber+1}
            \global\edef\patternnumber{\newpatternnumber}
        \fi
    \end{scope}
}

\usepackage[binary-units=true]{siunitx}
\DeclareSIUnit\flop{FLOP}

\usepackage{relsize}
\newcommand\CC{C\nolinebreak[4]\hspace{-.05em}\raisebox{.4ex}{\relsize{-3}{\textbf{++}}}}

\newcommand\ie{i.\,e.\ }

\newcommand\eg{e.\,g.\ }

\newcommand\egn{e.\,g.}

\newcommand\cf{c.\,f.\ }

\newcommand{\overbar}[1]{\mkern 1.5mu\overline{\mkern-1.5mu#1\mkern-1.5mu}\mkern 1.5mu}


%% file: Inputs/acronyms.tex
\usepackage[nolist]{acronym}

\begin{acronym}
\acro{MPI}[MPI]{Message Passing Interface}
\acro{MR}[MR]{multiresolution}
\acro{SIMD}[SIMD]{single instruction multiple data}
\acro{ENO}[ENO]{essentially non-oscillatory}
\acro{WENO}[WENO]{weighted essentially non-oscillatory}
\acro{WENOCU}[WENO-CU]{central-upwind weighted essentially non-oscillatory}
\acro{WENOAO}[WENO-AO]{adaptive order weighted essentially non-oscillatory}
\acro{TENO}[TENO]{targeted essentially non-oscillatory}
\acro{TVD}[TVD]{total variation diminishing}
\acro{OMP}[openMP]{Open Multi-Processing}
\acro{CPU}[CPU]{central processing unit}
\acro{HPC}[HPC]{high-performance computing}
\acro{LTS}{local time stepping}
\acro{ALTS}{adaptive local time stepping}
\acro{FVM}{finite-volume method}
\acrodefplural{FVM}[FVM]{finite-volume methods}
\acro{RK}[RK]{Runge-Kutta}
\acro{CFL}[CFL]{Courant-Friedrichs-Lewy}
\acro{CFD}[CFD]{computational fluid dynamics}
\acro{RTI}{Rayleigh-Taylor instability}
\acro{OOP}{object-oriented programming}
\acro{GCC}{GNU C++ compiler}
\acro{ICC}{Intel C++ compiler}
\acro{HW}{Haswell}
\acro{KNL}{Knights Landing Xeon Phi microarchitecture}
\acro{EOS}{equation of state}
\acro{MCDRAM}{multi-channel dynamic random-access memory}
\acro{HLLC}{Harten-Lax-van Leer-contact}
\acro{3D}{three-dimensional}
\acro{FD}{finite difference}
\acro{PDE}{partial differential equation}
\acro{SFC}{space filling curve}
\acro{AMR}{adaptive mesh refinement}
\acro{AVX}[AVX]{advanced vector extensions}
\acro{SSE}[SSE]{streaming SIMD extensions}
\acro{LLF}{local Lax-Friedrichs}
\acro{TBB}{Threading Building Blocks}
\acro{IC}{internal cells}
\acro{FLOP}{floating-point operations}
\acro{2D}{two-dimensional}
\acro{NSE}{Navier-Stokes equations}
\acro{CRTP}{curiously recurring template pattern}
\end{acronym}

%% file: Inputs/abstract.tex

Numerical investigation of compressible flows faces two main challenges. In order to accurately describe the flow characteristics, high-resolution nonlinear numerical schemes are needed to capture discontinuities and resolve wide convective, acoustic and interfacial scale ranges. The simulation of realistic \ac{3D} problems with state-of-the-art \acp{FVM} based on approximate Riemann solvers with weighted nonlinear reconstruction schemes requires the usage of \ac{HPC} architectures. Efficient compression algorithms reduce computational and memory load. Fully adaptive \ac{MR} algorithms with \ac{LTS} have proven their potential for such applications. While modern \acp{CPU} require multiple levels of parallelism to achieve peak performance, the fine grained \ac{MR} mesh adaptivity results in challenging compute/communication patterns. Moreover, \ac{LTS} incur for strong data dependencies which challenge a parallelization strategy.

We address these challenges with a block-based \ac{MR} algorithm, where arbitrary cuts in the underlying octree are possible. This allows for a parallelization on distributed-memory machines via the \ac{MPI}. We obtain neighbor relations by a simple bit-logic in a modified Morton Order. The block-based concept allows for a modular setup of the source code framework in which the building blocks of the algorithm, such as the choice of the Riemann solver or the reconstruction stencil, are interchangeable without loss of parallel performance. We present the capabilities of the modular framework with a range of test cases and scaling analysis with effective resolutions beyond one billion cells using $\mathcal{O}(10^3)$ cores.

%% file: Inputs/introduction.tex
\section{Introduction}

Physically accurate numerical investigation of compressible flows requires powerful flow solvers, both in terms of discretization techniques as well as computational efficiency. Key requirements for such a solver are vanishing artificial smearing of flow states, non-oscillatory treatment of discontinuities (\eg shocks), and enforcement of discrete conservation. \Acp{FVM} combined with high resolution \ac{WENO} type reconstruction stencils \cite{Jiang1996} have proven their suitability to study compressible flows \cite{Bell1989,Pember1995,Titarev2004,Johnsen2006} and are considered in this work. However, the numerical load of such solvers applied to practical compressible flow problems is high. Hence, the usage of compression algorithms and \ac{HPC} compute systems is inevitable. Aiming for a generalized simulations platform covering a broad range of state-of-the-art schemes, we have developed a modular massively parallel computing strategy. Our framework enables efficient simulation of compressible flow applications and encompasses a wide range of high-resolution numerical schemes and fluid models for a user to choose from.

Even with the compute power of latest \ac{HPC} clusters, fully resolved \ac{3D} simulations of compressible flows often are infeasible due to the high memory and computational requirements. Moreover, maximum resolution of the entire computational domain would be inefficient since resolution may be wasted in subdomains with little flow detail. Therefore, compression algorithms which locally adapt the mesh resolution have been developed. A well-known compression algorithm is the truncation-error based \ac{AMR} \cite{Berger1989} scheme. As its refinement is sensitive to the steepness of flow-field gradients, resolving regions of interest without strong gradient, \eg contact or smooth rarefaction waves, requires special care. Wavelet-based methods achieve such a refinement automatically and often lead to a better mesh compression \cite{Deiterding2020}. These algorithms are typically sorted into three groups: Wavelet-Galerkin, wavelet-collocation, and \ac{MR} methods. As the first group does not inherently maintain conservation, we do not elaborate on it further and refer to the review paper of Schneider and Vasilyev \cite{Schneider2010}. Collocation schemes are based on second-generation wavelets and require a full wavelet-transformation \cite{Vasilyev2000}. They have been applied to hyperbolic \cite{Regele2009} and parabolic \cite{Vasilyev2003} problems and have also been used to reach an adaptive compression in time \cite{Alam2006}. Regarding their performance on \ac{HPC} systems, collocation methods suffer from the wide support of active scales, which implies non-local communications. Current developments try to address this issue by an asynchronous wavelet transform \cite{Nejadmalayeri2015} at the cost of the vanishing moment property of the classic wavelet transform.

Interpolation-wavelet based \ac{MR} was initially proposed by Harten \cite{Harten1995} to reduce expensive flux computations in one dimension. This concept has been extended to viscous flows and to more dimensions \cite{Bihari1996,Bihari1997}. Harten's algorithm has been further augmented to also save memory, \eg, by the introduction of flux functions operating on non-equidistant points \cite{Kaibara2001} or by introducing graded-tree concepts for hyperbolic \cite{Cohen2003} and parabolic equations \cite{Roussel2003}. For the latter approaches, various modifications have been presented, \egn, using a binary tree structure for $n$-dimensional meshes \cite{Castro2016} or linking of reconstruction-stencil weights to the compression parameters \cite{Maulik2018}. Within the \ac{MR} algorithm, further compression can be achieved via \ac{LTS} schemes. Such methods were first introduced for single-stage methods \cite{Muller2007} and then extended to explicit multi-stage \ac{RK} schemes \cite{Domingues2008}. For the latter, also adaptive time step sizes have been introduced \cite{Domingues2009,Kaiser2019}. In contrast to collocation methods, the support of active scales in interpolation-based wavelet \ac{MR} is local. However, the number of communication partners per active scale is high and frequent neighbor-lookup can impede performance on \ac{HPC} clusters, as explained later in more detail. Shared-memory implementations of this traditional single-scale approach have been presented for \ac{2D} cases by Descombes et al. \cite{Descombes2017} and via \ac{MPI} parallelization by Brix et al. \cite{Brix2011}.

Optimizing the communication patterns of a compression algorithm alone, however, does not suffice to harvest strong \ac{HPC} performance \cite{Sutter2005}. For this, all layers of parallelism in the modern compute hardware need to be addressed, from distributed to shared-memory as well as \ac{SIMD} 'vectorization' parallelism \cite{Hager2011a}. Regarding vectorization, block-based schemes have proven their potential in the context of \ac{AMR} methods \cite{Ferreira2017} as well as for collocation methods \cite{Hejazialhosseini2010}. For \ac{MR} algorithms, blocking strategies have proven efficient for \ac{2D} cases using cell-centered \ac{FVM} with a \ac{TBB} shared-memory parallelization \cite{Han2011} and for point-based finite difference schemes via \ac{MPI} parallelization \cite{Sroka2019}.

In this paper, we present a \ac{3D} block-based and fully-adaptive \ac{MR} algorithm including \ac{ALTS} to solve hyperbolic and parabolic problems with \ac{FVM}. The algorithm is parallelized using \ac{MPI}. The implementation correctness, the achieved compression and the parallel performance of the algorithm are demonstrated on various test cases. Our algorithm is designed and implemented in a modular fashion, allowing to exchange the compute kernels easily without loss of parallel performance. We demonstrate this properly by combining different Riemann solvers (Roe \cite{Roe1981}, \ac{LLF} \cite{Rusanov1962} and \ac{HLLC} \cite{Toro1994}) with different reconstructions stencils (\ac{WENO} \cite{Jiang1996}, \ac{TENO} \cite{Fu2016}, \ac{WENOAO} \cite{Balsara2016}) and several equations of state. We evaluate the parallel performance of the combinations on a large \ac{HPC} compute cluster with more than a thousand cores and run simulations with more than a billion cells of effective resolution. Most \ac{MR} implementations found in Literature \cite{Roussel2003,Domingues2009,Descombes2017,Sroka2019} are limited to cubic computational domains. Our implementation naturally allows to also simulate channel-like domains.

We achieve the stated qualities via a combination of a modified Morton order \cite{Morton1966} for efficient neighbor lookup combined with a level-wise \ac{SFC} for dynamic load balancing. In the data structure, we separate heavy computation data from the lightweight topology tree. In the octree we allow arbitrary tree cuts both horizontally and vertically. The algorithm is implemented in state-of-the-art \CC 17 code and is available under open-source license\footnote{https://gitlab.lrz.de/nanoshock/ALPACA}.

The remainder of this paper is structured as follows. First, the governing equations, their numerical discretization and the key concepts of \ac{MR} and \ac{ALTS} are revised in \cref{sec:Preliminaries}. Next, in \cref{sec:Concepts}, our modular block-based \ac{MR} concept, the neighbor lookup and the load balancing scheme are presented. In \cref{sec:Algorithm} we describe the detailed algorithm and highlight important implementation details. \Cref{sec:Benchmarking} states the conducted numerical verification with error analysis. Concluding remarks are given in the end.

%% file: Inputs/preliminaries.tex
\section{Physical and numerical model}
\label{sec:Preliminaries}

\subsection{Governing Equations}

A general conservation law in symbolic notation for a state vector $\mathbf{U}$ is given by

\begin{equation}
\frac{\partial \mathbf{U}}{\partial t} + \nabla \mathbf{F}\left(\mathbf{U}\right) = S\left( \mathbf{U} \right),
\label{eq:Euler}
\end{equation}
where $\mathbf{F}\left( \mathbf{U} \right)$ and $S\left( \mathbf{U}, t \right)$ are the flux and source term vector, respectively. For the Euler equations, the state vector $\mathbf{U} = \left(\rho, \rho \mathbf{v}, E \right)$ contains the conserved quantities mass (concentration) $\rho$, momentum $\rho\mathbf{v}$ and total energy $E$. In these equations, $\mathbf{v}$ is the velocity vector and the total energy is given by $E = \rho e + \frac{1}{2} \rho\left| \mathbf{v} \right|^2$, with the internal energy per unit mass $e$ and the pressure $p$. To close the set of equations an additional \ac{EOS} is needed. For demonstration purpose, in this work the stiffened gas equation \cite{Harlow1971}

\begin{equation}
p(\rho,e) = \left( \gamma - 1 \right) \rho e - \gamma B
\label{eq:StiffenedGas}
\end{equation}
is used unless noted otherwise. A wide range of \ac{EOS} is implemented. In \cref{eq:StiffenedGas}, $\gamma$ and $B$ represent the ratio of specific heats and the background pressure, respectively.

\subsection{Numerical Scheme}

We solve the governing equations using finite volumes on Cartesian meshes with cubic cells in \ac{3D}. For lower dimensional problems only a single cell depth is used in the excess dimensions. The fluxes across the cell faces are computed via an approximate Riemann solver. For demonstration, we restrict ourselves to Roe \cite{Roe1981}, \ac{HLLC} \cite{Toro1994} and Rusanov-flux \cite{Rusanov1962} type solvers. In the latter, only the two cells sharing a face are considered for the eigenvalues. Within the Riemann solvers, different reconstruction stencils can be used to increase the order of the scheme. For the presented computations the stated Riemann solvers are used with fifth-order \ac{WENO} \cite{Liu1994,Jiang1996}, fifth-order \ac{TENO} \cite{Fu2016} or fifth-order \ac{WENOAO} \cite{Balsara2016} discretization schemes.

\subsection{Multiresolution Algorithm with Adaptive Local Time Stepping}

On dyadically refined grids with infinite levels of refinement $l_m \in \mathbb{Z}$, a function $u(x,t)$ can be exactly represented by its \ac{MR} representation in the form

\begin{equation}
u(x,t) = \sum\limits_k c_k \theta_k(x,t) + \sum\limits_m \sum\limits_k d_k^{l_m} \psi_k^{l_m}(x,t),
\label{eq:WaveletTrafo}
\end{equation}
where the subscript $k$ samples the function at grid point $x_k^{l_m} = 2^{-{l_m}}k \in \mathbb{Z}$. The first sum over the products of scaling coefficients $c_k$ and compact scaling functions $\theta_k$ defines the general shape of the function. The \emph{details} $d_k^l$ and the wavelet function $\psi_k^l$ define the local fluctuations on the m-th refinement level $l_m$. By using  interpolation wavelets, the details are given by mere difference of coarse and fine cell values

\begin{equation}
d^{l_m}(x) = \sum\limits_k d_k^{l_m}\psi_k^{l_m}(x) = u_{l_{m+1}}(x) - u_{l_m}(x).
\label{eq:DetailFunction}
\end{equation}
In his pioneering work, Harten \cite{Harten1995} used this derivation to reduce the amount of costly flux computations with interpolated coarse grid states in regions with vanishing details

\begin{equation}
\|d^{l_m}\| < \varepsilon_{l_m}.
\label{eq:Thresholding}
\end{equation}
If the level-dependent threshold $\varepsilon_{l_m}$ is chosen properly, the \ac{MR} representation converges with a desirable order $\alpha$. The convergence order of the applied numerical scheme thus is ensured by setting $\alpha$ equal to this convergence order considering both space and time. To do so, a reference level $l_{ref}$ and reference threshold $\varepsilon_{ref}$ need to be chosen. From these, $\varepsilon_{l_m}$ is computed as

\begin{align}
\varepsilon &= 2^{ - ( \alpha + 1 ) ( l_{max} - l_{ref} ) } \varepsilon_{ref}, \label{eq:UserEpsilon} \\
\varepsilon_{l_m} &= 2^{-D \left( l_{max} - l_m \right) } \varepsilon, \label{eq:LevelWiseEpsilon}
\end{align}
with the dimensionality $D$ of the considered problem.

Fully adaptive \ac{MR} schemes \cite{Cohen2003} remove cells with vanishing details from the grid to achieve high memory compression. These schemes naturally lead to octree data-structures. Following common notation, we call nodes in the tree \emph{parent} if they are refined with \emph{child} nodes on a higher level. Same-level neighboring nodes are called \emph{brothers} if they share a parent, or \emph{cousins} otherwise. A node is called a \emph{leaf} if it does not have children. If a node does not have a cousin on one side we call this side a \emph{resolution jump} or just \emph{jump} for brevity. Leaf \emph{refinement} is equivalent with creation of its finer grid structures in the tree. \emph{Coarsening} means elimination of leafs which turns their parent into a leaf. Note, nodes are only coarsened together with all their brothers. We propagate information from level $l_{m+1}$ down to level $l_{m}$ using the conservative averaging operator

\begin{equation}
\mathcal{A}\left(u_{l_{m+1}}\right) \rightarrow u_{l_m} \colon u_{l_m} = \frac{1}{N}\sum\limits_j^N u_{l_{m+1}}^j,
\label{eq:Averaging}
\end{equation}
where $N$ gives the number of children, \eg four in \ac{2D} or eight in \ac{3D}.

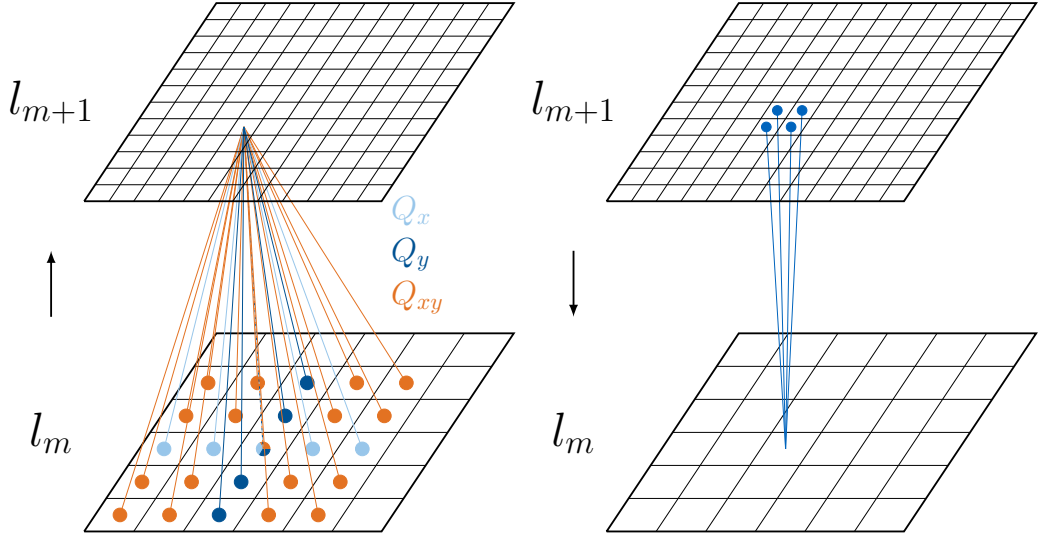
\begin{figure}
\begin{subfigure}[c]{0.5\textwidth}
\begin{center}
\begin{adjustbox}{width=\linewidth}
\begin{tikzpicture}
\draw[thick] (0,0) -- (4.5,0) -- (6.5,3.0) -- (2.0,3.0) -- (0,0);
\foreach \x in {1, ..., 5} {\draw (0.666667*0.5*\x,0.5*\x) -- (0.666667*0.5*\x+4.5,0.5*\x);}
\foreach \x in {1, ..., 5} {\draw (0.75*\x,0) -- (\x*0.75+2.0,3.0);}

\foreach \x in {1, 2, 4, 5} {\draw[draw =TUMBlauHell, fill=TUMBlauHell] (0.666667*1.25-0.375+0.75*\x,1.25) circle (.25em);}

\foreach \x in {1, 2, 4, 5} {\draw[draw= TUMBlauDunkel, fill=TUMBlauDunkel] (0.666667*0.5+1.375+0.666667*0.5*\x,\x*0.5-0.25) circle (.25em);}

\foreach \x in {1, 2, 4, 5} {
    \foreach \y in {1, 2, 4, 5} {
        \draw[draw= TUMOrange, fill=TUMOrange] (0.666667*0.5-0.875+0.75*\x+0.666667*0.5*\y,-0.25+0.5*\y) circle (.25em);
    }
}

\filldraw[TUMOrange] (2.7083334, 1.25) -- ++ (0:0.25em) arc [radius=.25em, start angle=0, end angle=119];
\filldraw[TUMBlauHell] (2.7083334, 1.25) -- ++(120:0.25em) arc [radius=.25em, start angle=120, end angle=239];
\filldraw[TUMBlauDunkel] (2.7083334, 1.25) -- ++(240:0.25em) arc [radius=.25em, start angle=240, end angle=359];

\draw[draw=TUMOrange, dash pattern = on 1pt off 2 pt] (2.7083334, 1.25) -- (1.75+0.666667,6.125);
\draw[draw=TUMBlauHell, dash pattern = on 1pt off 2 pt, dash phase = 2pt] (2.7083334, 1.25) -- (1.75+0.666667,6.125);
\draw[draw=TUMBlauDunkel, dash pattern = on 1pt off 2 pt, dash phase = 4pt] (2.7083334, 1.25) -- (1.75+0.666667,6.125);

\foreach \x in {1, 2, 4, 5} {
    \foreach \y in {1, 2, 4, 5} {
        \draw[draw=TUMOrange] (0.666667*0.5-0.875+0.75*\x+0.666667*0.5*\y,-0.25+0.5*\y) -- (1.75+0.666667,6.125);
    }
}
\foreach \x in {1, 2, 4, 5} {\draw[draw=TUMBlauHell] (0.666667*1.25-0.375+0.75*\x,1.25) -- (1.75+0.666667,6.125);}
\foreach \x in {1, 2, 4, 5} {\draw[draw=TUMBlauDunkel] (0.666667*0.5+1.375+0.666667*0.5*\x,\x*0.5-0.25) -- (1.75+0.666667,6.125);}

\draw[thick] (0,5) -- (4.5,5) -- (6.5,8.0) -- (2.0,8.0) -- (0,5);
\foreach \x in {1, ..., 11} {\draw (0.666667*0.25*\x,0.25*\x+5.0) -- (0.666667*0.25*\x+4.5,0.25*\x+5.0);}
\foreach \x in {1, ..., 11} {\draw (0.375*\x,5) -- (\x*0.375+2.0,8.0);}

\draw (-0.5,1.5) node {\LARGE $l_m$};
\draw (-0.5,6.5) node {\LARGE $l_{m+1}$};

\draw[TUMBlauHell]   ( 4.5, 4.85 ) node[anchor=west] {\large $Q_x$};
\draw[TUMBlauDunkel] ( 4.5, 4.20 ) node[anchor=west] {\large $Q_y$};
\draw[TUMOrange]     ( 4.5, 3.55 ) node[anchor=west] {\large $Q_{xy}$};

\draw[-latex, thick] (-0.5, 3.25) -- (-0.5,4.25);
\end{tikzpicture}
\end{adjustbox}
\end{center}
\end{subfigure}
\begin{subfigure}[c]{0.5\textwidth}
\begin{center}
\begin{adjustbox}{width=\linewidth}
\begin{tikzpicture}
\draw[thick] (0,0) -- (4.5,0) -- (6.5,3.0) -- (2.0,3.0) -- (0,0);
\foreach \x in {1, ..., 5} {\draw (0.666667*0.5*\x,0.5*\x) -- (0.666667*0.5*\x+4.5,0.5*\x);}
\foreach \x in {1, ..., 5} {\draw (0.75*\x,0) -- (\x*0.75+2.0,3.0);}

\foreach \x in {1, 2} {
    \foreach \y in {1, 2} {
        \draw[draw= TUMBlau, fill=TUMBlau] (1.875 + 0.375*\x+0.666667*0.25*\y,-0.125+0.25*\y+6) circle (.17677766953em);
        \draw[draw= TUMBlau] (1.875 + 0.375*\x+0.666667*0.25*\y,-0.125+0.25*\y+6) -- (1.875+0.666667*1.25,1.25);
    }
}

\draw[thick] (0,5) -- (4.5,5) -- (6.5,8.0) -- (2.0,8.0) -- (0,5);
\foreach \x in {1, ..., 11} {\draw (0.666667*0.25*\x,0.25*\x+5.0) -- (0.666667*0.25*\x+4.5,0.25*\x+5.0);}
\foreach \x in {1, ..., 11} {\draw (0.375*\x,5) -- (\x*0.375+2.0,8.0);}

\draw (-0.5,1.5) node {\LARGE $l_m$};
\draw (-0.5,6.5) node {\LARGE $l_{m+1}$};
\draw[latex-, thick] (-0.5, 3.25) -- (-0.5,4.25);
\end{tikzpicture}
\end{adjustbox}
\end{center}
\end{subfigure}
\caption{Prediction (left) and averaging (right) operators illustrated for a \ac{2D} grid. The finer mesh is plotted above the coarser one as it overlays this region in the computational grid.}
\label{fig:AvgPre}
\end{figure}

Likewise, we define a prediction operation to propagate information form the lower level $l_m$ to the higher level $l_{m+1}$. The prediction is defined locally and consistently, \ie conservative with respect to the coarse grid cell averages. Mathematically, this is expressed as $\mathcal{A} \left( \mathcal{P} \left( u_{l_m} \right) \right)= u_{l_m}$. We use the fifth-order accurate prediction

\begin{equation}
\begin{aligned}
    \mathcal{P}\left(u_{l_m}\right) \rightarrow u_{l_{m+1}} &\colon \\
    u_{l_{m+1}}^{ijk} &= u_{l_m}^{ijk} + (-1)^i Q_x + (-1)^j Q_y + (-1)^k Q_z \\
    &+ (-1)^i (-1)^j Q_{xy} + (-1)^i (-1)^k Q_{xz} + (-1)^j (-1)^k Q_{yz} \\
    &+ (-1)^i (-1)^j (-1)^k Q_{xyz}.
\end{aligned}
\label{eq:Prediction}
\end{equation}
The respective $Q$-terms are given in the Appendix. Both operators are illustrated for a \ac{2D} mesh in \cref{fig:AvgPre}. The locality of the operators is clearly visible.

In addition to spatial compression, compute resources can also be saved by employing adaptive time stepping schemes. Therein, coarser cells are advanced with fewer, but larger time steps than fine ones. \ac{LTS} schemes for \ac{MR} \cite{Domingues2008} leverage the dyadic grid setup and obtain a level dependent time step size

\begin{equation}
\Delta t_{l_m} = 2^{l_{max}-l_m} \Delta t_{l_{max}}
\label{eq:LtsTimestep}
\end{equation}
for the employed explicit multi-stage \ac{RK} schemes. We call a time step on the finest level \emph{micro time step} and a step on the coarsest level \emph{macro time step}. In addition, \ac{ALTS} schemes \cite{Domingues2009} adjust the time step sizes $\Delta t_{l_{max}}$ once all levels have advanced to the same point in time.

Here, we employ an improved \ac{ALTS} scheme \cite{Kaiser2019}, where the local time-steps are adjusted after every micro time step. This is achieved by applying the averaging and prediction operation not only to conserved quantities, but also to the divergence of the numerical flux function directly.

%% file: Inputs/modular_block_mr.tex
\section{Modular block-based MR algorithm}
\label{sec:Concepts}

\ac{MR} algorithms achieve high compression rates \cite{Cohen2003,Roussel2003,Domingues2008} when fine resolutions are required only in a fraction of the entire computational domain. In case of unsteady flows, however, the grid needs to be updated according to the given threshold $\varepsilon_l$ at every micro time step. In a straightforward implementation, a single cell or scale may be coarsened or refined in each such time step. This, however, inhibits efficient execution of the algorithm on modern \ac{HPC} \ac{CPU} as it impedes two levels of parallelism: \ac{SIMD} and distributed-memory. For both, the problem originates from  (wide) stencil reconstruction. In a single-scale \ac{MR} algorithm, \ie where single cells are refined or coarsened, such stencils would cross resolution jumps, which implies either non-regular and possibly remote memory accesses, or introduction of \emph{halo} cell buffers holding copies of the required values. Within a single-scale approach each cell potentially needs to be accompanied by a halo buffer, whose size is proportional to the stencil width. Allocating, filling and deallocating such single-scale halos, in particular by distributed-memory communication, is costly \cite{Brix2011} and should be minimized.

\begin{figure}
\begin{center}
\begin{adjustbox}{width=\linewidth}
\begin{tikzpicture}
\node at (0 , 6.5) {level}; \node at (0.8, 6.475) {$l_1$};
\node at (0.8 , 1.975) {$l_0$};
\node at (0.8 , 9.975) {$l_2$};

\fill[pattern=crosshatch,pattern color=TUMBlau] (8.5,0.5) rectangle (10.0, 3.5);

\fill[pattern=crosshatch,pattern color=TUMBlau] (9.5,5) rectangle (10, 8);

\fill[pattern=north west lines,pattern color=TUMBlauHell] (6.5,5.0) rectangle (7.5, 6.0);
\fill[pattern=north west lines,pattern color=TUMBlauHell] (6.5,6.5) rectangle (7.5, 7.0);
\fill[pattern=north west lines,pattern color=TUMBlauHell] (8.0,6.5) rectangle (8.5, 7.0);
\fill[pattern=north west lines,pattern color=TUMBlauHell] (8.0,5.0) rectangle (8.5, 6.0);

\fill[pattern=horizontal lines,pattern color=TUMGruen] (8.5,5.0) rectangle (9.5, 6.0);
\fill[pattern=horizontal lines,pattern color=TUMGruen] (8.5,6.5) rectangle (9.5, 7.0);
\fill[pattern=horizontal lines,pattern color=TUMGruen] (7.5,6.5) rectangle (8.0, 7.0);
\fill[pattern=horizontal lines,pattern color=TUMGruen] (7.5,5.0) rectangle (8.0, 6.0);

\fill[pattern=north east lines,pattern color=TUMBlauDunkel] (6.5,7) rectangle (7.5, 8);
\fill[pattern=north east lines,pattern color=TUMBlauDunkel] (6.5,6) rectangle (7.5, 6.5);
\fill[pattern=north east lines,pattern color=TUMBlauDunkel] (8.0,6) rectangle (8.5, 6.5);
\fill[pattern=north east lines,pattern color=TUMBlauDunkel] (8.0,7) rectangle (8.5, 8);

\fill[pattern=vertical lines,pattern color=TUMOrange] (8.5,7) rectangle (9.5, 8);
\fill[pattern=vertical lines,pattern color=TUMOrange] (8.5,6) rectangle (9.5, 6.5);
\fill[pattern=vertical lines,pattern color=TUMOrange] (7.5,6) rectangle (8.0, 6.5);
\fill[pattern=vertical lines,pattern color=TUMOrange] (7.5,7) rectangle (8.0, 8);

\draw[color=TUMGrauZwanzig,very thin,step=0.25] (1.5,0) grid (5.5,4);
\draw[color=black,dashed,thick] (1.5,0) -- (5.5,0) -- (5.5,4) -- (1.5,4) -- cycle;
\draw[color=black, thick]  (2.5,1) -- (4.5,1) -- (4.5,3) -- (2.5,3) -- cycle;

\draw[color=TUMGrauZwanzig,very thin,step=0.25] (6.0,0) grid (10,4);
\draw[color=black,dashed,thick] (6.0,0) -- (10,0) -- (10,4) -- (6.0,4) -- cycle;
\draw[color=black, thick]  (7,1) -- (9,1) -- (9,3) -- (7,3) -- cycle;

\draw[color=TUMGrauZwanzig,very thin,step=0.25] (10.5,0) grid (14.5,4);
\draw[color=black,dashed,thick] (10.5,0) -- (14.5,0) -- (14.5,4) -- (10.5,4) -- cycle;
\draw[color=black, thick]  (11.5,1) -- (13.5,1) -- (13.5,3) -- (11.5,3) -- cycle;

\draw[color=TUMGrauZwanzig,very thin,step=0.125] (6.0,4.5) grid (10,8.5);
\draw[color=black,dashed,thick] (6.0,4.5) -- (10,4.5) -- (10,8.5) -- (6.0,8.5) -- cycle;
\draw[color=black,dashed,thick] (6.0,6.5) -- (10,6.5);
\draw[color=black,dashed,thick] (8.0,4.5) -- (8.0,8.5);
\draw[color=black, thick] (6.5,5.0) -- (7.5,5.0) -- (7.5,6.0) -- (6.5,6.0)-- cycle;
\draw[color=black, thick] (8.5,5.0) -- (9.5,5.0) -- (9.5,6.0) -- (8.5,6.0)-- cycle;
\draw[color=black, thick] (6.5,7.0) -- (7.5,7.0) -- (7.5,8.0) -- (6.5,8.0)-- cycle;
\draw[color=black, thick] (8.5,7.0) -- (9.5,7.0) -- (9.5,8.0) -- (8.5,8.0)-- cycle;

\draw[color=TUMGrauZwanzig,very thin,step=0.0625] (8.0,9.0) grid (10.0,11.0);
\draw[color=black,dashed,thick] (8.0,9.0) -- (10.0,9.0) -- (10.0,11.0) -- (8,11) -- cycle;
\draw[color=black,dashed,thick] (8.0,10.0) -- (10.0,10.0);
\draw[color=black,dashed,thick] (9.0,9.0) -- (9.0,11.0);
\draw[color=black, thick] (8.25,9.25)  -- (8.75,9.25) -- (8.75,9.75) -- (8.25,9.75) -- cycle;
\draw[color=black, thick] (9.25,9.25)  -- (9.75,9.25) -- (9.75,9.75) -- (9.25,9.75) -- cycle;
\draw[color=black, thick] (8.25,10.25)  -- (8.75,10.25) -- (8.75,10.75) -- (8.25,10.75) -- cycle;
\draw[color=black, thick] (9.25,10.25)  -- (9.75,10.25) -- (9.75,10.75) -- (9.25,10.75) -- cycle;
\end{tikzpicture}
\end{adjustbox}
\end{center}
\caption{Internal data layout for a locally refined computational domain with $l_{max} = l_2$ with communication patterns for $l_1$ indicated.}
\label{fig:MrCommunicationMesh}
\end{figure}
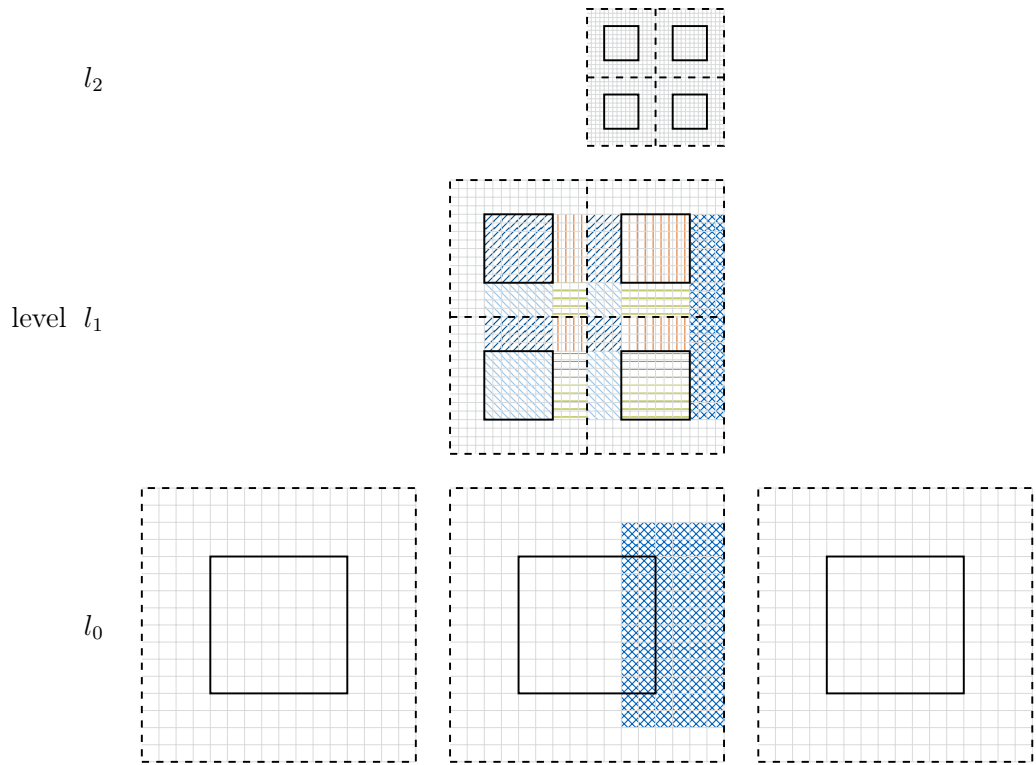

For octree-based \ac{AMR} and with wavelet collocation methods, block-based schemes have shown remarkable \ac{SIMD} capabilities and overall application speedups \cite{Ferreira2017,Rossinelli2011}. Hence, we introduce a similar blocking concept to our \ac{MR} tree. Therein, each node in the octree holds one block with a fixed number of cells. Blocks at different levels differ only by their cell size, namely $\Delta x_{l_m} = 2^{-p} \Delta x_{l_{m+p}}$ ($\Delta y$ and $ \Delta z$ respectively). To allow parallel computations on blocks we also add halo cells around each block. The amount of halo cells is dependent on the width of the used stencil. In this work, we use four halo cells at each block boundary which suffices for the employed 5th-order reconstruction schemes. In contrast to the single-scale halos, the ratio of compute-to-halo cells is thereby immensely improved and cell-neighbor lookups require only local array accesses. In a first analysis, we found our block-based approach capable to effectively utilize modern \ac{CPU} \ac{SIMD} capabilities \cite{Hoppe2019}. In the physical domain, halo cells overlap the \emph{\ac{IC}} of the neighboring blocks. If two blocks with different cell sizes touch, the respective finer halo cells are filled with predicted values from the parent's block, otherwise the brother's or cousin's cell values are copied. \Cref{fig:MrCommunicationMesh} illustrates these communication patterns together with the block-mapping at the coarsest level for a simple \ac{2D} mesh, when we illustrate a channel-like domain of three blocks at level $l_0$. The  middle node is refined and one of the children is refined further. Each block has eight \ac{IC} per dimension. Between the right most block and the blocks on $l_{max}$, a resolution jump of height two is present. Cells inside the thick lines are \ac{IC}, the ones around are the halo cells. For the blocks on level $l_1$ the different communication patterns are indicated. The crosshatch shows how halo cells facing a jump are filled via prediction, whereas the other cells are filled with copies from the respective brother node. Halo cells not highlighted on $l_1$ are part of the external boundary condition and are filled accordingly.

In our block-based algorithm, nodes are refined or coarsened whenever the used error norm for this block - including halos - exceeds the given threshold, \cf \cref{eq:Thresholding} \eg, the $l_\infty$ norm is chosen, the refinement or coarsening of a block is triggered as soon as any cell in the block does not meet the respective threshold. This ensures that approaching or steepening discontinuities are detected and the grid is refined accordingly as in single-scale \ac{MR} algorithms. In addition, this halo-concept allows for arbitrary resolution jumps, without violating the graded-tree requirement \cite{Cohen2003} of single-scale \ac{MR} algorithms.

Besides the favorable \ac{SIMD} capabilities, our block-based approach is beneficial also with respect to distributed-memory parallelization. Concordant to our implementation, we follow the \ac{MPI} terminology with \emph{ranks} being the parallel workers running an instance of the program. The frequency of topological changes is reduced by treating a block as smallest inter-rank parallel unit. This also leads to larger grain sizes of the overall load distribution, which in turn reduce sequential operations \cite{Hejazialhosseini2010}. With fewer topology changes, fewer load-balance operations are required, reducing parallelization overhead. Nevertheless, a well-distributed load is key to parallel performance. For the domain partitioning, we aim at high locality and compactness. Ideally, all nodes of one rank should be continuously grouped together and at the same time require communication with a limited set of other ranks only. When employing \ac{LTS} these requirements are complemented by the demand, that the load needs to be well-balanced within each level of refinement. Otherwise, ranks could idle for several micro time steps if no blocks of the corresponding level are present.

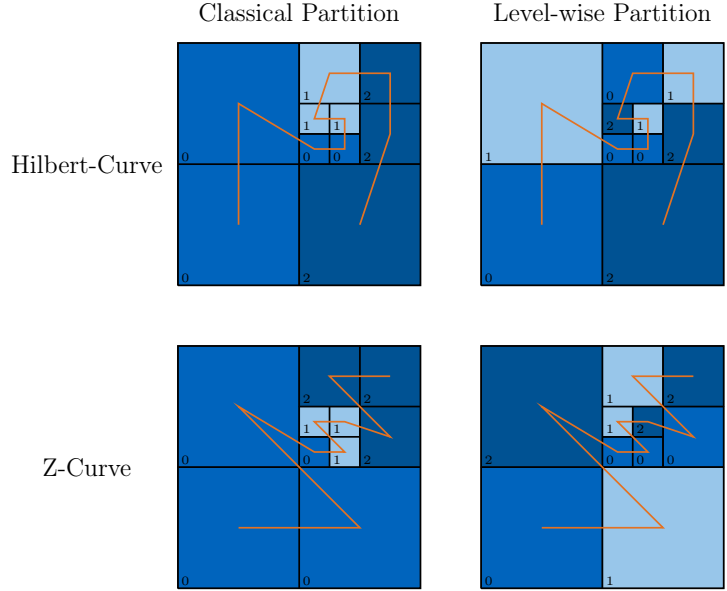
\begin{figure}
\begin{center}
\begin{adjustbox}{width=0.7\linewidth}
\begin{tikzpicture}
\filldraw[fill=TUMBlau] (0,0) rectangle (4,2);
\filldraw[fill=TUMBlau] (0,2) rectangle (2,4);
\filldraw[fill=TUMBlau] (2,2) rectangle (2.5,2.5);
\filldraw[fill=TUMBlauHell] (2.5,2) rectangle (3,3);
\filldraw[fill=TUMBlauHell] (2,2.5) rectangle (2.5,3);
\filldraw[fill=TUMBlauDunkel] (2,3) rectangle (3,4);
\filldraw[fill=TUMBlauDunkel] (3,2) rectangle (4,4);

\filldraw[fill=TUMBlau] (5,0) rectangle (7,2);
\filldraw[fill=TUMBlau] (7,2) rectangle (8,2.5);
\filldraw[fill=TUMBlau] (8,2) rectangle (9,3);
\filldraw[fill=TUMBlauHell] (7,0) rectangle (9,2);
\filldraw[fill=TUMBlauHell] (7,2.5) rectangle (7.5,3);
\filldraw[fill=TUMBlauHell] (7,3) rectangle (8,4);
\filldraw[fill=TUMBlauDunkel] (5,2) rectangle (7,4);
\filldraw[fill=TUMBlauDunkel] (7.5,2.5) rectangle (8,3);
\filldraw[fill=TUMBlauDunkel] (8,3) rectangle (9,4);

\filldraw[fill=TUMBlau] (0,5) rectangle (2,9);
\filldraw[fill=TUMBlau] (2,7) rectangle (3,7.5);
\filldraw[fill=TUMBlauHell] (2,7.5) rectangle (3,9);
\filldraw[fill=TUMBlauDunkel] (3,7) rectangle (4,9);
\filldraw[fill=TUMBlauDunkel] (2,5) rectangle (4,7);

\filldraw[fill=TUMBlau] (5,5) rectangle (7,7);
\filldraw[fill=TUMBlau] (7,7) rectangle (8,7.5);
\filldraw[fill=TUMBlau] (7,8) rectangle (8,9);
\filldraw[fill=TUMBlauHell] (5,7) rectangle (7,9);
\filldraw[fill=TUMBlauHell] (7.5,7.5) rectangle (8,8);
\filldraw[fill=TUMBlauHell] (8,8) rectangle (9,9);
\filldraw[fill=TUMBlauDunkel] (7,7.5) rectangle (7.5,8);
\filldraw[fill=TUMBlauDunkel] (8,7) rectangle (9,8);
\filldraw[fill=TUMBlauDunkel] (7,5) rectangle (9,7);

\draw[thick] (0,0) -- (4,0) -- (4,4) -- (0,4) -- (0,0);
\draw[thick] (2,0) -- (2,4);
\draw[thick] (0,2) -- (4,2);
\draw[thick] (3,2) -- (3,4);
\draw[thick] (2,3) -- (4,3);
\draw[thick] (2.5,2) -- (2.5,3);
\draw[thick] (2,2.5) -- (3,2.5);

\draw[thick] (5,0) -- (9,0) -- (9,4) -- (5,4) -- (5,0);
\draw[thick] (7,0) -- (7,4);
\draw[thick] (5,2) -- (9,2);
\draw[thick] (8,2) -- (8,4);
\draw[thick] (7,3) -- (9,3);
\draw[thick] (7.5,2) -- (7.5,3);
\draw[thick] (7,2.5) -- (8,2.5);

\draw[thick] (0,5) -- (4,5) -- (4,9) -- (0,9) -- (0,5);
\draw[thick] (2,5) -- (2,9);
\draw[thick] (0,7) -- (4,7);
\draw[thick] (3,7) -- (3,9);
\draw[thick] (2,8) -- (4,8);
\draw[thick] (2.5,7) -- (2.5,8);
\draw[thick] (2,7.5) -- (3,7.5);

\draw[thick] (5,5) -- (9,5) -- (9,9) -- (5,9) -- (5,5);
\draw[thick] (7,5) -- (7,9);
\draw[thick] (5,7) -- (9,7);
\draw[thick] (8,7) -- (8,9);
\draw[thick] (7,8) -- (9,8);
\draw[thick] (7.5,7) -- (7.5,8);
\draw[thick] (7,7.5) -- (8,7.5);

\draw node at (0.125,0.125) {\tiny 0}; \draw node at (2.125,0.125) {\tiny 0}; \draw node at (0.125,2.125) {\tiny 0}; \draw node at (2.125,2.125) {\tiny 0};
\draw node at (2.625,2.125) {\tiny 1}; \draw node at (2.625,2.625) {\tiny 1}; \draw node at (2.125,2.625) {\tiny 1};
\draw node at (3.125,2.125) {\tiny 2}; \draw node at (3.125,3.125) {\tiny 2}; \draw node at (2.125,3.125) {\tiny 2};

\draw node at (0.125,5.125) {\tiny 0}; \draw node at (0.125,7.125) {\tiny 0}; \draw node at (2.125,7.125) {\tiny 0}; \draw node at (2.625,7.125) {\tiny 0};
\draw node at (2.625,7.625) {\tiny 1}; \draw node at (2.125,7.625) {\tiny 1}; \draw node at (2.125,8.125) {\tiny 1};
\draw node at (3.125,7.125) {\tiny 2}; \draw node at (3.125,8.125) {\tiny 2}; \draw node at (2.125,5.125) {\tiny 2};

\draw node at (5.125,0.125) {\tiny 0}; \draw node at (8.125,2.125) {\tiny 0}; \draw node at (7.125,2.125) {\tiny 0}; \draw node at (7.625,2.125) {\tiny 0};
\draw node at (7.125,2.625) {\tiny 1}; \draw node at (7.125,0.125) {\tiny 1}; \draw node at (7.125,3.125) {\tiny 1};
\draw node at (5.125,2.125) {\tiny 2}; \draw node at (7.625,2.625) {\tiny 2}; \draw node at (8.125,3.125) {\tiny 2};

\draw node at (5.125,5.125) {\tiny 0}; \draw node at (7.125,7.125) {\tiny 0}; \draw node at (7.625,7.125) {\tiny 0}; \draw node at (7.125,8.125) {\tiny 0};
\draw node at (5.125,7.125) {\tiny 1}; \draw node at (7.625,7.625) {\tiny 1}; \draw node at (8.125,8.125) {\tiny 1};
\draw node at (7.125,7.625) {\tiny 2}; \draw node at (8.125,7.125) {\tiny 2}; \draw node at (7.125,5.125) {\tiny 2};

\draw[thick,TUMOrange] (1,1) -- (3,1) -- (1,3) -- (2.25,2.25) -- (2.75, 2.25) -- (2.25,2.75) -- (2.75, 2.75) -- (3.5, 2.5) -- (2.5, 3.5) -- (3.5, 3.5);
\draw[thick, TUMOrange] (6,1) -- (8,1) -- (6,3) -- (7.25,2.25) -- (7.75, 2.25) -- (7.25,2.75) -- (7.75, 2.75) -- (8.5, 2.5) -- (7.5, 3.5) -- (8.5, 3.5);

\draw[thick, TUMOrange] (1,6) -- (1,8) -- (2.25, 7.25) -- (2.75, 7.25) -- (2.75, 7.25) -- (2.75, 7.75) -- (2.25,7.75) -- (2.5, 8.5) -- (3.5, 8.5) -- (3.5,7.5) -- (3,6);
\draw[thick, TUMOrange] (6,6) -- (6,8) -- (7.25, 7.25) -- (7.75, 7.25) -- (7.75, 7.25) -- (7.75, 7.75) -- (7.25,7.75) -- (7.5, 8.5) -- (8.5, 8.5) -- (8.5,7.5) -- (8,6);

\node[align=center] at (-1.5,2) {Z-Curve};
\node[align=center] at (-1.5,7) {Hilbert-Curve};

\node[align=center] at (2,9.5) {Classical Partition};
\node[align=center] at (7,9.5) {Level-wise Partition};
\end{tikzpicture}
\end{adjustbox}
\end{center}
\caption{Domain partitioning via Hilbert- (top) or Z-curves (bottom) and their level-wise counterparts.}
\label{fig:LevelWiseSFC}
\end{figure}

We generate such partitions using a \ac{SFC} \cite{Bader2013}, but do not simply bisect the curve into equal parts to distribute the nodes to ranks. Instead, a list of leaves-per-level is generated and ordered according to the \ac{SFC}. These nodes in the list are then distributed to the available ranks. Thereby, an even load per level is obtained, while maintaining neighbor locality within a level. Exemplarily, \cref{fig:LevelWiseSFC}, illustrates the difference between the partitioning obtained using traditional Hilbert- \cite{Hilbert1891} or Z-curves \cite{Morton1966}, and our level-wise modifications. In the figure, the domain matches a \ac{2D} quadtree refined once everywhere and two more times in the upper quadrant. The color scale and the small number in each node indicates on which rank the node resides. The \ac{SFC} is illustrated which remains the same for the level-wise partitioning. In all cases, nodes are assigned to ranks by traversing the curve. On the right, also the level of the node is considered during the assignment to a rank. Clearly, the level-wise curve leads to a more scattered distribution, but in the traditional distribution whole ranks would idle for multiple micro time steps. For example, in the classical Z-curve distribution rank ``two'' holds only nodes at level $l_1$ and thus would idle for half of all micro time steps. The local regions within a level are also clearly visible in the figure. Once, the leaves are distributed as described, parents are recursively assigned to the rank where the majority of their children resides on.

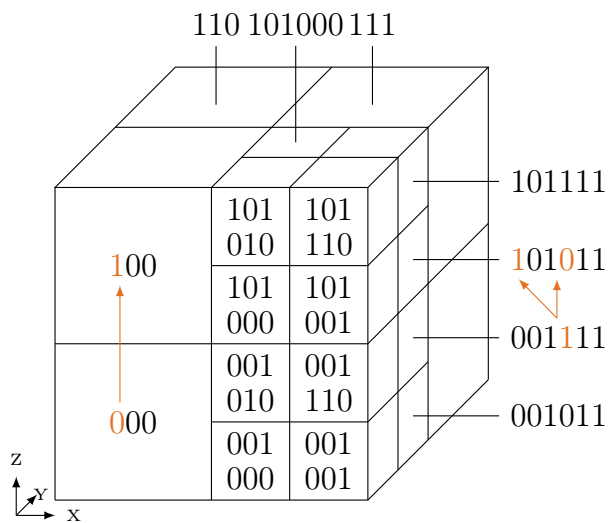
\begin{figure}
\begin{center}
\begin{adjustbox}{width=0.6\linewidth}
\begin{tikzpicture}
   \draw[-latex] (-0.15,-0.1,0) -- (-0.15,-0.1,-0.7) node {\tiny ~Y};
   \draw[-latex] (-0.15,-0.1,0) -- (-0.15,0.4,0) node[above] {\tiny Z};
   \draw[-latex] (-0.15,-0.1,0) -- (0.35,-0.1,0) node[right] {\tiny X};
   \draw (0.25,0,-0.25) -- (4.25,0,-0.25) -- (4.25,0,-4.25);
   \draw (0.25,4,-0.25) -- (4.25,4,-0.25) -- (4.25,4,-4.25) -- (0.25,4,-4.25) -- cycle;
   \draw (0.25,0,-0.25) -- (0.25,4,-0.25);
   \draw (4.25,0,-0.25) -- (4.25,4,-0.25);
   \draw (4.25,0,-4.25) -- (4.25,4,-4.25);
   \draw (2.25,0,-0.25) -- (2.25,4,-0.25) -- (2.25,4,-4.25);
   \draw (4.25,0,-2.25) -- (4.25,4,-2.25) -- (0.25,4,-2.25);
   \draw (0.25,2,-0.25) -- (4.25,2,-0.25) -- (4.25,2,-4.25);
   \draw (3.25,2,-0.25) -- (3.25,4,-0.25) -- (3.25,4,-2.25);
   \draw (4.25,2,-1.25) -- (4.25,4,-1.25) -- (2.25,4,-1.25);
   \draw (2.25,3,-0.25) -- (4.25,3,-0.25) -- (4.25,3,-2.25);

   \draw node at (1.25,1,-0.25) {{\color{TUMOrange}0}00};
   \draw node[align=left] at (2.75,0.5,-0.25) {001\\[-9pt]000};
   \draw node[align=left] at (3.75,0.5,-0.25) {001\\[-9pt]001};
   \draw node[align=left] at (2.75,1.5,-0.25) {001\\[-9pt]010};
   \draw node[align=left] at (3.75,1.5,-0.25) {001\\[-9pt]110};
   \draw node at (1.25,3,-0.25) {{\color{TUMOrange}1}00};
   \draw node[align=left] at (2.75,2.5,-0.25) {101\\[-9pt]000};
   \draw node[align=left] at (3.75,2.5,-0.25) {101\\[-9pt]001};
   \draw node[align=left] at (2.75,3.5,-0.25) {101\\[-9pt]010};
   \draw node[align=left] at (3.75,3.5,-0.25) {101\\[-9pt]110};

   \draw (1.25,4,-3) -- (1.25,4.75,-3) node[above] {110};
   \draw (3.25,4,-3) -- (3.25,4.75,-3) node[above] {111};
   \draw (2.75,4,-1.75) -- (2.75,5.235,-1.75) node[above] {101000};

   \draw (4.25,2.5,-1.75) -- (5.35,2.5,-1.75) node[right] {{\color{TUMOrange}1}01{\color{TUMOrange}0}11};
   \draw (4.25,3.5,-1.75) -- (5.35,3.5,-1.75) node[right] {101111};

   \draw (3.25,0,-0.25) -- (3.25,2,-0.25);
   \draw (4.25,0,-1.25) -- (4.25,2,-1.25);
   \draw (2.25,1,-0.25) -- (4.25,1,-0.25) -- (4.25,1,-2.25);
   \draw (4.25,0.5,-1.75) -- (5.35,0.5,-1.75) node[right] {001011};
   \draw (4.25,1.5,-1.75) -- (5.35,1.5,-1.75) node[right] {001{\color{TUMOrange}1}11};

   \draw[TUMOrange, -latex] (6.09,1.75,-1.75) -- (6.09,2.25,-1.75);
   \draw[TUMOrange, -latex] (6.09,1.75,-1.75) -- (5.6,2.25,-1.75);

   \draw[TUMOrange, -latex] (1.08,1.25,-0.25) -- (1.08,2.75,-0.25);
\end{tikzpicture}
\end{adjustbox}
\end{center}
\caption{Morton Ordering for a cube. On the left and the right side the neighbor-lookup bit flipping is illustrated in orange.}
\label{fig:NeigboorLookup}
\end{figure}

Halo cells adjacent to partition boarders are updated via communication. In order to identify communication partners and to store received data in the correct location, efficient neighbor lookups are necessary. For this purpose, a mapping which allows cheap identification of the neighboring nodes and fast rank-lookup is introduced. For octree structures, the Morton Order \cite{Morton1966} offers the required functionality. Originally, each node is assigned a unique id in form of a bitstream like

\begin{equation}
\mathtt{0...0}\underbrace{\mathtt{001}}_{\mathclap{l_2\text{-group}}}\overbrace{\mathtt{110}}^{\mathclap{z\text{- and }y\text{-offset in }l_1\text{-group}}}\underbrace{\mathtt{010}}_{\mathclap{l_0\text{-group}}}.
\label{eq:MortonBits}
\end{equation}
Note, this stream has the position on the respective level in the tree encoded. In \ac{3D}, each three-bit long group indicates its positioning along the $x$, $y$ and $z$-axis, respectively. As an example, the node at the origin of the coordinate axis on level $l_0$ is labeled $\mathtt{000},$ and the node with id $\mathtt{010}$ is positioned on top of the previous node in the $y$-axis without an offset in the other two directions. Parents of nodes are identified by simply shifting the id stream three bits to the right. Respectively, child ids are found by shifting the parent id three bits to the left and adding the bits of the new group to the shifted parent id. Furthermore, the respective bitstream $\mathtt{000}$ to $\mathtt{111}$ is addded depending on the $x$-,$y-$ and $z$-offset of the child within the new group. Neighbor lookups with similar simplicity are based on (concatenated) bitwise logic operations. We list the instructions to find the neighbor in positive direction along any axis as \cref{eq:NeigboorLookup}. In case the opposite neighbor on the same axis is to be found the last line changes to $f \gets \overbar{id_n[i]} \vee f$. This formulation can be seen as operating from right to left on every third bit of the stream thereby only changing the bits corresponding to one particular axis. If a one is encountered, the current and the next bits are flipped. Once a zero is flipped, the process terminates. An example is shown in \Cref{fig:NeigboorLookup}. Diagonal neighbor ids are determined by concatenating the instruction for the respective directions.

\begin{algorithm}
   \caption{Calculation of neighbor $id_n$ form $id$}
   \label{eq:NeigboorLookup}
   \begin{algorithmic}
      \State $N \gets$ id bit length.
      \State $d \gets$ 0,1, or 2 for $x$-,$y$- or $z$-axis, respectively.
      \State $id_n \gets id$
      \State $f \gets$ 0
      \For{ $i = d;\, i < N;\, i \mathrel{{+}{=}} 3$ }
         \State $id_n[i] \gets id_n[i]\, \overbar{\veebar}\, f$
         \State $f \gets id_n[i] \vee f$
      \EndFor
   \end{algorithmic}
\end{algorithm}

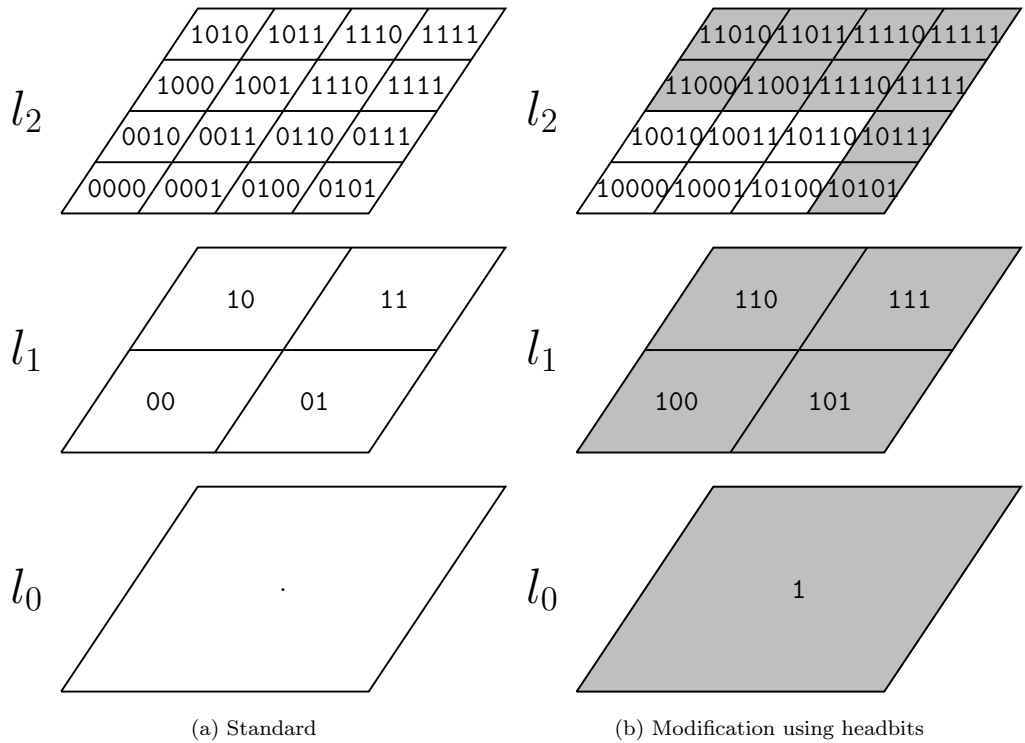
\begin{figure}
\begin{subfigure}[c]{0.49\textwidth}
\begin{center}
\begin{adjustbox}{width=\linewidth}
\begin{tikzpicture}
\draw (-0.5,1.5) node {\LARGE $l_0$};
\draw (-0.5,5.0) node {\LARGE $l_1$};
\draw (-0.5,8.5) node {\LARGE $l_2$};

\draw[thick] (0,7) -- (4.5,7) -- (6.5,10) -- (2.0,10) -- (0,7);
\draw[thick] (1.125,7) -- (3.125,10);
\draw[thick] (2.25,7) -- (4.25,10);
\draw[thick] (3.375,7) -- (5.375,10);
\draw[thick] (0.75*0.666667,7.75) -- (4.5+0.75*0.666667,7.75);
\draw[thick] (1.5*0.666667,8.5) -- (4.5+1.5*0.666667,8.5);
\draw[thick] (2.25*0.666667,9.25) -- (4.5+2.25*0.666667,9.25);

\draw (0.25*0.666667+0.6525,7.375) node {$\mathtt{0000}$};
\draw (0.25*0.666667+1.7775,7.375) node {$\mathtt{0001}$};
\draw (0.25*0.666667+2.9025,7.375) node {$\mathtt{0100}$};
\draw (0.25*0.666667+4.0275,7.375) node {$\mathtt{0101}$};

\draw (1.0*0.666667+0.6525,8.125) node {$\mathtt{0010}$};
\draw (1.0*0.666667+1.7775,8.125) node {$\mathtt{0011}$};
\draw (1.0*0.666667+2.9025,8.125) node {$\mathtt{0110}$};
\draw (1.0*0.666667+4.0275,8.125) node {$\mathtt{0111}$};

\draw (1.75*0.666667+0.6525,8.875) node {$\mathtt{1000}$};
\draw (1.75*0.666667+1.7775,8.875) node {$\mathtt{1001}$};
\draw (1.75*0.666667+2.9025,8.875) node {$\mathtt{1110}$};
\draw (1.75*0.666667+4.0275,8.875) node {$\mathtt{1111}$};

\draw (2.5*0.666667+0.6525,9.625) node {$\mathtt{1010}$};
\draw (2.5*0.666667+1.7775,9.625) node {$\mathtt{1011}$};
\draw (2.5*0.666667+2.9025,9.625) node {$\mathtt{1110}$};
\draw (2.5*0.666667+4.0275,9.625) node {$\mathtt{1111}$};

\draw[thick] (0,3.5) -- (4.5,3.5) -- (6.5,6.5) -- (2.0,6.5) -- (0,3.5);
\draw[thick] (2.25,3.5) -- (4.25,6.5); 
\draw[thick] (1,5) -- (5.5,5); 

\draw (0.5*0.666667+1.125,4.25) node {$\mathtt{00}$};
\draw (0.5*0.666667+3.375,4.25) node {$\mathtt{01}$};
\draw (2.25*0.666667+1.125,5.75) node {$\mathtt{10}$};
\draw (2.25*0.666667+3.375,5.75) node {$\mathtt{11}$};

\draw[thick] (0,0) -- (4.5,0) -- (6.5,3) -- (2.0,3) -- (0,0);
\draw (3.25,1.5) node {$\mathtt{.}$};

\end{tikzpicture}
\end{adjustbox}
\caption{Standard}
\label{fig:MortonOrderTreeLeft}
\end{center}
\end{subfigure}
\begin{subfigure}[c]{0.49\textwidth}
\begin{center}
\begin{adjustbox}{width=\linewidth}
\begin{tikzpicture}
\draw (-0.5,1.5) node {\LARGE $l_0$};
\draw (-0.5,5.0) node {\LARGE $l_1$};
\draw (-0.5,8.5) node {\LARGE $l_2$};

\filldraw[fill opacity = 0.5, fill=gray] (3.375,7) -- (4.5,7) -- (4.5+1.5*0.666667,8.5) -- (3.375+1.5*0.666667,8.5) -- (3.375,7);
\filldraw[fill opacity = 0.5, fill=gray] (1.5*0.666667,8.5) -- (4.5+1.5*0.666667,8.5) -- (6.5,10.0) -- (2.0,10.0) -- (1.5*0.666667,8.5);

\draw[thick] (0,7) -- (4.5,7) -- (6.5,10.0) -- (2.0,10.0) -- (0,7);
\draw[thick] (1.125,7) -- (3.125,10);
\draw[thick] (2.25,7) -- (4.25,10);
\draw[thick] (3.375,7) -- (5.375,10);
\draw[thick] (0.75*0.666667,7.75) -- (4.5+0.75*0.666667,7.75);
\draw[thick] (1.5*0.666667,8.5) -- (4.5+1.5*0.666667,8.5);
\draw[thick] (2.25*0.666667,9.25) -- (4.5+2.25*0.666667,9.25);

\draw (0.25*0.666667+0.6525,7.375) node {$\mathtt{10000}$};
\draw (0.25*0.666667+1.7775,7.375) node {$\mathtt{10001}$};
\draw (0.25*0.666667+2.9025,7.375) node {$\mathtt{10100}$};
\draw (0.25*0.666667+4.0275,7.375) node {$\mathtt{10101}$};

\draw (1.0*0.666667+0.6525,8.125) node {$\mathtt{10010}$};
\draw (1.0*0.666667+1.7775,8.125) node {$\mathtt{10011}$};
\draw (1.0*0.666667+2.9025,8.125) node {$\mathtt{10110}$};
\draw (1.0*0.666667+4.0275,8.125) node {$\mathtt{10111}$};

\draw (1.75*0.666667+0.6525,8.875) node {$\mathtt{11000}$};
\draw (1.75*0.666667+1.7775,8.875) node {$\mathtt{11001}$};
\draw (1.75*0.666667+2.9025,8.875) node {$\mathtt{11110}$};
\draw (1.75*0.666667+4.0275,8.875) node {$\mathtt{11111}$};

\draw (2.5*0.666667+0.6525,9.625) node {$\mathtt{11010}$};
\draw (2.5*0.666667+1.7775,9.625) node {$\mathtt{11011}$};
\draw (2.5*0.666667+2.9025,9.625) node {$\mathtt{11110}$};
\draw (2.5*0.666667+4.0275,9.625) node {$\mathtt{11111}$};

\filldraw[fill opacity = 0.5, fill=gray] (0,3.5) -- (4.5,3.5) -- (6.5,6.5) -- (2.0,6.5) -- (0,3.5);
\draw[thick] (0,3.5) -- (4.5,3.5) -- (6.5,6.5) -- (2.0,6.5) -- (0,3.5);
\draw[thick] (2.25,3.5) -- (4.25,6.5); 
\draw[thick] (1,5) -- (5.5,5); 

\draw (0.5*0.666667+1.125,4.25) node {$\mathtt{100}$};
\draw (0.5*0.666667+3.375,4.25) node {$\mathtt{101}$};
\draw (2.25*0.666667+1.125,5.75) node {$\mathtt{110}$};
\draw (2.25*0.666667+3.375,5.75) node {$\mathtt{111}$};

\filldraw[fill opacity = 0.5, fill=gray] (0,0) -- (4.5,0) -- (6.5,3) -- (2.0,3) -- (0,0);
\draw[thick] (0,0) -- (4.5,0) -- (6.5,3) -- (2.0,3) -- (0,0);
\draw (3.25,1.5) node {$\mathtt{1}$};
\end{tikzpicture}
\end{adjustbox}
\caption{Modification using headbits}
\label{fig:MortonOrderTreeRight}
\end{center}
\end{subfigure}
\caption{2D tree indexed by standard morton order (a) and our proposed modification (b). Shadow nodes are indicated in gray. Note the double assignment of the id ``zero'' on the left.}
\label{fig:MortonOrderTree}
\end{figure}

With the Morton order in-place, tree searches can be replaced by efficient hash-based lookups. This allows to split the 'light-weight' topology data and the 'heavy' computational data and map them by the Morton id. In case of inter-rank communication, the id-logic can be used to identify the communication partner and to tag asynchronous communication messages. Partitioning a mesh via Morton ordering, however, has its drawbacks. First, only cubic domains can be partitioned directly due to the embedded octree structure. Second, the id $\mathtt{000}$ is not unique unless the size of the stream is measured in addition. We overcome both limitations by introducing the headbit $\mathtt{1}$, which is added from the left to the normal Morton index

\begin{equation}
\mathtt{0..0}\underset{ \underset{\mathclap{\text{\scriptsize  headbit}}}{\text{\tiny |}} }{1}\underbrace{101 001 010 111}_{\text{as before}}.
\label{eq:AlpacaBits}
\end{equation}
With the headbit in place, every zero-id is uniquely defined. The neighbor lookups still follow \cref{eq:NeigboorLookup}, yet the headbit is simply removed before the computation and then re-added to the final result. The level of an id is obtained by shifting the id one level-group to the right until only the headbit remains. Now the count of performed shifts minus one gives the level number. Additionally, introducing the headbit generalizes the Morton order for arbitrarily tiled geometries like, \eg, channels or T-shapes. Such domains are embedded in a coarse zero-level octree with deactivated nodes. We call such deactivated nodes \textit{shadow nodes}. \Cref{fig:MortonOrderTree} shows a comparison of the two Morton ordering schemes for a \ac{2D} tree with three levels. As illustrated, using the headbit and two levels of shadow nodes allows to represent a channel with $6\times 2$ nodes. The shadow levels do not need to be computed or stored; not even their light-weight topology data. Instead, a forest of trees starting on the first non-shadow level can be used. In this work, we use fixed-size 64-bit long ids with seven shadow levels. This allows us to simulate channels with length ratios up to $128\times 1\times 1$ and 13 levels of refinement.

%% file: Inputs/algorithm_implementation.tex
\section{Algorithm and Implementation}
\label{sec:Algorithm}

We have developed a modular, thus flexible, \ac{OOP} framework in \CC 11 to solve the previously described block-based \ac{MR} algorithm. The core principle is separation of data containers and data manipulators. For simplicity, we restrict ourselves to blocks with the same amount of cubic cells per dimension, \ie $\Delta x = \Delta y = \Delta z$. The \texttt{Block} class is the heavy data container, holding buffers for conservative and prime states. Note, storing prime states proved to outperform re-computations from conservative states for our implementation. In addition, six buffers for the fluxes across each surface of a block are reserved to ensure conservation at resolution jumps \cite{Roussel2003}. For the chosen fifth-order accurate prediction operator \cref{eq:Prediction} we use a multiple of four cells as \ac{IC} in each block.

As proposed earlier, we split the topology and the compute heavy data. The topology is programmed as standard octree, which eases the execution of the \ac{SFC} partitioning algorithm. Using the modified Morton order in a hash-map, the heavy data is connected to its respective topology data. The heavy data is local to one \ac{MPI} rank. We employ template programming using the \ac{CRTP} to change the Riemann solver or the reconstruction stencil at compile time. Similarly, the number of \ac{IC} per block, the error norm \cref{eq:Thresholding}, the \ac{SFC} and the activation of source terms in \cref{eq:Euler} are specified at compile time. We applied design patterns like the proxy, the factory and the strategy pattern to change the initial and boundary conditions, the \ac{EOS}, the fluid parameters and the \ac{MR} threshold $\varepsilon_{l_m}$ at runtime.

Ideally, the \ac{MR} compression would be applied to the initial condition on a fully-refined grid. For the considered huge resolutions, however, this increases the needed compute memory manifold and contradicts employed fully-adaptive \ac{MR} compression. Hence, we instead build the initial topology bottom-up: the given initial condition is discretized level-by-level and \ac{MR} compression is applied successively after each level. Our initialization routine is listed in \cref{algo:Init}. During node refinement, we allocate the memory of all children on the rank of the parent. Remember, the thresholding norm is evaluated for whole blocks. From \cref{algo:Init} it is clear that no node on level $l_1$ is ever coarsened, \ie a homogeneous mesh is created if the user sets $l_{max} = l_{1}$. This is necessary, as refinements to higher levels are based on the details computed between $l_0$ and $l_1$, see \cref{eq:DetailFunction}.

\begin{algorithm}
\caption{Initialization}
\label{algo:Init}
\begin{algorithmic}
\State $l_{max} \gets$ user input
\State $\varepsilon_{l_m} \gets$ from user inputs $\varepsilon_{ref}$ and $l_{ref}$ according to \cref{eq:UserEpsilon,eq:LevelWiseEpsilon}
\State Initialize topology for level $l_0$ on all ranks
\State Allocate $l_0$-blocks on respective rank
\State Impose initial condition on all $l_0$-blocks
\For{$m = 1; l_m < l_{max} ; ++m$}
  \State Refine all blocks of $l_{m-1}$
  \State Impose initial condition on all $l_{m}$-blocks
  \If{$m > 1$}
    \For{ All nodes on $l_m$}
      \If{$\|d^l\| < \varepsilon_{l_m}$}
        \State Coarsen respective node
      \EndIf
    \EndFor
  \EndIf
  \State Apply load balancing to re-distribute nodes
\EndFor
\end{algorithmic}
\end{algorithm}

The compute loop is summarized in \cref{algo:ComputeLoop}. We define the ordered sets of levels $\mathbb{L}_{all} = \{l_{max}, \dotsc, l_0\}$ and $\mathbb{L}$, with the first holding all levels and the latter holding the \emph{active} levels (which change over time). The number of \ac{RK} stages $s$, the integration weights $b_i$ and the increments $k_i$ depend on the order of the used \ac{RK} scheme \cite{Kaiser2019}. The maximum time step on the finest level  is limited by the global \ac{CFL} criterion

\begin{equation}
\Delta t_{l_{max}} = k \cdot \frac{\Delta x}{ \sum\limits_i \| |v_i| + c \|_\infty},
\label{eq:Cfl}
\end{equation}
with the \ac{CFL} number $k \le 1$, the velocity in $i$-direction $v_i$ and the cell-local speed-of-sound $c$. Note, \cref{eq:Cfl} can be extended to consider source terms if present in \cref{eq:Euler}. For a \ac{RK} method of second-order, two sets of buffers for the conservatives variables are sufficient. For higher orders, however, a third permanent buffer needs to be included. The second averaging per \ac{RK}-stage and the integration in halo cells, \cf \cref{algo:Integration}, is attributed to the \ac{ALTS} scheme \cite{Kaiser2019}.

\begin{algorithm}
\caption{Compute Loop}
\label{algo:ComputeLoop}
\begin{algorithmic}
  \State Final simulation time: $t_{end} \gets$ user input
    \While{$t < t_{end}$}
    \State $\mathbb{L} = \mathbb{L}_{all}$
    \For{Micro time steps $r = 0;\; r < 2^{l_{max}};\; \mathrel{{+}{+}{r}}$}
      \ForAll{ $l_m \in \mathbb{L}_{all}$ }
        \State $\Delta t_{l_m} \gets$ according to \cref{eq:LtsTimestep,eq:Cfl}
      \EndFor
      \For{\ac{RK} stages $s = 0;\; s < s_{max};\; \mathrel{{+}{+}{s}}$}

        \ForAll{leaves on $l_m \in \mathbb{L}$}
          \State $f\left(u_{l_m}\right) = \Delta x_{l_m} F\left(u_{l_m}\right)$, according to \cref{eq:Euler}
        \EndFor

        \ForAll{nodes on $l_m \in \mathbb{L}\setminus{\{0\}}$ in descending order}
          \State $\mathcal{A}\left(f\left(u_{l_m}\right)\right) \rightarrow f\left(u_{l_{m-1}}\right)$
        \EndFor

        \ForAll{nodes on $l_m \in \mathbb{L}_{all}$}
          \If{resolution jump}
            \State Fill halo cells by $\mathcal{P}\left(F\left(u_{l_{m-1}}\right)\right) \rightarrow F\left(u_{l_{m}}\right)$
          \Else
            \State Fill halo cells with copies of neighbor values
          \EndIf
        \EndFor

        \State $\mathbb{L} = \left\{l_{max}, \dots, l_{max + 1 - \sum \left( (r+1) \veebar r \right)}\right\}$

        \State Integration according to \cref{algo:Integration}
      \algstore{computeloop}
\end{algorithmic}
\end{algorithm}

\begin{algorithm}
  \begin{algorithmic}
    \algrestore{computeloop}
      \ForAll{node on $l_m \in \mathbb{L}\setminus{\{ \min{ \left( \mathbb{L} \right) } \}}$}
          \State $\mathcal{A}\left(F\left(u_{l_m}\right)\right) \rightarrow F\left(u_{l_{m-1}}\right)$
        \EndFor

        \If{$s = s_{max} - 1$}
          \ForAll{leaves on $l_m \in \mathbb{L}$}
            \If{neigbor is on finer level}
              \State Adjust coarse cell values:
              \State $u_{l_{m-1}} = u_{l_{m-1}} - \Delta t_{l_m} F\left(u_{l_{m-1}}\right)  + \Delta t_{l_m} \mathcal{A}\left(F\left(u_{l_m}\right)\right)$
            \EndIf
          \EndFor
        \EndIf

        \ForAll{nodes on $l_m \in \mathbb{L}$}
          \If{!resolution jump}
            \State Fill halo cells with copies of neighbor values
          \EndIf
        \EndFor

        \State $\mathcal{A}\left(u_{l_m}\right) \rightarrow u_{{l_m}-1} \forall m \in \mathbb{L}\setminus{\{0\}}$

      \EndFor
      \State refine and coarse mesh according to \acs{MR} analysis \cref{eq:DetailFunction,eq:Thresholding,eq:LevelWiseEpsilon}
        \State re-distribute nodes according to load balancing strategy
        \State $t \mathrel{{+}{=}} \Delta t$

      \EndFor
      \State Write time step output (if desired)
    \EndWhile
  \end{algorithmic}
\end{algorithm}

\begin{algorithm}
  \caption{Integration according to Kaiser et al., 2019.}
  \label{algo:Integration}
  \begin{algorithmic}
    \ForAll{nodes on $l_m \in \mathbb{L}$}
    \State Integrate according to \acs{RK} scheme: $u_{l_m} = u + \Delta t_{l_m} \sum\limits_{i=0}^{i \leq s} b_i k_i$
    \If{resolution jump}
      \State Integrate in halo cells: $u_{l_m} = u + \Delta t_{l_m} \sum\limits_{i=0}^{i \leq s} b_i k_i$
    \EndIf
  \EndFor
  \end{algorithmic}
\end{algorithm}

%% file: Inputs/Benchmarking/benchmarking.tex
\section{Benchmarking}
\label{sec:Benchmarking}

We use an automated test suite to asses the physical correctness of our implementation. For demonstration, we show two such verifications in the following for a Sod shock tube and \acp{RTI} problem \cref{sec:RefinementConvergence} before we give evaluate the (parallel) numerical performance of our implementation. All these simulations were conducted on the CoolMUC2 Cluster at LRZ, consisting of 28-way Intel Xeon E5-2697 v3 codename ``Haswell'' based nodes with FDR14 Infiniband interconnect\footnote{https://doku.lrz.de/display/PUBLIC/CoolMUC-2}\footnote{https://www.lrz.de/services/compute/linux-cluster/overview}. We have varied the choice of the Riemann solver, the reconstruction stencil, the number of \ac{IC}, the \ac{MR} thresholding norm, and the \ac{EOS} to demonstrate the flexibility of our framework. We introduce a ``standard configuration'' which uses the approximate Riemann solver of Roe with \ac{WENO}-5 reconstruction stencil and $16^3$ \ac{IC} per block. The followings runtime settings are chosen for this configuration: the l-infinity norm is used for \ac{MR} thresholding with $l_{ref} = l_{max-1}$ and $\varepsilon_{ref} = 0.01$. The stiffened gas \ac{EOS} with $\gamma = 1.4$ and $B = 0$ defines the gas phase. Time is integrated with a \ac{CFL} number of 0.6 using a second-order \ac{RK} scheme. This standard configuration is used unless otherwise stated.

\input{./Inputs/Benchmarking/convergance.tex}
\input{./Inputs/Benchmarking/rti_growth_rates.tex}
\input{./Inputs/Benchmarking/single_core_performance.tex}
\input{./Inputs/Benchmarking/parallel_perfromance.tex}
\input{./Inputs/Benchmarking/compression.tex}

%% file: Inputs/Benchmarking/convergance.tex
\subsection{Sod shock tube}
\label{sec:RefinementConvergence}

\begin{figure}
   \begin{subfigure}[c]{0.49\textwidth}
      \begin{center}
         \begin{adjustbox}{width=\linewidth}
            \begin{tikzpicture}
               \begin{axis}[xlabel = {$x_{~}$}, ylabel = {Prime state value}, cycle list name = TUMcyclelistNomark]
                  \addplot+[thick] table [x=XCoordinate, y=Density, col sep=comma, mark=none]  {./ExtractedData/sod_single_line_l5.csv};
                  \addplot+[thick] table [x=XCoordinate, y=Pressure, col sep=comma, mark=none] {./ExtractedData/sod_single_line_l5.csv};
                  \addplot+[thick] table [x=XCoordinate, y=Velocity, col sep=comma, mark=none] {./ExtractedData/sod_single_line_l5.csv};
                  \legend{$\rho$,$p$,$v$}
               \end{axis}
            \end{tikzpicture}
         \end{adjustbox}
         \caption{}
         \label{fig:SodConverganceLeft}
      \end{center}
   \end{subfigure}
   \begin{subfigure}[c]{0.49\textwidth}
      \begin{center}
         \begin{adjustbox}{width=\linewidth}
            \begin{tikzpicture}
               \begin{axis}[ymode=log, cycle list name = TUMcyclelist, xlabel = {$l_{max}$}, ylabel = {Error}]
                  \addplot[thick, dashed] table [x=level, y=ideal, col sep=comma, forget plot] {./ExtractedData/sodnorms.csv};
                  \addplot+[thick] table [x=level, y=l1-abs-density, col sep=comma] {./ExtractedData/sodnorms.csv};
                  \addplot+[thick] table [x=level, y=l1-abs-velocity, col sep=comma] {./ExtractedData/sodnorms.csv};
                  \legend{$\| \rho \|_1$, $\| \mathbf{v} \|_1$}
               \end{axis}
            \end{tikzpicture}
         \end{adjustbox}
         \caption{}
         \label{fig:SodConverganceRight}
      \end{center}
   \end{subfigure}
   \caption{Prime state profiles of the Sod shock tube at $ t - 0.2$ (a) and corresponding error analysis (right).}
\label{fig:Convergence}
\end{figure}
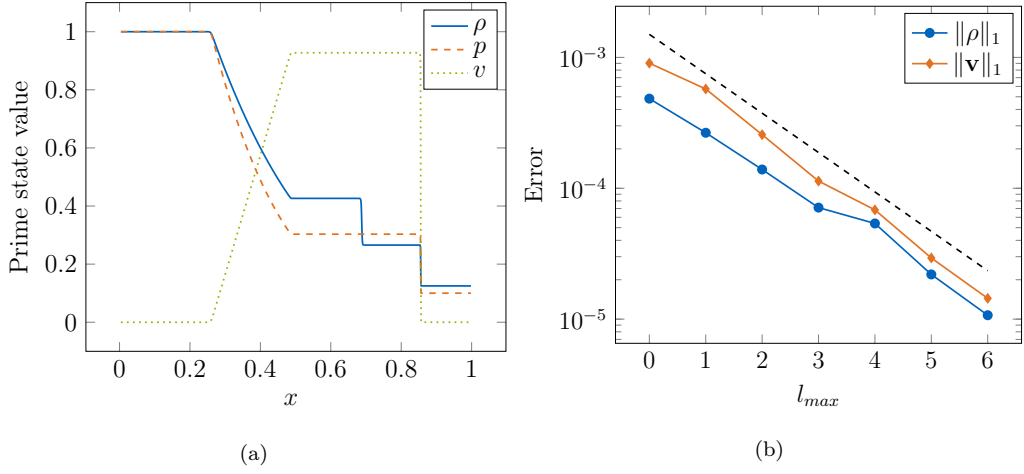

We asses the correctness of the implemented algorithm by studying the well-known one-dimensional Sod shock tube problem \cite{Sod1978}. The Sod test case is characterized by evolution of three distinct waves along the $x$-axis. Therefore, an initial discontinuity with left $\left(\rho = 1.0, \mathbf{v} = 0, p = 1.0 \right)_{x \leq 0.5}$ and right states $\left(\rho = 0.25, \mathbf{v} = 0, p = 0.125 \right)_{x > 0.5}$, defines a Riemann problem in the computational domain of size $[0, 1.0] \times [0, 0.25] \times [0, 0.25]$. The final state at $t_{end} = 0.2$ is shown in \Cref{fig:SodConverganceLeft}. We simulate this problem using $l_{max} = l_i \forall i \in [0,6]$ while keeping $l_{ref} = 4$ and $\varepsilon_{ref} = 0.01$ fixed. Thus, the resolution varies from 64 to 4096 (effective) cells in the coarsest and finest setup, respectively. We ensure that no spurious $y$- or $z$-momenta occur and compare our results against the known analytic solution. In \Cref{fig:SodConverganceRight}, we plot the $l_1$ error-norm for density and velocity. The observed second-order convergence agrees with the employed order of the time-integration scheme. For a more detailed convergence study we refer to our previous work \cite{Kaiser2019}.

%% file: Inputs/Benchmarking/rti_growth_rates.tex
\subsection{\acl{RTI}}

In the following section we analyze the \ac{RTI}. The source-term implementation of the governing equation is validated with the analysis of the growth-rate prediction.

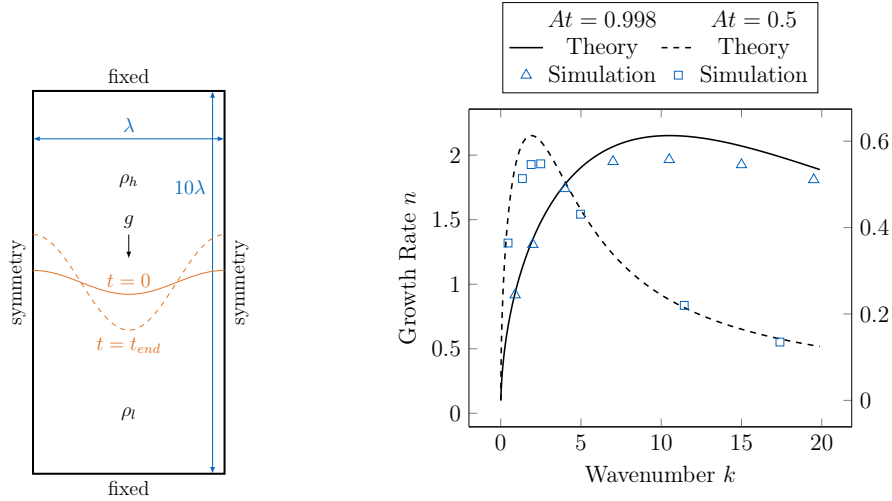
\begin{figure}
   \begin{subfigure}[c]{0.49\textwidth}
   \begin{center}
      \begin{adjustbox}{width=0.5\linewidth}
         \begin{tikzpicture}
            \draw[very thick] (0.0, 0.0) -- node[anchor=north,midway] {fixed} (4.0, 0.0) -- node[anchor=north,midway,rotate=90] {symmetry} (4.0, 8.0)
                              -- node[anchor=south,midway] {fixed} (0.0, 8.0) -- node[anchor=south,midway,rotate=90] {symmetry} cycle;
            \draw[TUMBlau,>=latex, <->,] (0.0,7.0) --node[anchor=south,midway] {$\lambda$} (4.0,7.0);
            \draw[TUMBlau,>=latex, <->,] (3.75,0.0) --node[anchor=east,near end] {$10\lambda$} (3.75,8.0);
            \draw[TUMOrange,domain=0:4, samples=50] plot (\x, {4+0.25*cos(0.5*pi*\x r)});
            \draw[TUMOrange,dashed, domain=0:4, samples=50] plot (\x, {4+cos(0.5*pi*\x r)});
            \draw[TUMOrange] node[anchor=south] at (2.0,3.75) {$t = 0$};
            \draw[TUMOrange] node[anchor=north] at (2.0,3.0) {$t = t_{end}$};
            \draw node at (2.0,1.25) {$\rho_l$};
            \draw node at (2.0,6.125) {$\rho_h$};
            \draw[-latex] (2.0, 5.0) node[anchor=south] {$g$} -- (2.0, 4.5);
         \end{tikzpicture}
      \end{adjustbox}
      \caption{Generic case setup used in the growth-rate study.}
      \label{fig:RtiGrowthRatesLeft}
      \end{center}
   \end{subfigure}
   \begin{subfigure}[c]{0.49\textwidth}
      \begin{center}
         \begin{adjustbox}{width=\linewidth}
            \begin{tikzpicture}
               \begin{axis}[axis y line* = right, xtick = \empty]
                  \addplot[thick, dashed, black, no marks] table [col sep=comma] {./ExtractedData/atwood05g1mu01analytic.csv};
                  \addplot[TUMBlau,only marks, mark=square] table [x=wavenumber,   y=fixed, col sep=comma] {./ExtractedData/atwood_05_results.csv};
               \end{axis}
               \begin{axis}[axis y line* = left, legend style={ at={(axis description cs:0.5,1.05)}, anchor=south}, legend columns = 2, xlabel = {Wavenumber  $k$},ylabel = {Growth Rate $n$}]
                  \addlegendimage{empty legend}
                  \addlegendimage{empty legend}
                  \addplot[thick, black, no marks] table [col sep=comma] {./ExtractedData/atwood0998g1mu05analytic.csv};
                  \addlegendimage{thick, dashed, no markers, black}
                  \addplot[TUMBlau,only marks, mark=triangle, mark options = {scale=1.5}] table [x=wavenumber, y=fixed, col sep=comma] {./ExtractedData/atwood_0998_results.csv};
                  \addlegendimage{TUMBlau, mark=square, only marks}
                  \legend{$At = 0.998$, $At = 0.5$, Theory, Theory, Simulation, Simulation}
               \end{axis}
            \end{tikzpicture}
         \end{adjustbox}
         \caption{Growth rate over the wavenumber of the initial disturbance. The analytic curve is obtained by solving \cref{eq:SinglePhaseGrowth} for each $k = 0, 0.01, \dots, 20.0$.}
         \label{fig:RtiGrowthRatesRight}
   \end{center}
   \end{subfigure}
   \caption{Case setup for the Rayleigh-Taylor instability (left) and the obtained growth rates in the linear regime (right).}
   \label{fig:RtiGrowthRates}
\end{figure}

A heavy fluid with density $\rho_h$ is placed on top of a lighter one with density $\rho_l$. If the interface between the fluids is disturbed, a mushroom-like instability can develop under the action of gravity. It is known \cite{Harrison1908,Chandrasekhar1961} that the growth rate $n$ depends on the wavenumber $k$ of the disturbance. In the linear regime, $n$ fulfills

\begin{multline}
   - \left\{ \frac{gk}{n^2} \left[ \left( \alpha_l - \alpha_h \right) + \frac{k^2T}{g\left( \rho_l + \rho_h \right)} \right] + 1 \right\} \left( \alpha_h q_l + \alpha_l q_h - k \right) - 4k\alpha_l\alpha_h \\
   + \frac{4k^2}{n} \left( \alpha_l \nu_l - \alpha_h \nu_h \right) \left\{ \left( \alpha_h q_l - \alpha_l q_h \right) + k \left( \alpha_l - \alpha_h \right) \right\} \\
   + \frac{4k^3}{n^2} \left( \alpha_l \nu_l - \alpha_h \nu_h \right)^2 \left( q_l - k \right) \left( q_h - k\right) = 0.
   \label{eq:RtiGrowth}
\end{multline}
Therein, $q_i = \sqrt{k^2 + \frac{n}{\nu_i}}$ and $\alpha_i = \frac{\rho_i}{\rho_l + \rho_h}$. Here, the kinematic viscosity of each fluid is denoted by $\nu_i$, gravity by $g$ and surface tension by $T$. We follow the setup of Shi et al. \cite{Shi2003}, hence $T = 0$ and $\mu_l = \rho_l \nu_l = \rho_h \nu_h = \mu_h$. Together with $\alpha_l \nu_l = \frac{\rho_l \nu_l}{\rho_l + \rho_h} = \frac{\rho_h \nu_h}{\rho_l + \rho_h} = \alpha_h \nu_h$  \cref{eq:RtiGrowth} reduces to

\begin{equation}
- \left\{ \frac{gk}{n^2} \left( \alpha_l - \alpha_h \right) + 1 \right\} \left( \alpha_h q_l + \alpha_l q_h - k \right) - 4k\alpha_l\alpha_h = 0.
\label{eq:SinglePhaseGrowth}
\end{equation}
We varied the wavenumber $k \in [0, 20]$ for two Atwood numbers $At = \alpha_h - \alpha_l = 0.5$ and $At = 0.998$, which resemble a gas-gas or liquid-gas system, respectively. The detailed setting for each configuration is given in \cref{tab:RtiInputTables}.
\begin{table}
   \centering
   \caption{Fluid properties for both Atwood numbers used in the \acs{RTI} growth-rate study.}
   \label{tab:RtiInputTables}
   \begin{tabular}{l|c|c}
      & $At = 0.5$ & $At = 0.998$ \\
      \hline
      $\rho_h$   & 1.5 & 99.9 \\
      $\rho_l$   & 0.5 & 0.1 \\
      $\mu$      & 0.1 & 0.5 \\
      $g$       & 1.0 & 1.0 \\
      $l_{max}$ &  $l_3$  &  $l_3$
   \end{tabular}
\end{table}
For all simulations, we adjust the domain size such that the short side spans one respective wavelength $\lambda = 2\pi / k$ while keeping a fixed aspect ratio of 1 to 10 between the short and the long axis. We illustrated the setup in \Cref{fig:RtiGrowthRatesLeft}. This yields an effective resolution of 1280 cells for the long side. We disturb the interface initially with a deflection of $\lambda$ and an amplitude of $0.05 / k$. We apply fixed-value boundary conditions at the top and bottom and symmetry conditions otherwise. We monitor the interface displacement $d(t)$ along the vertical centerline, where the position of the maximum density gradient denotes the apparent interface position. The growth rate $n$ thus becomes

\begin{equation}
   n = \frac{\partial}{\partial t} \ln{ \frac{d(t)}{d(0)} }.
   \label{eq:NumericGrowthRate}
\end{equation}
\Cref{fig:RtiGrowthRatesRight} shows a comparison of the growth rates obtained from our simulations and the analytical results from \cref{eq:SinglePhaseGrowth} for both Atwood numbers.

We see that our framework predicts the growth rates very well, which verifies the correctness of the implementation of the viscous source terms and body force contributions.

%% file: Inputs/Benchmarking/single_core_performance.tex
\subsection{Single-core performance}
\label{sec:SingleCore}

The single-core performance is analyzed for the Sod shock tube problem in a cubic box of size $[0,1]^3$. We run this test case for ten macro time steps and profile the run with Intel Advisor 2018 Update 4. The test runs take 0.78\si{\second} wall-clock time. We found that about 91.2\% of the runtime is spent in the approximate Riemann solver, which also accounts for 91.4\% of the overall \ac{FLOP}. The compute-kernel \ie the main compute-loop within the Riemann solver runs at 13.71\si{\giga\flop\per\second} and an arithmetic intensity of 0.22\si{\flop\per\byte}.

The analysis with Advisor revealed that auto-vectorization by the compiler results only in \ac{SSE} parallelization. Hence, we manually enabled \ac{AVX} instructions using the \texttt{-xCORE-AVX2} compiler flag. With the higher instruction set, the runtime reduces to 0.68s (-13\%) and now the main compute-loop accounts for 79\% of the overall runtime. The performance of the main loop increased to 22.82 \si{\giga\flop\per\second}, obviously still at 0.22 \si{\flop\per\byte}.

\begin{figure}
\begin{center}
\begin{subfigure}[c]{0.49\textwidth}
\begin{adjustbox}{width=1.0\linewidth}
\includegraphics{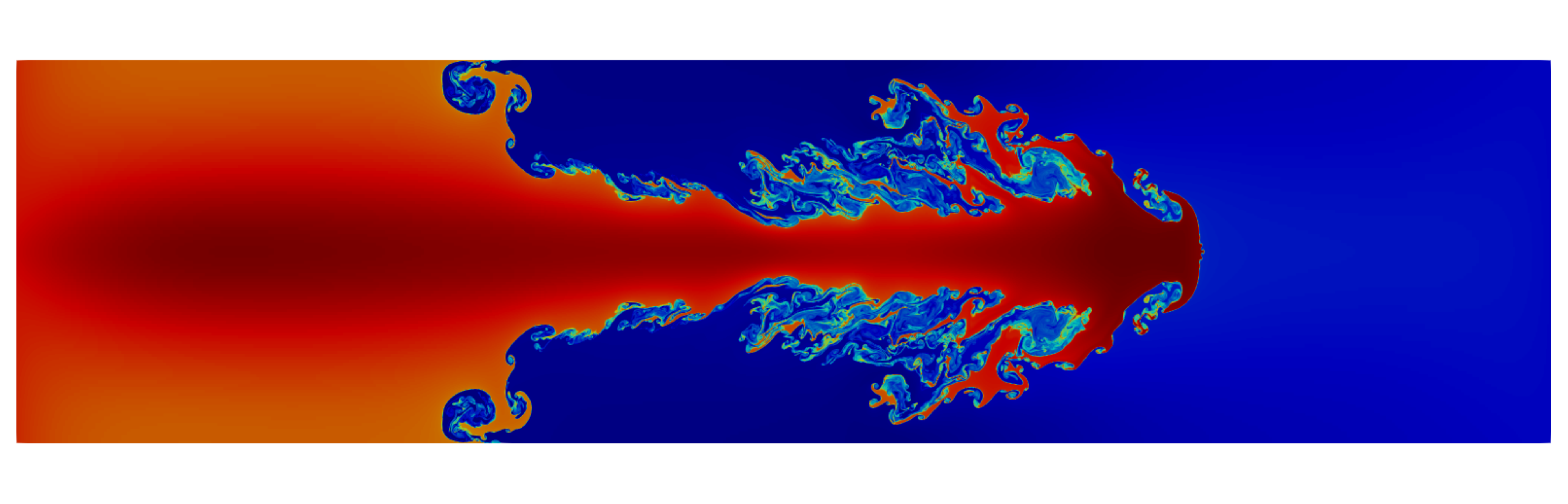}
\end{adjustbox}
\end{subfigure}
\begin{subfigure}[c]{0.49\textwidth}
\begin{adjustbox}{width=1.0\linewidth}
\includegraphics{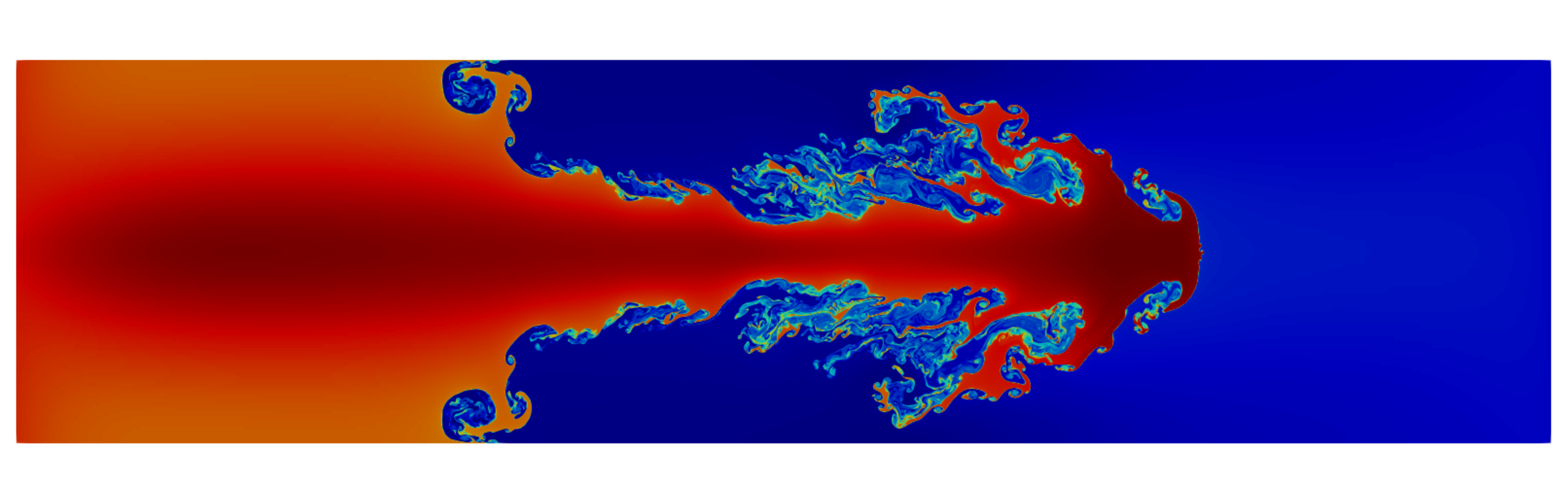}
\end{adjustbox}
\end{subfigure}
\end{center}
\caption{Influence of vectorization on the symmetry of a \acs{RTI} simulation. In the standard configuration (left) the initial symmetry prevails, whereas for the configuration compiled with \acs{AVX} instructions (right) it does not.}
\label{fig:CompilerFlagRti}
\end{figure}

However, it is important to note that the \ac{AVX} vectorization changes the truncation error of floating-point operations, \eg by changing the summation order. For our studies of hyperbolic problems with low-dissipation high-resolution schemes, this can strongly affect the simulation results \cite{Fleischmann2019}. This effect can be visualized using an inviscid \ac{RTI} setup with gravity $g = 1$ in y-direction, see \cite{Shi2003}. The physical simulation time is $t_{end} = 1.95$ using a \ac{CFL} number of 0.6. The effective resolution along the y-axis is set to 8192 cells with $l_{max} = 6$. \Cref{fig:CompilerFlagRti} shows the final time step of this test case with and without the \texttt{-xCORE-AVX2} compiler flag. Although we ensured exact symmetry in the initial condition according to \cite{Fleischmann2019}, we see nonphysical symmetry breaking for the \ac{AVX}-instructions enabled case.

Aiming for exact reproducibility avoiding numerical artifacts as mus as possible, here we sacrifice this performance optimization for the remainder of this work. Note, as shown in \cite{Hoppe2019} the scaling behavior is anyway unaffected by node-level \ac{SIMD} performance.

%% file: Inputs/Benchmarking/parallel_perfromance.tex
\subsection{Parallel Performance}

We evaluate the parallel performance for different numerical setups with both strong and weak scaling analyses. Therein, hyper-threading is disabled and we pin one \ac{MPI} rank to one physical core. Note, we always pin an \ac{MPI} rank to one physical core,. and, hence, use these terms interchangeably. At first we look at the performance implications of choosing the more commonly changed Riemann solvers and reconstruction stencils. Later, we also analyze the effects of more edge-case changes on the parallel performance such as changing the \ac{EOS}, the \ac{SFC} or the number of internal cells. We test the latter, configurations in smaller scaling setups using a maximum of eight
nodes.

Throughout this section we report averages over three runs for each configuration. Therein, we measure the runtimes using \texttt{MPI\_Wtime} and neglect time needed for initialization or input/output routines. The standard deviation for all runs was less than 3\%. The purpose of this systematic performance analysis for numerical setup permutations is to demonstrate both the flexibility of the presented software framework and the general code efficiency without specific single-purpose tuning.

\input{./Inputs/Benchmarking/ParallelPerformance/weak_scaling.tex}

\input{./Inputs/Benchmarking/ParallelPerformance/strong_scaling.tex}

\input{./Inputs/Benchmarking/ParallelPerformance/worst_case.tex}

\input{./Inputs/Benchmarking/ParallelPerformance/space_filling_curve.tex}

\input{./Inputs/Benchmarking/ParallelPerformance/equation_of_state.tex}

\input{./Inputs/Benchmarking/ParallelPerformance/error_norms.tex}

\input{./Inputs/Benchmarking/ParallelPerformance/cells.tex}

%% file: Inputs/Benchmarking/ParallelPerformance/weak_scaling.tex
\subsection{Weak Scaling}
\label{sec:Weakscale}

\begin{figure}
\begin{center}
\begin{adjustbox}{width=0.7\linewidth}
\begin{tikzpicture}
\begin{axis}[axis y line* = right, axis x line = none , ylabel = {Pressure}, cycle list name = TUMcyclelistNomark, cycle list shift = 2]
\addplot+[thick] table [x=Ycoordinates, y=Pressure, col sep=comma] {./ExtractedData/weak_final_extract.csv};
\end{axis}
\begin{axis}[axis y line* = left, xlabel = {$y$}, ylabel = {Density/Velocity}, cycle list name = TUMcyclelistNomark, legend pos = north west]
\addplot+[thick] table [x=Ycoordinates, y=Density, col sep=comma] {./ExtractedData/weak_final_extract.csv};
\addplot+[thick] table [x=Ycoordinates, y=Velocity, col sep=comma] {./ExtractedData/weak_final_extract.csv};
\addplot+[thick, draw=none] coordinates {(0.5,0.0)};
\legend{Density,Velocity, Pressure}
\end{axis}
\end{tikzpicture}
\end{adjustbox}
\end{center}
\caption{Density, pressure and velocity distribution along the $x$-axis for the weak scaling case at $t_{end} = 0.38$ using the standard configuration for the resolution described above.}
\label{fig:WeakStates}
\end{figure}
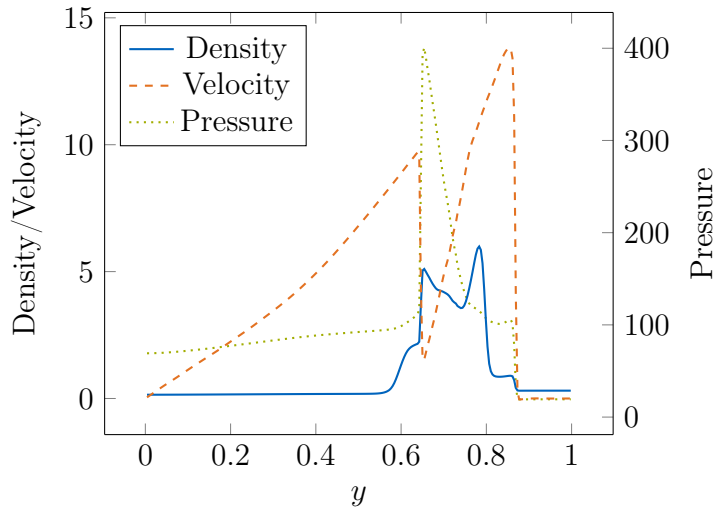

We conduct a weak scaling analysis to measure the influence of changing the Riemann solver or reconstruction stencil on the parallel performance of our implementation. We use the quasi-one-dimensional blast wave case \cite{Woodward1984} in $y$-direction as tests case and extend the domain in the other two dimensions linearly with the core count. We set $l_{max} = 3$, with an effective resolution of $128 \times 512 \times 128$ cells in the one-core configuration. The test case is typically ran until $t_{end} = 0.38$. The final states at this point using the standard configuration are depicted in \cref{fig:WeakStates}.

For our scaling runs we, only compute the first 25 macro time steps. We set $CFL = 0.9$, which allows to cover a maximum of physical time in these 25 steps, and, in turn give a more realistic numerical load as more mesh adaptations are stimulated. Using multiple macro time steps allows to take changing load due to re-meshing into account which is ubiquitous in real-world physical simulations. We performed the scaling runs using four different setups. The standard configuration, a configuration where only the reconstruction stencil is changed from \ac{WENO}5 to \ac{TENO}5, one where only the Riemann solver is changed to \ac{LLF}, and one where both were changed to \ac{HLLC} and \ac{WENO}-WAO53, respectively.

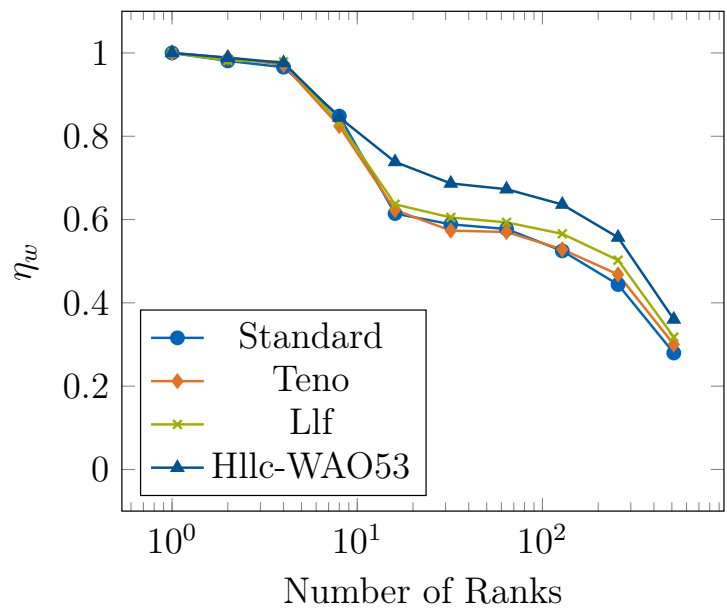
\begin{figure}
\begin{center}
\begin{adjustbox}{width=0.7\linewidth}
\begin{tikzpicture}
\begin{axis}[xmode = log, cycle list name = TUMcyclelist, ymin = -0.1, ymax = 1.1, xlabel = {Number of Ranks}, ylabel = {$\eta_w$}, legend pos = south west]
\addplot+[thick] table [x=Ranks, y=Efficiency-Std, col sep=comma] {./ExtractedData/weakscale.csv};
\addplot+[thick] table [x=Ranks, y=Efficiency-Teno, col sep=comma] {./ExtractedData/weakscale.csv};
\addplot+[thick] table [x=Ranks, y=Efficiency-Llf, col sep=comma] {./ExtractedData/weakscale.csv};
\addplot+[thick] table [x=Ranks, y=Efficiency-Hllc-W53, col sep=comma] {./ExtractedData/weakscale.csv};
\legend{Standard, Teno, Llf, Hllc-WAO53}
\end{axis}
\end{tikzpicture}
\end{adjustbox}
\end{center}
\caption{Weak scaling efficiency for different configurations of Riemann solvers, reconstruction stencils and internal cells per block.}
\label{fig:WeakEfficency}
\end{figure}

All configurations are run with 1 to 512 cores. The overall runtimes of the single-core configurations were comparable, except for the \ac{HLLC} one which ran about 30\% slower compared to the standard configuration. Note however, that for 512 cores this penalty decreased to only 1\% as the overall scaling is best in this configuration. The recorded efficiency is plotted in \cref{fig:WeakEfficency}. The scaling behaviors of all configurations are nearly identical with a steep drop in the efficiency when going from 2 to 16 cores. Nevertheless, all configurations show an efficiency plateau from 16 to 256 core, which indicating almost ideal scaling. For all rank numbers \ac{HLLC}-WAO53 shows the best efficiency. Going above 256 cores we observe a clear drop below an efficiency of 50\%.

We see, that modular setup of our solver enabled us to obtain comparable runtimes and weak scaling behaviors across different compute kernels. Nevertheless, configurations with larger compute to communication ratios yield higher efficiencies. So does the more complex reconstruction stencil in the \ac{HLLC}-WAO53 configuration allows to overlap computation and communication more effectively. Furthermore, we see, a decrease in performance when scaling within one node, \ie when less than 32 cores are used, but the efficiency remains almost constant when scaling across nodes. We conclude that our \ac{MPI}-only implementation handles inter-node parallelization more efficiently than intra-node parallelization.

%% file: Inputs/Benchmarking/ParallelPerformance/strong_scaling.tex
\subsection{Strong Scaling}
\label{sec:Strongscale}

\begin{figure}
\begin{center}
\begin{adjustbox}{width=0.7\linewidth}
\includegraphics[width=\linewidth]{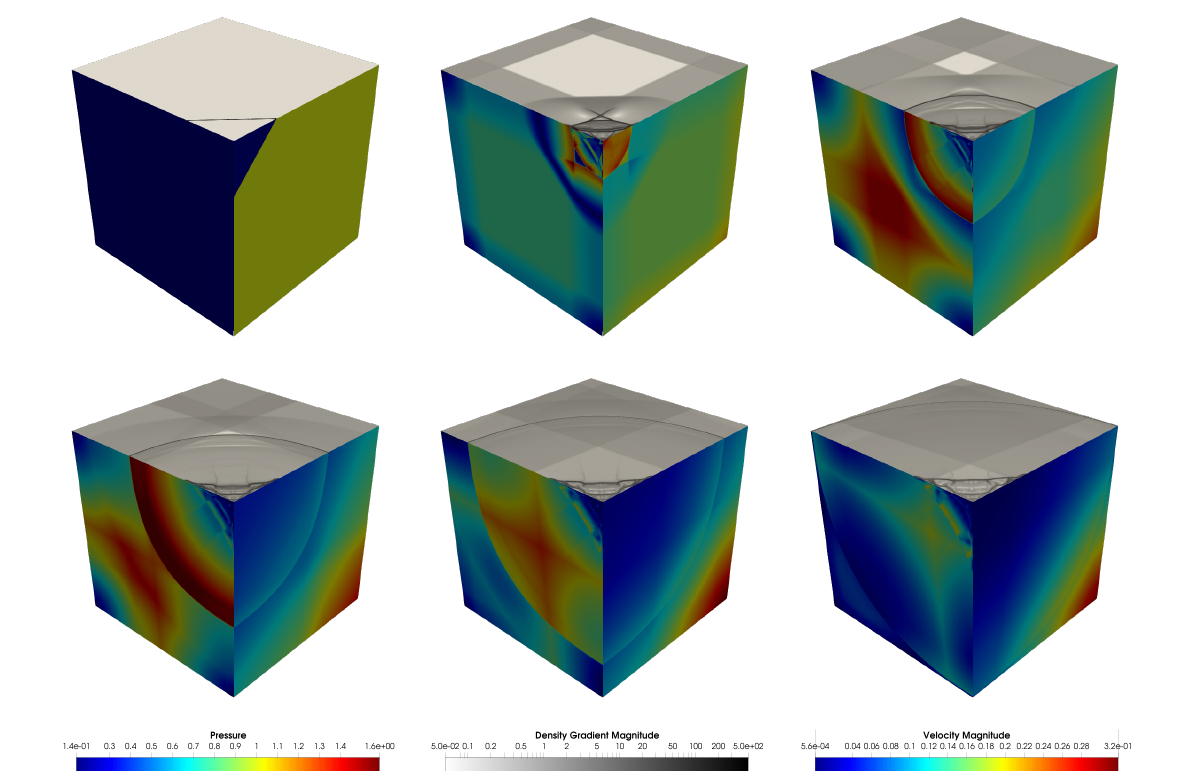}
\end{adjustbox}
\end{center}
\caption{Temporal evolution of the strong-scaling implosion case. Form left to right and top to bottom the images show give the time from $t = 0.0$ to $t_{end} = 0.5$ in steps of $0.1$. Each depicted cube shows the numerical schlieren, the velocity magnitude and the pressure on the top, left and right face, respectively. The last row gives the used color scales.}
\label{fig:StrongVisualization}
\end{figure}

Here, we analyse the strong scaling performance for the four numerical configurations used in the weak scaling above. Now, we use a variant of the implosion case of \cite{Liska2003} by extending the problem to \ac{3D}, giving us a domain of $[0,0.6]^3$ with symmetry boundary conditions on all sides. The initial discontinuity is placed at $x + y + z = 0.15$. Secondly, we model the fluid to be viscous with $\mu = 1.0e-6$, and impose a body force acceleration $g = \left( 1, 1, 1 \right)$. The temporal evolution of the pressure, velocity and density field is shown in \cref{fig:StrongVisualization}.

\begin{figure}
\begin{subfigure}[c]{0.49\textwidth}
\begin{center}
\begin{adjustbox}{width=\linewidth}
\begin{tikzpicture}
\begin{axis}[xmode = log, ymode = log, cycle list name = TUMcyclelist, xlabel = {Number of Ranks}, ylabel = {Speedup}, legend pos = north west]
\addplot[dashed] table [x=Ranks-Small, y=Ranks-Small, col sep=comma, forget plot] {./ExtractedData/strongscale_small.csv};
\addplot+[thick] table [x=Ranks-Small, y=Speedup-Std, col sep=comma] {./ExtractedData/strongscale_small.csv};
\addplot+[thick] table [x=Ranks-Small, y=Speedup-Teno, col sep=comma] {./ExtractedData/strongscale_small.csv};
\addplot+[thick] table [x=Ranks-Small, y=Speedup-Llf, col sep=comma] {./ExtractedData/strongscale_small.csv};
\addplot+[thick] table [x=Ranks-Small, y=Speedup-Hllc-WAO53, col sep=comma] {./ExtractedData/strongscale_small.csv};
\addplot[very thick, dotted, gray, no marks] table [x=Ranks-Small, y=Amdahl-998, col sep=comma] {./ExtractedData/strongscale_small.csv};
\legend{Standard, Teno, Llf, Hllc-WAO53, Amdahl}
\end{axis}
\end{tikzpicture}
\end{adjustbox}
\end{center}
\end{subfigure}
\begin{subfigure}[c]{0.49\textwidth}
\begin{center}
\begin{adjustbox}{width=\linewidth}
\begin{tikzpicture}
\begin{axis}[xmode = log, ymode = log, cycle list name = TUMcyclelist, xlabel = {Number of Ranks}, ylabel = {Speedup}, legend pos = north west]
\addplot[dashed] table [x=Ranks, y=Ideal, col sep=comma, forget plot] {./ExtractedData/strongscale_large.csv};
\addplot+[thick] table [x=Ranks, y=Speedup-Std, col sep=comma] {./ExtractedData/strongscale_large.csv};
\addplot+[thick] table [x=Ranks, y=Speedup-Teno, col sep=comma] {./ExtractedData/strongscale_large.csv};
\addplot+[thick] table [x=Ranks, y=Speedup-Llf, col sep=comma] {./ExtractedData/strongscale_large.csv};
\addplot+[thick] table [x=Ranks, y=Speedup-Hllc-WAO53, col sep=comma] {./ExtractedData/strongscale_large.csv};
\addplot[very thick, dotted, gray, no marks] table [x=Ranks, y=Amdahl-998, col sep=comma] {./ExtractedData/strongscale_large.csv};
\legend{Standard, Teno, Llf, Hllc-WAO53, Amdahl}
\end{axis}
\end{tikzpicture}
\end{adjustbox}
\end{center}
\end{subfigure}
\caption{Speedups for the coarse (left) and fine (right) strong scaling cases. The dashed line shows the ideal speedup, and the grey dotted one shows the scaling according to Amdahl's law with a parallel fraction of 99.8\%}
\label{fig:StrongSpeedup}
\end{figure}
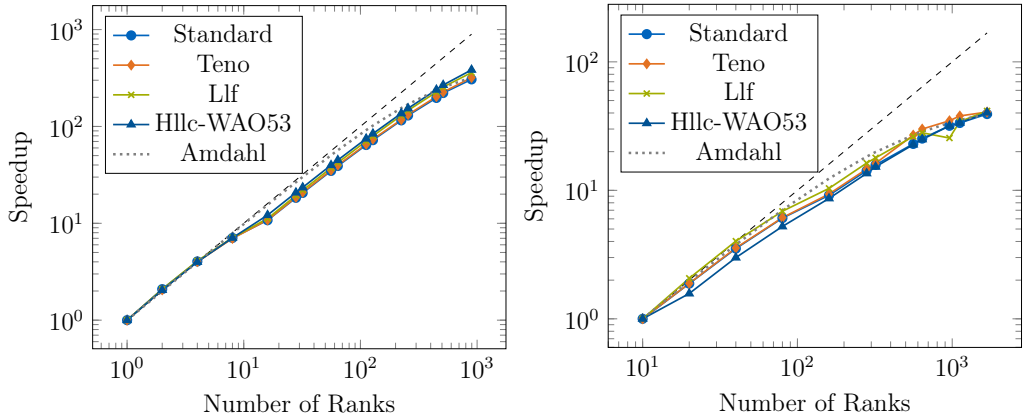

For the strong scaling runs, we record the compute time needed to advance the simulation from $t = 0.4$ to $t_{end} = 0.5$. We ran the test case in two different resolutions. The lower resolved case holds $512^3$ effective cells with $l_{max} = l_4$. The resolution is chosen such that it fits into the memory of a single compute node. The finer-resolved case has a resolution of $1536^3$ effective cells with $l_{max} = l_5$. This case requires the memory of at least 10 compute nodes. The finer resolution is needed as the coarser one suffers from under-subscription problems for large core counts. This under-subscription is analyzed in detail in \cref{sec:WorstCase}. The obtained speedups are shown in  \cref{fig:StrongSpeedup}. For the fine resolution case we take the runtime on ten ranks (spread onto ten nodes) as initial point.

As for the weak scaling, we see that the parallel performance of the different kernels is very similar. The obtained speedups, also drop going from 4 to 16 cores. For higher core counts, \ie from 32 to 256, a near-ideal speedup is seen until it finally levels off. The observed speedups indicate a parallel program fraction of $>99.8\%$ according to Amdahl's law. Furthermore, we can see that scalability is unaffected whether a node is completely packed or if the MPI ranks are spread out, \cf points for 28 and 32 or 112 and 128 cores.

%% file: Inputs/Benchmarking/ParallelPerformance/worst_case.tex
\subsection{Worst Case Setup}
\label{sec:WorstCase}

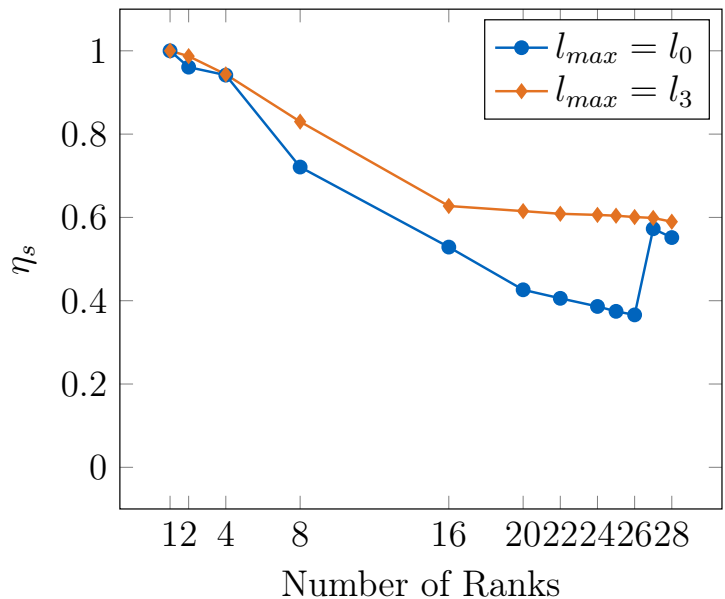
\begin{figure}
   \begin{center}
      \begin{adjustbox}{width=0.7\linewidth}
         \begin{tikzpicture}
            \begin{axis}[
               cycle list name = TUMcyclelist,
               ymin = -0.1,
   ymax = 1.1,
   xlabel = {Number of Ranks},
   ylabel = {$\eta_s$},
   xtick={1,2,4,8,16,20,22,24,26,28}
   ]
   \addplot+[thick] table [x=Ranks, y=Efficiency-L0, col sep=comma] {./ExtractedData/worstcase.csv};
\addplot+[thick] table [x=Ranks, y=Efficiency-L3, col sep=comma] {./ExtractedData/worstcase.csv};
\legend{$l_{max} = l_0$, $l_{max} = l_3$}
\end{axis}
\end{tikzpicture}
\end{adjustbox}
\end{center}
\caption{Strong scaling efficiency for a spherical Sod shock tube case. The domain decomposition is artificially set to a worst case scenario for $l_{max} = l_0$. An improved decomposition is obtained by increasing the allowed refinement to $l_{max} = l_3$.}
\label{fig:WorstCase}
\end{figure}

As we have demonstrated the performance of our solver in the last sections, here we look at the performance given an artificial worst case setup. Therefore, we simulate the spherical explosion case of Toro \cite[Section 17.1]{Toro2009}. It basically resembles a spherical Sod shock tube, \cf \cref{sec:RefinementConvergence}. Therein, the initial high pressure and density states are spherically centered in a cubic domain of side length $[0,2]^3$. As for the Sod shock tube case, we simulate the problem until $t_{end} = 0.2$. We partitioned the domain into three blocks per dimension and set $l_{max} = l_0$. We run this setup using different number of ranks $r = [1;28]$. In this setup, unfavorable latencies and/or communication patterns are inevitable for certain $r$. As an example,  consider the case of $r = 16$, where eleven ranks hold two blocks, while the remaining five ranks hold only one block. The latter five are, hence, idle half the time. In the case of $r = 27$, each block has a maximum of communication partners; the center block even reaches the theoretical maximum of $27$ same-level communication partners. \Cref{fig:WorstCase} shows the obtained strong scaling efficiency $\eta_s$ over $r$. We see a continuous decrease in efficiency until 26 cores are used where the efficiency drops below 50\% as soon as more than 16 cores are used. When using 27 cores, however, $\eta_s$ jumps back to 57\%. Confirming our previous assessment. Since we suggest that the performance increase if the blocks are more evenly distributed to the available cores. Therefore, we rerun the case with an increased $l_{max} = l_3$, which creates at least 218 blocks. The obtained efficiency with increased $l_{max}$ is also plotted in \cref{fig:WorstCase}. We observe that in this case the efficiency is higher for all core counts and remains above 50\% at all times, and is in agreement with our earlier assessment.

The conducted test shows the limitations of our block-based \ac{MR} algorithm for low block to core rations. When simulation physical problems, however, the block to core ratio is typically one or more orders of magnitude higher \cite{Kaiser2019,Winter2019} than in the cases presented in this section. Hence, these performance limitations are of little significance.

%% file: Inputs/Benchmarking/ParallelPerformance/space_filling_curve.tex
\subsection{Influence of Space Filling Curve}
\label{sec:CurvesPerformance}

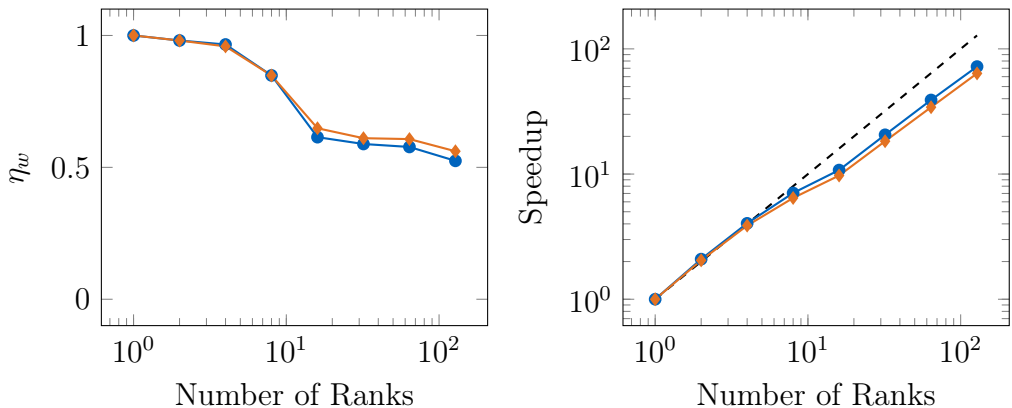
\begin{figure}
\begin{subfigure}[c]{0.49\textwidth}
\begin{center}
\begin{tikzpicture}
\begin{axis}[width=\linewidth, xmode = log, ymin = -0.1, ymax = 1.1, cycle list name = TUMcyclelist, xlabel = {Number of Ranks}, ylabel = {$\eta_w$}]
\addplot+[thick] table [x=Ranks, y=Efficiency-Weak-Std, col sep=comma] {./ExtractedData/curves.csv}; \label{plot:curves:standard}
\addplot+[thick] table [x=Ranks, y=Efficiency-Weak-Zcurve, col sep=comma] {./ExtractedData/curves.csv}; \label{plot:curves:z}
\end{axis}
\end{tikzpicture}
\end{center}
\end{subfigure}
\begin{subfigure}[c]{0.49\textwidth}
\begin{center}
\begin{tikzpicture}
\begin{axis}[width=\linewidth, xmode = log, ymode = log, cycle list name = TUMcyclelist, xlabel = {Number of Ranks}, ylabel = {Speedup}]
\addplot[thick, dashed] table [x=Ranks, y=Ranks, col sep=comma, forget plot] {./ExtractedData/curves.csv};
\addplot+[thick] table [x=Ranks, y=Speedup-Strong-Std, col sep=comma] {./ExtractedData/curves.csv};
\addplot+[thick] table [x=Ranks, y=Speedup-Strong-Zcurve, col sep=comma] {./ExtractedData/curves.csv};
\end{axis}
\end{tikzpicture}
\end{center}
\end{subfigure}
\caption{Comparison of weak (left) and strong (right) scaling using two different space-filling curves, namely the Hilbert-curve \ref{plot:curves:standard} and the Z-curve \ref{plot:curves:z}.}
\label{fig:Curves}
\end{figure}

In this section, we analyze the performance impact of different \ac{SFC}, \ie the level-wise Hilbert curve versus the level-wise Z-Curve. For both configurations, we compare the weak scaling on the test case described in \cref{sec:Weakscale} (channel), and the strong scaling on the test case described in \cref{sec:Strongscale} (cubic geometry). Following our earlier discussion, here, we limit the core counts from 1 to 128. \Cref{fig:Curves} shows the obtained scaling for the weak (efficiency) and strong (speedup) scaling tests.

We see that the performance of the two types of curves are comparable. However, the Hilbert curve partitioning yields slightly better strong scaling, while the opposite is true for the weak scaling. Despite the scaling behavior being very similar, the Z-curve is always slower in absolute wall-clock runtime. We recorded relative differences from 9.6\% to 25\% for increasing core counts in the strong-scaling case and, form 11\% to 4\% in the weak-scaling setup. The obtained results motivate our default selection of the level-wise Hilbert-curve for its runtime advantage over the  level-wise Z-curve.

%% file: Inputs/Benchmarking/ParallelPerformance/equation_of_state.tex
\subsection{Influence of Equation of State}
\label{sec:InfluenceEos}

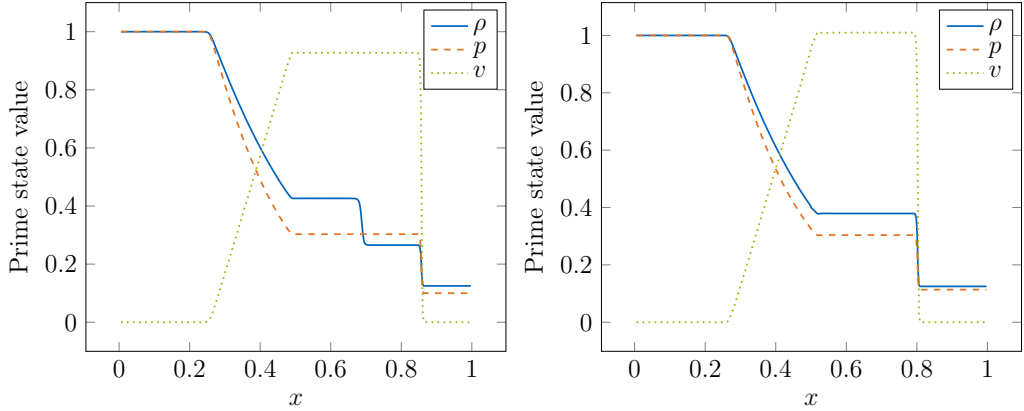
\begin{figure}
\begin{subfigure}[c]{0.49\textwidth}
\begin{center}
\begin{adjustbox}{width=\linewidth}
\begin{tikzpicture}
\begin{axis}[xlabel = {$x$}, ylabel = {Prime state value}, cycle list name = TUMcyclelistNomark]
\addplot+[thick] table [x=Xcoordinates, y=Density, col sep=comma, mark=none] {./ExtractedData/stiff_extract.csv};
\addplot+[thick] table [x=Xcoordinates, y=Pressure, col sep=comma, mark=none] {./ExtractedData/stiff_extract.csv};
\addplot+[thick] table [x=Xcoordinates, y=Velocity, col sep=comma, mark=none] {./ExtractedData/stiff_extract.csv};
\legend{$\rho$,$p$,$v$}
\end{axis}
\end{tikzpicture}
\end{adjustbox}
\end{center}
\end{subfigure}
\begin{subfigure}[c]{0.49\textwidth}
\begin{center}
\begin{adjustbox}{width=\linewidth}
\begin{tikzpicture}
\begin{axis}[xlabel = {$x$}, ylabel = {Prime state value}, cycle list name = TUMcyclelistNomark]
\addplot+[thick] table [x=Xcoordinates, y=Density, col sep=comma, mark=none] {./ExtractedData/tait_extract.csv};
\addplot+[thick] table [x=Xcoordinates, y=Pressure, col sep=comma, mark=none] {./ExtractedData/tait_extract.csv};
\addplot+[thick] table [x=Xcoordinates, y=Velocity, col sep=comma, mark=none] {./ExtractedData/tait_extract.csv};
\legend{$\rho$,$p$,$v$}
\end{axis}
\end{tikzpicture}
\end{adjustbox}
\end{center}
\end{subfigure}
\caption{Spatial prime state profile at $t = 0.2$ for the Sod sock tube simulation using different equations of state. The stiffened-gas and Tait's equation are shown on the left and right, respectively.}
\label{fig:EosStates}
\end{figure}

Based on the reference results from \cref{sec:RefinementConvergence}, we test the influence of changing the \ac{EOS} from stiffened-gas to Tait's \cite{Fedkiw1999} on the parallel performance. We restrict the analysis to the case of $l_{max} = 3$. The material parameters are $\gamma = 1.4$, $A = 1.0$, $B = 0.937315$ and $\rho_0 = 1.0$. The prime state profiles at $t_{end} = 0.2$ are shown for both \ac{EOS} in \cref{fig:EosStates}.

\begin{figure}
\begin{center}
\begin{adjustbox}{width=0.7\linewidth}
\begin{tikzpicture}
\begin{axis}[
   ymax = 3.25,
   xmode = log,
   axis y line* = right,
   axis x line = none,
   ylabel = {Runtime [h]},
   every y tick scale label/.style={at={(1.15,1)},xshift=1pt,anchor=north east,inner sep=0pt},
]
\addplot[TUMBlau, thick, dashed, mark = square*, mark options = {solid}, y filter/.code={\pgfmathparse{#1/3600}\pgfmathresult}] table [x=Ranks, y=Stiff-Runtime, col sep=comma] {./ExtractedData/equationsofstate.csv};
\addplot[TUMOrange, thick, dashed, mark = triangle*, mark options = {solid, scale=1.2}, y filter/.code={\pgfmathparse{#1/3600}\pgfmathresult}] table [x=Ranks, y=Tait-Runtime, col sep=comma] {./ExtractedData/equationsofstate.csv};
\end{axis}
\begin{axis}[
   xmode = log,
   ymin = 0,
   axis y line* = left,
   xlabel = {Number of Ranks},
   ylabel = {$\eta_w$},
]
\addplot[TUMBlau, thick, mark = square*, mark options = {solid}] table [x=Ranks, y=Stiff-Efficiency, col sep=comma] {./ExtractedData/equationsofstate.csv};
\addplot[TUMOrange, thick, mark = triangle*, mark options = {solid, scale=1.2} ] table [x=Ranks, y=Tait-Efficiency, col sep=comma] {./ExtractedData/equationsofstate.csv};
\legend{Stiffened gas, Tait}
\end{axis}
\end{tikzpicture}
\end{adjustbox}
\end{center}
\caption{Weak scaling efficiency $\eta_w$ and absolute runtime for the Sod case using different equations of state. Note, coloring is identical for both graphs. The dashed lines indicates the wall clock time and the solid curves denote $\eta_w$.}
\label{fig:EosSpeedups}
\end{figure}
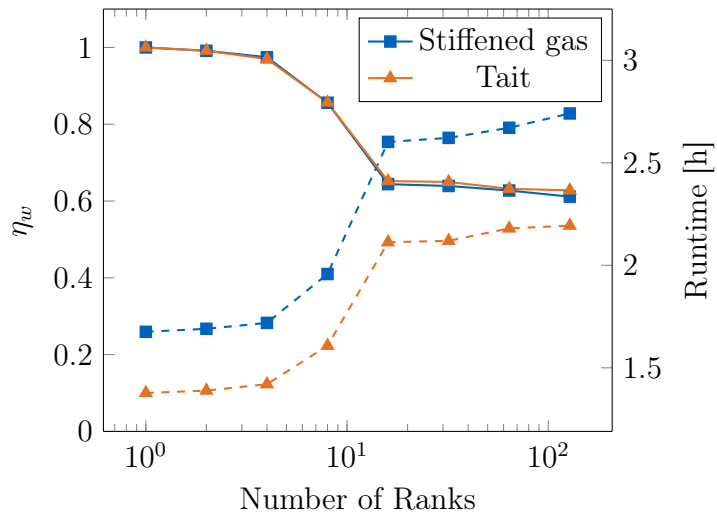

With both \ac{EOS} we conducted a weak-scaling analysis form 1 to 224 cores. The obtained parallel performance is summarized in \cref{fig:EosSpeedups}, where we plot the weak-scaling efficiency $\eta_w$ together with the respective runtimes for different core count. Clearly, the scaling behavior is similar for both \ac{EOS}, despite a runtime difference of about 18\%. As in the large weak scaling study \cf \cref{sec:Weakscale}, the efficiency drops to 64\% on 16 cores, but remains near ideal for successively higher core counts. As discussed later in \cref{sec:Compression}, also the compression for both \ac{EOS} is similar. Overall, the difference in runtime is explained by the reduced amount of computational work as in Tait's formulation the energy in \cref{eq:Euler} becomes a dependent variable.

%% file: Inputs/Benchmarking/ParallelPerformance/error_norms.tex
\subsection{Influence of Error Norm}
\label{sec:InfluenceErrorNorm}

\begin{figure}
\begin{center}
\begin{subfigure}[c]{0.29\textwidth}
\includegraphics[height=\linewidth,trim={0 6.5cm 0 0},clip,angle=270]{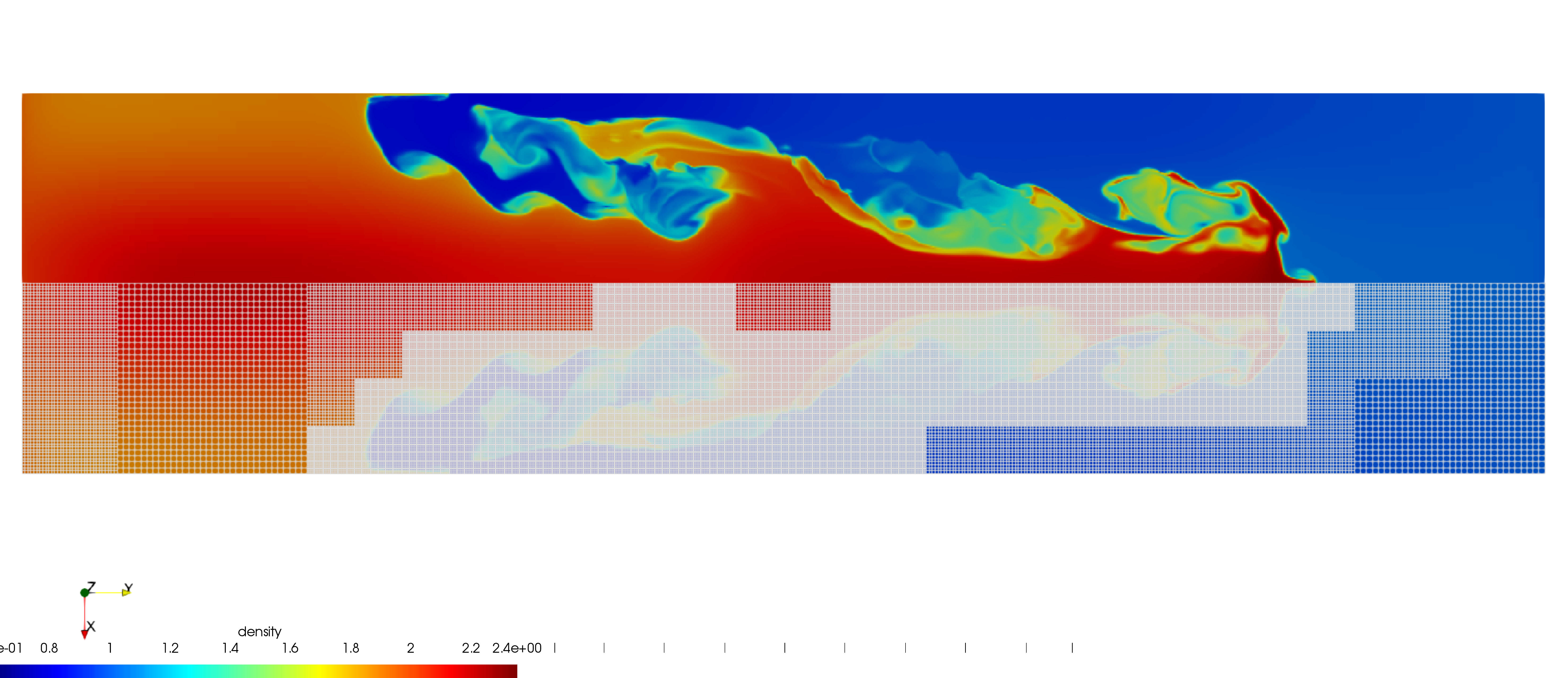}
\end{subfigure}
\begin{subfigure}[c]{0.29\textwidth}
\includegraphics[height=\linewidth,trim={0 6.5cm 0 0},clip,angle=270]{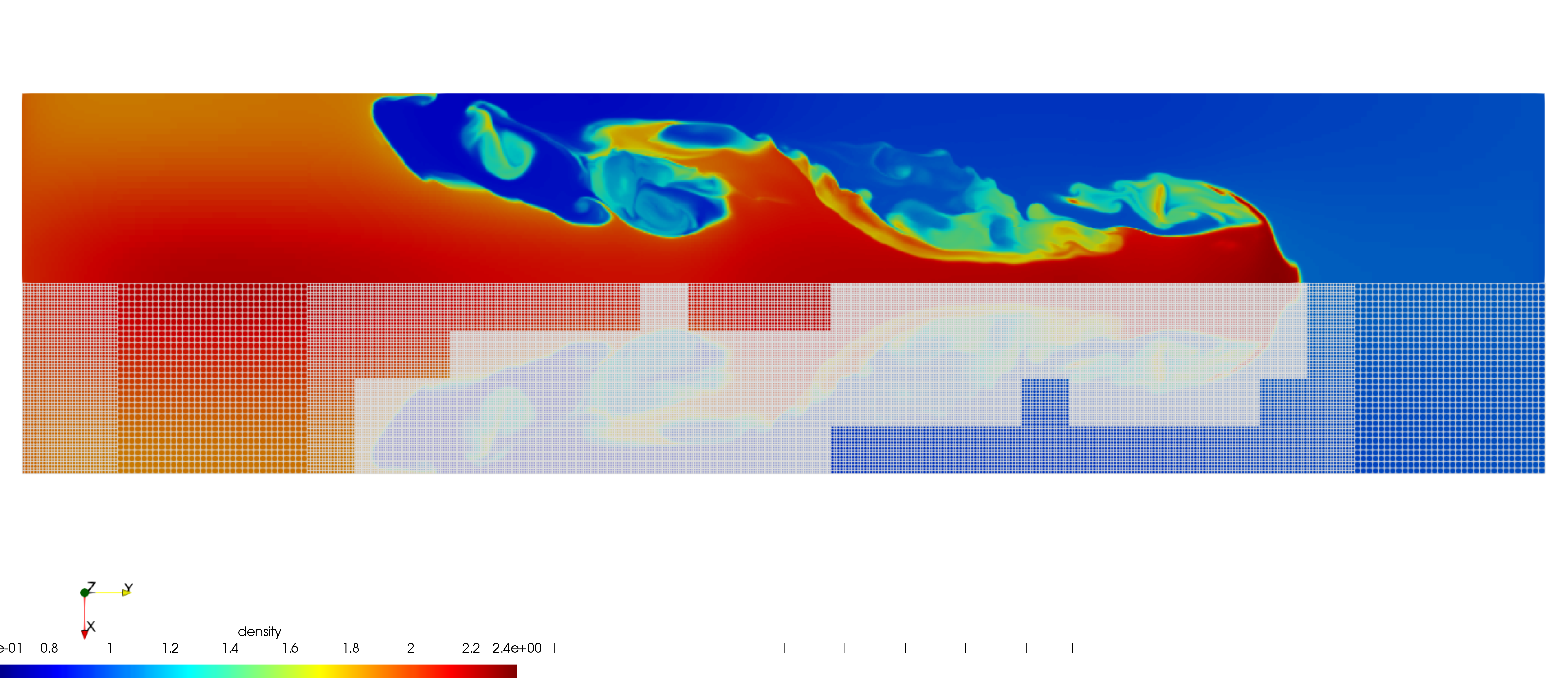}
\end{subfigure}
\begin{subfigure}[c]{0.29\textwidth}
\includegraphics[height=\linewidth,trim={0 6.5cm 0 0},clip,angle=270]{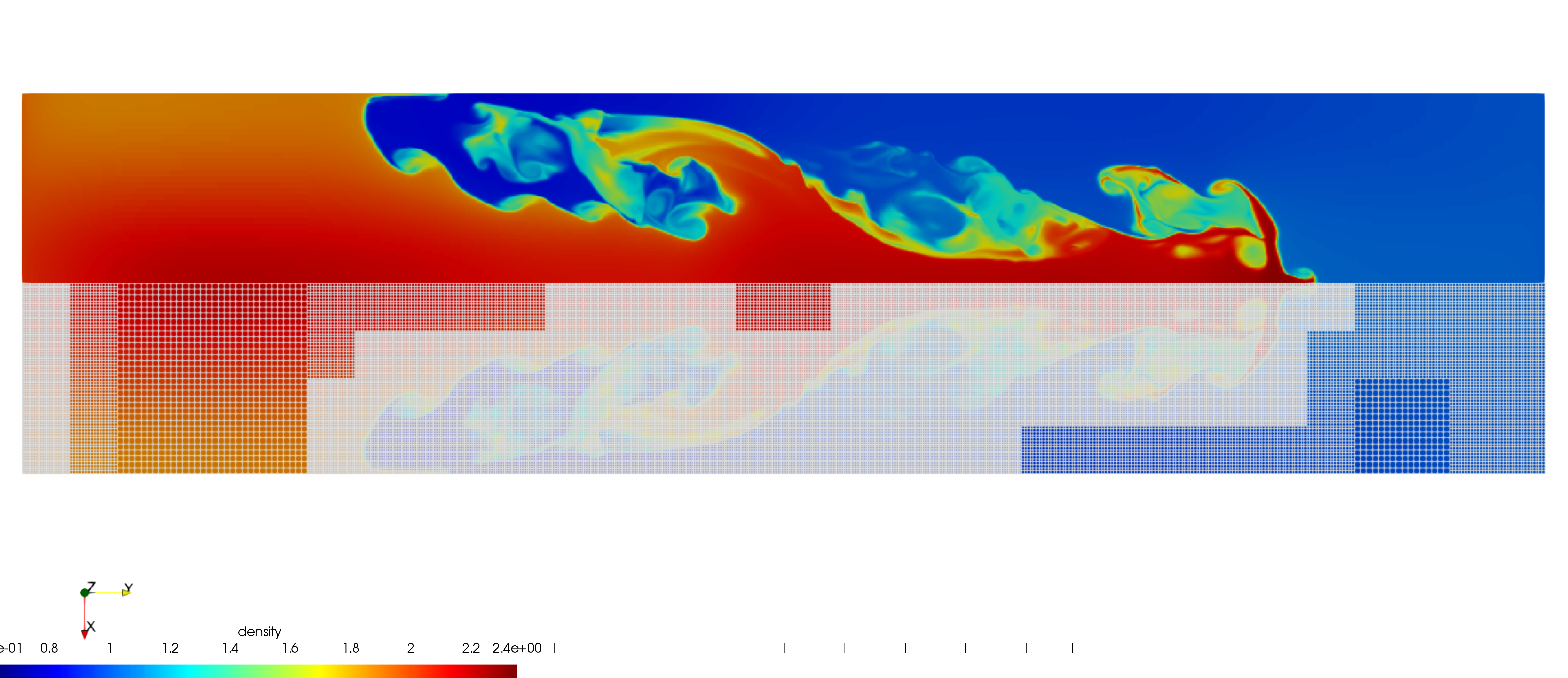}
\end{subfigure}
\end{center}
\caption{Density field and corresponding mesh at $t = 1.95$ for the inviscid \acs{RTI} case in the $x-y$ center plane. The color scale ranges from blue $\rho = 0.64$ to red $\rho = 2.4$. From left to right the $l_\infty$-, $l_1$- and $l_2$-norms are used, respectively.}
\label{fig:ErrorNorms}
\end{figure}

We stated earlier that the error norm guiding the \ac{MR} compression, \cf \cref{eq:Thresholding}, can be chosen at compile-time. In this section, we want to analyze the performance impact of different norms. We re-use the \ac{RTI} case given in \cref{sec:SingleCore}, but we limit $l_{max}$ to four levels. We chose the inviscid \ac{RTI} case as it is sensitive to the employed error norm, see \cref{fig:ErrorNorms}. Here, we show the density field at $t_{end} = 1.95$ and the resulting mesh.

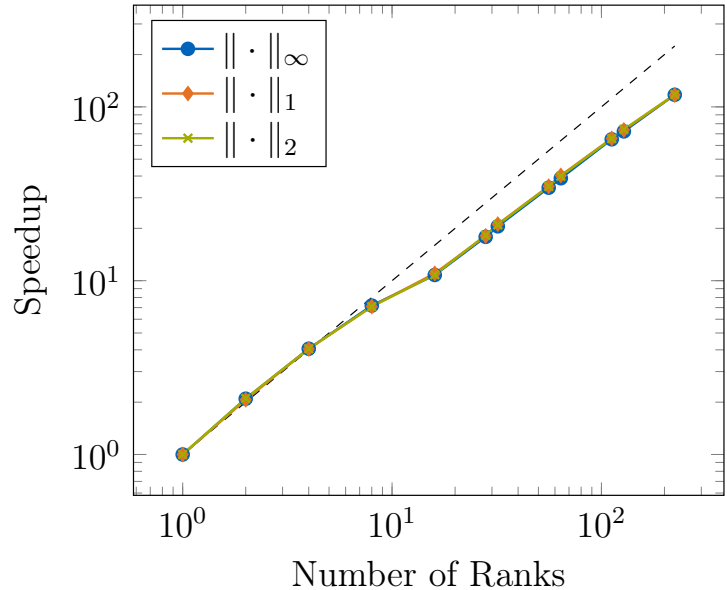
\begin{figure}
\begin{center}
\begin{adjustbox}{width=0.7\linewidth}
\begin{tikzpicture}
\begin{loglogaxis}[xlabel = {Number of Ranks}, ylabel = {Speedup}, cycle list name = TUMcyclelist, legend pos = north west]
\addplot[dashed] table [x=Ranks, y=Ranks, col sep=comma, forget plot] {./ExtractedData/scalingnorms.csv};
\addplot+[thick] table [x=Ranks, y=Speedup-Linf, col sep=comma] {./ExtractedData/scalingnorms.csv};
\addplot+[thick] table [x=Ranks, y=Speedup-Lone, col sep=comma] {./ExtractedData/scalingnorms.csv};
\addplot+[thick] table [x=Ranks, y=Speedup-Ltwo, col sep=comma] {./ExtractedData/scalingnorms.csv};
\legend{$\| \cdot \|_\infty$, $\| \cdot \|_1$ ~, $\| \cdot \|_2$ ~}
\end{loglogaxis}
\end{tikzpicture}
\end{adjustbox}
\end{center}
\caption{Speedup for the strong scaling analysis of configuration using different refinement norms. The dashed line indicates the ideal speedup.}
\label{fig:SpeedUpErrorNorms}
\end{figure}

In order to compare the computational performance of the three setups, we tune the respective reference errors $\varepsilon_{ref}$ for each case to obtain a comparable number of leaves. For each norm, we conducted a strong scaling analysis from one to eight compute nodes. Therein, we record the wall clock time for the simulation of a developed \ac{RTI} from $t = 1.90$ until $t_{end}$. As for the weak-scaling test, we again see in \cref{fig:SpeedUpErrorNorms} almost ideal speedup using more than 16 cores. For all cases the efficiency stays above 50\%, which implies a parallel fraction of the code $>0.99$ according to Amdahl's law. Hence, the error norm does not affect the parallel performance itself, despite its strong influence on the resulting mesh topology.

%% file: Inputs/Benchmarking/ParallelPerformance/cells.tex
\subsection{Influence of Internal Cells per Block}
\label{sec:InfluenceInternalCells}

\begin{figure}
   \begin{center}
      \begin{subfigure}[c]{0.49\textwidth}
         \centering
         \begin{adjustbox}{width=\linewidth}
            \begin{tikzpicture}
            \begin{axis}[axis y line* = right, xlabel = {x}, cycle list name = TUMcyclelistNomark, legend pos = north west, axis line style = {TUMBlau}, every tick label/.append style={TUMBlau},name=lleft plot]
               \addplot+[thick] table [x=CellID, y=density, col sep=comma] {./ExtractedData/toro3.csv}; \label{plot:torothree:rho}
               \addplot+[thick] table [x=CellID, y=velocityX, col sep=comma] {./ExtractedData/toro3.csv}; \label{plot:torothree:v}
            \end{axis}
            \begin{axis}[axis y line* = left, ylabel = {{\textcolor{TUMBlau}{$\rho$}, \textcolor{TUMBlau}{$v$}} and $p$}, cycle list name = TUMcyclelistNomark, cycle list shift=2, legend pos = north west, name=rleft plot]
               \addplot+[thick] table [x=CellID, y=pressure, col sep=comma] {./ExtractedData/toro3.csv}; \label{plot:torothree:p}
            \end{axis}
            \end{tikzpicture}
         \end{adjustbox}
         \caption{Density \ref{plot:torothree:rho}, pressure \ref{plot:torothree:p} and velocity \ref{plot:torothree:v} along the $x$-axis for ``Test 3'' of Toro at $t = 0.012$.}
         \label{fig:Toro3Solution}
      \end{subfigure}
      \begin{subfigure}[c]{0.49\textwidth}
         \centering
         \begin{adjustbox}{width=\linewidth}
            \begin{tikzpicture}
               \begin{axis}[xmode = log, ymode = log, ymin = 20, xlabel = {Number of Ranks}, ylabel = {Runtime}, cycle list name = TUMcyclelist, legend columns = 3, legend pos = south west, name=right plot]
                   \addplot+[thick] table [x=Ranks, y=Avg-Comp-8, col sep=comma] {./ExtractedData/cells_strong.csv};
                   \addplot+[thick] table [x=Ranks, y=Avg-Comp-16, col sep=comma] {./ExtractedData/cells_strong.csv};
                   \addplot+[thick] table [x=Ranks, y=Avg-32cpb, col sep=comma] {./ExtractedData/cells_strong.csv};
                   \addplot+[thick] table [x=Ranks, y=Avg-32cpb, col sep=comma] {./ExtractedData/cells_weak.csv};
                   \addplot+[thick] table [x=Ranks, y=Avg-Comp-8, col sep=comma] {./ExtractedData/cells_weak.csv};
                   \addplot+[thick] table [x=Ranks, y=Avg-Comp-16, col sep=comma] {./ExtractedData/cells_weak.csv};
                   \legend{S: $8^3$, S: $16^3$, S: $32^4$, W: $8^3$, W: $16^3$, W: $32^4$}
               \end{axis}
            \end{tikzpicture}
         \end{adjustbox}
         \caption{Runtimes for the case with same effective resolutions using different number of internal cells per block. The type of scaling weak (W) and strong (S) is abbreviated in the legend.}
         \label{fig:CellsRuntime}
      \end{subfigure}
      \hfill
   \begin{subfigure}[c]{0.49\textwidth}
      \centering
      \begin{adjustbox}{width=\linewidth}
         \begin{tikzpicture}
         \begin{axis}[xmode = log, xlabel = {Number of Ranks}, ylabel = {$\eta_w$}, ymin = -0.1, ymax = 1.1, cycle list name = TUMcyclelist, legend pos = south west]
         \addplot+[thick] table [x=Ranks, y=Eff-8cpb, col sep=comma] {./ExtractedData/cells_weak.csv};
         \addplot+[thick] table [x=Ranks, y=Eff-Comp-8, col sep=comma] {./ExtractedData/cells_weak.csv};
         \addplot+[thick] table [x=Ranks, y=Eff-16cpb, col sep=comma] {./ExtractedData/cells_weak.csv};
         \addplot+[thick] table [x=Ranks, y=Eff-Comp-16, col sep=comma] {./ExtractedData/cells_weak.csv};
         \addplot+[thick] table [x=Ranks, y=Eff-32cpb, col sep=comma] {./ExtractedData/cells_weak.csv};
         \legend{$8^3\,l_2$,$8^3\,l_4$,$16^3\,l_2$,$16^3\,l_3$,$32^4\,l_2$}
         \end{axis}
         \end{tikzpicture}
      \end{adjustbox}
      \caption{Weak scaling efficiency for different number of internal cells per block. The used maximum level is denoted in the legend.}
      \label{fig:CellsWeak}
   \end{subfigure}
   \begin{subfigure}[c]{0.49\textwidth}
      \centering
      \begin{adjustbox}{width=\linewidth}
         \begin{tikzpicture}
            \begin{axis}[xmode = log, ymode = log, xlabel = {Number of Ranks}, ylabel = {Speedup}, cycle list name = TUMcyclelist, legend pos = north west]
               \addplot[thick, dashed] table [x=Ranks, y=Ranks, col sep=comma, forget plot] {./ExtractedData/cells_strong.csv};
               \addplot+[thick] table [x=Ranks, y=Speedup-8cpb, col sep=comma] {./ExtractedData/cells_strong.csv};
               \addplot+[thick] table [x=Ranks, y=Speedup-Comp-8, col sep=comma] {./ExtractedData/cells_strong.csv};
               \addplot+[thick] table [x=Ranks, y=Speedup-16cpb, col sep=comma] {./ExtractedData/cells_strong.csv};
               \addplot+[thick] table [x=Ranks, y=Speedup-Comp-16, col sep=comma] {./ExtractedData/cells_strong.csv};
               \addplot+[thick] table [x=Ranks, y=Speedup-32cpb, col sep=comma] {./ExtractedData/cells_strong.csv};
               \legend{$8^3\,l_2$,$8^3\,l_4$,$16^3\,l_2$,$16^3\,l_3$,$32^4\,l_2$}
            \end{axis}
         \end{tikzpicture}
      \end{adjustbox}
      \caption{Strong scaling speedups for different number of internal cells per block. The used maximum level is denoted in the legend.}
      \label{fig:CellsStrong}
   \end{subfigure}
   \end{center}
\caption{Final states, scaling efficiency, speedup and runtimes when using different number of internal cells per block.}
\label{fig:CellScaling}
\end{figure}
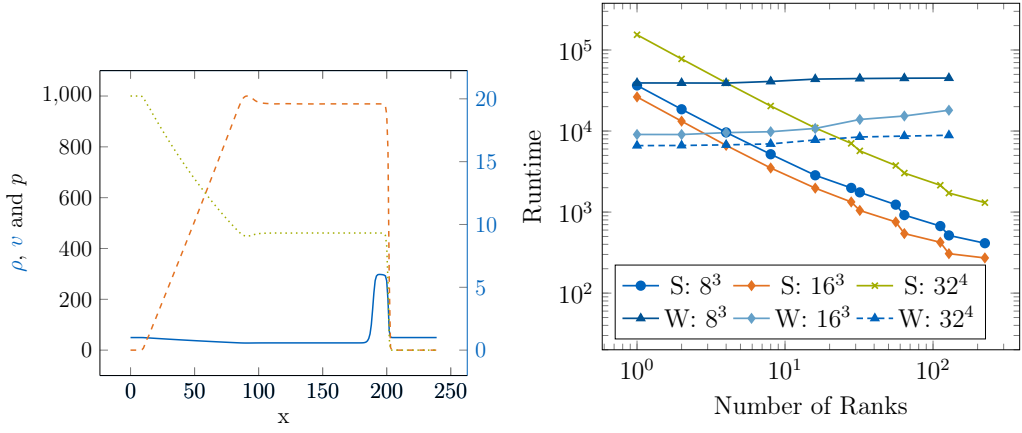
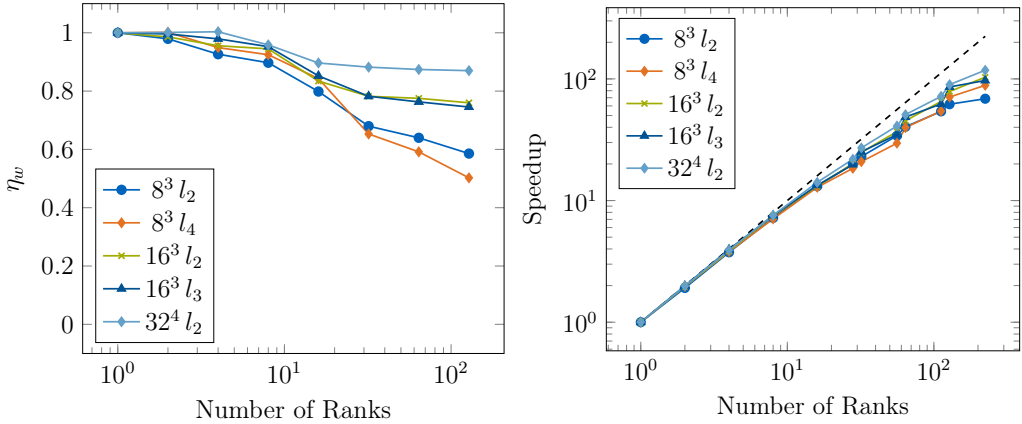

In this section we analyze the effect of the grain size, \ie the number \ac{IC} per block. Therefore, we run a weak and strong scaling analysis for ``Test 3'' of Toro \cite[Section 4.3.3]{Toro2009}. The final states at $t = 0.012$ are given in \cref{fig:Toro3Solution}. For the strong scaling and the initial configuration of the weak scaling analysis, we use a domain of size $1 \times 0.25 \times 0.25$ and discretized it using four blocks on $l_0$ with $l_{max} = 2$. For the weak scaling analysis, we increase the superfluous y- and z-dimensions. We use the standard configuration and vary only the \ac{IC} per block from $8^3$ to $16^3$ and $32^3$. This results in a four times higher resolved domain for the $32^3$ cells per block case compared to the $8^3$ one, but the number of nodes and leaves is comparable. In contrast, we also test the configurations with effectively the same resolution. We obtain these same resolutions by increasing $l_{max}$ until the cases with $16^3$ and $8^3$ \ac{IC} per block are as fine resolved as the case with $32^3$ \ac{IC} per block. The obtained weak scaling efficiency $\eta_w$ and strong scaling speedups and the absolute runtimes are given in \cref{fig:CellScaling}.

We observe that the scaling efficiency/speedup is better the more internal cells per block are used. This is expected as more internal cells improve the ratio of computation-to-communication. However, a closer look at the absolute runtimes shows, that for a case of comparable resolution the configuration using $16^3$ \ac{IC} per block is fastest. On average it outperforms the one using $8^3$ \ac{IC} per block by over 50\%. We attribute this speedup to the improved computation-to-communication ratio. In comparison to $32^3$ \ac{IC} per block the performance improves even by 450\%. The reason for this tremendous difference can be found in the compression, see the following \cref{sec:Compression}.

%% file: Inputs/Benchmarking/compression.tex
\subsection{Compression}
\label{sec:Compression}

Finally we analyze at the compression of our block-based \ac{MR} algorithm for the presented cases. We define the memory-compression $C$ as the ratio between the number of cells actually present in the simulation, and the \emph{effective} number of cells via

\begin{equation}
  C = 1 - \frac{\#\text{cells}}{\#\text{effective cells}}.
  \label{eq:Compression}
\end{equation}
Note that the compression of flux function computations is proportional to the memory compression.

\begin{figure}
  \begin{center}
    \begin{subfigure}[c]{0.49\textwidth}
      \begin{adjustbox}{width=\linewidth}
        \begin{tikzpicture}
          \begin{axis}[cycle list name = TUMcyclelistNomark, ymin = -0.1, ymax = 1.1, ylabel = {$C$}, xlabel = {Relative time}, legend columns=3, legend style=
          {at={(0.5,1.02)},anchor=south}]
            \addplot+[thick] table [x=TimeSodL0, y=CompressionSodL0, col sep=comma] {./ExtractedData/compression.csv};
            \addplot+[thick] table [x=TimeSodL4, y=CompressionSodL3, col sep=comma] {./ExtractedData/compression.csv};
            \addplot+[thick] table [x=TimeSodL6, y=CompressionSodL4, col sep=comma] {./ExtractedData/compression.csv};
            \addplot+[thick] table [x=TimeEosStiff, y=CompressionEosStiff, col sep=comma] {./ExtractedData/compression.csv};
            \addplot+[thick] table [x=TimeRtiInfer, y=CompressionRtiInfer, col sep=comma] {./ExtractedData/compression.csv};
            \legend{Sod $l_0$, Sod $l_3$, Sod $l_4$, Stiff $l_3$, RTI $l_4$}
          \end{axis}
        \end{tikzpicture}
      \end{adjustbox}
    \end{subfigure}
    \begin{subfigure}[c]{0.49\textwidth}
      \begin{adjustbox}{width=\linewidth}
        \begin{tikzpicture}
          \begin{axis}[cycle list name = TUMcyclelistNomark, ymin = -0.1, ymax = 1.1, ylabel = {$C$}, xlabel = {Relative time}, legend columns=3, legend style=
            {at={(0.5,1.02)},anchor=south}]
            \addplot+[thick] table [x=TimeCells-8, y=CompressionCells-8, col sep=comma] {./ExtractedData/compression.csv};
            \addplot+[thick] table [x=TimeCells-8, y=CompressionCells-16, col sep=comma] {./ExtractedData/compression.csv};
            \addplot+[thick] table [x=TimeCells-8, y=CompressionCells-32, col sep=comma] {./ExtractedData/compression.csv};
            \addplot+[thick] table [x=TimeEosTait, y=CompressionEosTait, col sep=comma] {./ExtractedData/compression.csv};
            \addplot+[thick] table [x=TimeStrongscale, y=CompressionStrongscale, col sep=comma] {./ExtractedData/compression.csv};
            \addplot+[thick] table [x=TimeWoodward, y=CompressionWoodward, col sep=comma] {./ExtractedData/compression.csv};
            \legend{$8^3\,l_4$, $16^3\,l_3$, $32^3\,l_2$, Tait $l_3$, Weak $l_3$, Strong $l_4$}
          \end{axis}
        \end{tikzpicture}
      \end{adjustbox}
    \end{subfigure}
  \end{center}
\caption{Compression of different test cases over time. The compression values are sampled after each completed macro time step. For better readability, the cases are split into two figures. The used $l_{max}$ for each case is indicated.}
\label{fig:Compression}
\end{figure}
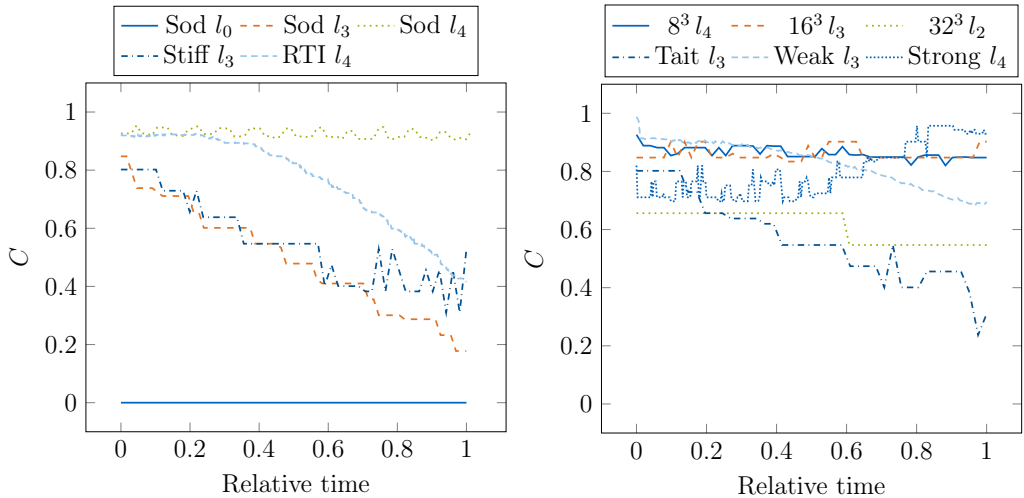

According to this definition we strive for high values of $C$. We plot the achieved compression of the cases used above as function over the relative simulation time in \Cref{fig:Compression}. Besides the test case name also the respective $l_{max}$ is given. The case names are abbreviated as follows: `Weak', `Strong', `\acs{RTI}' and `Sod' correspond to the cases of \cref{sec:Weakscale,sec:Strongscale,sec:SingleCore,sec:RefinementConvergence}. `Stiff' and `Tait' refer to the used \ac{EOS} in \cref{sec:InfluenceEos} and $x^3$ with $x \in 8,16,32$ stands for the number of \ac{IC} per block as defined and used in \cref{sec:InfluenceInternalCells}. If we look at the Sod cases, we see how the compression increases with $l_{max}$. For $l_{max} = l_0$, obviously, no compression is achieved. With $l_{max} = l_3$, the average compression is already at 50\%. Nevertheless, in the final time steps most of the domain gets refined, reducing compression significantly. With a further increase of $l_{max}$, the compression increases drastically. In the `\ac{RTI}' case the compression reduces monotonically as the initial disturbance grows and generates more and more finer scales. Nevertheless, an average compression of 75\% is achieved. For the different \ac{EOS}, the achieved compression is very similar. Hence, differences in the runtimes as stated above may not be credited to unequal compression. In the case of varying \ac{IC} per block, the effect of increasing the level of refinement is again clearly visible: While the cases using $l_{max} = l_3$ and $l_4$ achieved similar compressions above 80\%, the case with $l_{max} = l_2$ gives much smaller compression. Furthermore, the lack of levels prohibits effective adaptation to changes in the flow field and the compression remains almost constant over the whole simulation time. In contrast to the other cases, here the compression fluctuates, showing that the mesh quickly adapts to changes in the flow. In the `weak' case we also observe similar characteristics as for the Sod cases with the same maximum level of refinement. The compression rate is high on average, but in the case where multiple distinct waves travel trough the domain the compression drops. For the weak case this is visible at the beginning of the simulation. Once the waves collide in the middle of the domain, the compression increases again. The `strong' case shows a combination of the effects observed in the other cases. In the beginning, compression is reduced by the outward travelling of different shock waves, then additionally, smaller scales are formed at the initial discontinuity. Towards the end, the compression reaches a deflection point as no more strong shocks are generated and the viscous effects start to smooth the flow field.

Overall, detailed analysis demonstrates that our block-based implementation achieves high compression rates if $l_{max}$ is chosen sufficiently large. In particular for \ac{3D} simulations, the compression of our block-based scheme is comparable to those reported for cell-based \ac{MR} algorithms \cite{Harten1995, Bihari1996, Cohen2003, Roussel2003, Descombes2017}.

%% file: Inputs/conclusion.tex
\section{Conclusion}

We have presented a \ac{MPI} parallelization strategy for a \ac{3D} block-based \ac{MR} algorithm with \ac{ALTS} for the simulation of the compressible Euler or Navier-Stokes equations. Our scheme allows to bisect the underlying octree structure arbitrarily. This is achieved by introducing a modified Morton order and level-wise \ac{SFC}-based load balancing. The achieved compression in \ac{3D} of our block-based approach is comparable to cell-based approaches. The presented strategy was implemented in a modular \CC framework, a detailed description of the implementation is given. The modular framework was validated with multiple test cases and its (parallel) performance was assessed thoroughly for varying compute kernels by means of weak and strong scaling analyses up to 1680 cores. Our implementation shows reasonable parallel performance and is able to deliver similar performance for a range of differing  kernels. The presented framework is available under open-source licence.

In future work the performance of the scheme is to be improved further. This could be achieved by further optimizing the load balancing strategy and exploiting further \ac{SIMD}-capabilities, or by reducing the \ac{MPI} overhead via an additional shared-memory hybrid parallelization, \egn, with \ac{OMP}.

%% file: Inputs/acknowledgment.tex
\section{Acknowledgment}

The authors would like to thank Felix Sp\"ath for his help in the implementation of the \ac{SFC} and the communication cache. The first author would like to further thank Vladimir Bogdanov for helping with the bit-wise logic operations.

The authors have received funding from the European Research Council (ERC) under the European Union's Horizon 2020 research and innovation programme (grant agreement No. 667483).

The authors gratefully acknowledge the Gauss Centre for Supercomputing e.V. for funding this project by providing computing time on the CoolMUC-2 Linux-Cluster at Leibniz Supercomputing Centre.

The authors acknowledge the Bavarian State Ministry of Science and the Arts for partial funding through the Competence Network for Scientific High Performance Computing in Bavaria (KONWIHR).

%% file: Inputs/appendix.tex
\section{Definitions of $Q$-terms}

In the prediction equation \cref{eq:Prediction} we skipped the $Q$-terms for brevity. Here, their definition is given in full for a 5th-order accurate Prediction as used throughout this paper. In the definitions $u_{ijk}$ refers to the parent that is the base of the prediction and we drop the level-indicating subscript for better readability.

\begin{align}
c_0 &= - \frac{22}{128}, \, \quad c_1 = \frac{3}{128}, \nonumber \\
c_{pq} &= c_p c_q \forall (p,q) \in [0,1] \nonumber \nonumber \\
c_{rpq} &= c_r c_p c_q \forall (r,p,q) \in [0,1] \nonumber \\
\nonumber \\
Q_x &= c_0 \left(u_{i+1,j,  k  } - u_{i-1,j  ,k  } \right) + c_1 \left( u_{i+2,j  ,k  } - u_{i-2,j  ,k  } \right) \nonumber \\
Q_y &= c_0 \left(u_{i,  j+1,k  } - u_{i  ,j-1,k  } \right) + c_1 \left( u_{i  ,j+2,k  } - u_{i  ,j-2,k  } \right) \nonumber \\
Q_z &= c_0 \left(u_{i,  j,  k+1} - u_{i  ,j  ,k-1} \right) + c_1 \left( u_{i+2,j  ,k+2} - u_{i  ,j  ,k-2} \right) \nonumber \\
\nonumber
\end{align}
\begin{align}
Q_{xy} &= c_{00} \left[ \left( u_{i+1,j+1,k} + u_{i-1,j-1,k} \right) - \left( u_{i-1,j+1,k} + u_{i+1,j-1,k} \right) \right] \nonumber \\
       &+ c_{11} \left[ \left( u_{i+2,j+2,k} + u_{i-2,j-2,k} \right) - \left( u_{i-2,j+2,k} + u_{i+2,j-2,k} \right) \right] \nonumber \\
       &+ c_{01} \left[ \left( u_{i+1,j+2,k} + u_{i-1,j-2,k} \right) - \left( u_{i-1,j+2,k} + u_{i+1,j-2,k} \right) \right] \nonumber \\
       &+ c_{10} \left[ \left( u_{i+2,j+1,k} + u_{i-2,j-1,k} \right) - \left( u_{i-2,j+1,k} + u_{i+2,j-1,k} \right) \right] \nonumber \\
\nonumber \\
Q_{xz} &= c_{00} \left[ \left( u_{i+1,j,k+1} + u_{i-1,j,k-1} \right) - \left( u_{i-1,j,k+1} + u_{i+1,j,k-1} \right) \right] \nonumber \\
       &+ c_{11} \left[ \left( u_{i+2,j,k+2} + u_{i-2,j,k-2} \right) - \left( u_{i-2,j,k+2} + u_{i+2,j,k-2} \right) \right] \nonumber \\
       &+ c_{01} \left[ \left( u_{i+1,j,k+2} + u_{i-1,j,k-2} \right) - \left( u_{i-1,j,k+2} + u_{i+1,j,k-2} \right) \right] \nonumber \\
       &+ c_{10} \left[ \left( u_{i+2,j,k+1} + u_{i-2,j,k-1} \right) - \left( u_{i-2,j,k+1} + u_{i+2,j,k-1} \right) \right] \nonumber \\
\nonumber \\
Q_{yz} &= c_{00} \left[ \left( u_{i,j+1,k+1} + u_{i,j-1,k-1} \right) - \left( u_{i,j-1,k+1} + u_{i,j+1,k-1} \right) \right] \nonumber \\
       &+ c_{11} \left[ \left( u_{i,j+2,k+2} + u_{i,j-2,k-2} \right) - \left( u_{i,j-2,k+2} + u_{i,j+2,k-2} \right) \right] \nonumber \\
       &+ c_{01} \left[ \left( u_{i,j+1,k+2} + u_{i,j-1,k-2} \right) - \left( u_{i,j-1,k+2} + u_{i,j+1,k-2} \right) \right] \nonumber \\
       &+ c_{10} \left[ \left( u_{i,j+2,k+1} + u_{i,j-2,k-1} \right) - \left( u_{i,j-2,k+1} + u_{i,j+2,k-1} \right) \right] \nonumber
\end{align}
\begin{alignat}{3}
Q_{xyz} &= c_{000}   \big[ \left( u_{i+1,j+1,k+1} - u_{i-1,j-1,k-1} \right) &&+ \left( u_{i+1,j-1,k-1} - u_{i-1,j+1,k+1} \right) \nonumber \\
        &\qquad \; \; \! + \left( u_{i-1,j-1,k+1} - u_{i+1,j+1,k-1} \right) &&+ \left( u_{i-1,j+1,k-1} - u_{i+1,j-1,k+1} \right) \big] \nonumber \\
        &+ c_{111}   \big[ \left( u_{i+2,j+2,k+2} - u_{i-2,j-2,k-2} \right) &&+ \left( u_{i-2,j-2,k+2} - u_{i+2,j+2,k-2} \right) \nonumber \\
        &\qquad \; \; \! + \left( u_{i+2,j-2,k-2} - u_{i-2,j+2,k+2} \right) &&+ \left( u_{i-2,j+2,k-2} - u_{i+2,j-2,k-2} \right) \big] \nonumber \\
        &+ c_{001}   \big[ \left( u_{i+1,j+1,k+2} - u_{i-1,j-1,k-2} \right) &&+ \left( u_{i-1,j-1,k+2} - u_{i+1,j+1,k-2} \right) \nonumber \\
        &\qquad \; \; \! + \left( u_{i+1,j-1,k-2} - u_{i-1,j+1,k+2} \right) &&+ \left( u_{i-1,j+1,k-2} - u_{i+1,j-1,k+2} \right) \big] \nonumber \\
        &+ c_{110}   \big[ \left( u_{i+2,j+2,k+1} - u_{i-2,j-2,k-1} \right) &&+ \left( u_{i-2,j-2,k+1} - u_{i+2,j+2,k-1} \right) \nonumber \\
        &\qquad \; \; \! + \left( u_{i+2,j-2,k-1} - u_{i-2,j+2,k+1} \right) &&+ \left( u_{i-2,j+2,k-1} - u_{i+2,j-2,k+1} \right) \big] \nonumber \\
        &+ c_{011}   \big[ \left( u_{i+1,j+2,k+2} - u_{i-1,j-2,k-2} \right) &&+ \left( u_{i-1,j-2,k+2} - u_{i+1,j+2,k-2} \right) \nonumber \\
        &\qquad \; \; \! + \left( u_{i+1,j-2,k-2} - u_{i-1,j+2,k+2} \right) &&+ \left( u_{i-1,j+2,k-2} - u_{i+1,j-2,k+2} \right) \big] \nonumber \\
        &+ c_{100}   \big[ \left( u_{i+2,j+1,k+1} - u_{i-2,j-1,k-1} \right) &&+ \left( u_{i-2,j-1,k+1} - u_{i+2,j+1,k-1} \right) \nonumber \\
        &\qquad \; \; \! + \left( u_{i+2,j-1,k-1} - u_{i-2,j+1,k+1} \right) &&+ \left( u_{i-2,j+1,k-1} - u_{i+2,j-1,k+1} \right)\big] \nonumber \\
        &+ c_{010}   \big[ \left( u_{i+1,j+2,k+1} - u_{i-1,j-2,k-1} \right) &&+ \left( u_{i-1,j-2,k+1} - u_{i+1,j+2,k-1} \right) \nonumber \\
        &\qquad \; \; \! + \left( u_{i+1,j-2,k-1} - u_{i-1,j+2,k+1} \right) &&+ \left( u_{i-1,j+2,k-1} - u_{i+1,j-2,k+1} \right) \big] \nonumber \\
        &+ c_{101}   \big[ \left( u_{i+2,j+1,k+2} - u_{i-2,j-1,k-2} \right) &&+ \left( u_{i-2,j-1,k+2} - u_{i+2,j+1,k-2} \right) \nonumber \\
        &\qquad \; \; \! + \left( u_{i+2,j-1,k-2} - u_{i-2,j+1,k+2} \right) &&+ \left( u_{i-2,j+1,k-2} - u_{i+2,j-1,k+2} \right) \big] \nonumber
\end{alignat}

%% file: main.bbl
\begin{thebibliography}{10}
\expandafter\ifx\csname url\endcsname\relax
  \def\url#1{\texttt{#1}}\fi
\expandafter\ifx\csname urlprefix\endcsname\relax\def\urlprefix{URL }\fi
\expandafter\ifx\csname href\endcsname\relax
  \def\href#1#2{#2} \def\path#1{#1}\fi

\bibitem{Jiang1996}
G.-S. Jiang, Shu, {Efficient implementation of weighted ENO schemes}, Journal
  of Computational Physics 126 (1996) 202--228.

\bibitem{Bell1989}
J.~B. Bell, P.~Colella, J.~A. Trangenstein, Higher order godunov methods for
  general systems of hyperbolic conservation laws, Journal of Computational
  Physics 82~(2) (1989) 362--397.

\bibitem{Pember1995}
R.~B. Pember, J.~B. Bell, P.~Colella, W.~Y. Curtchfield, M.~L. Welcome, An
  adaptive cartesian grid method for unsteady compressible flow in irregular
  regions, Journal of Computational Physics 120~(2) (1995) 278--304, cited By
  :242.

\bibitem{Titarev2004}
V.~A. Titarev, E.~F. Toro, {Finite-volume WENO schemes for three-dimensional
  conservation laws}, Journal of Computational Physics 201~(1) (2004) 238--260.
\newblock \href {https://doi.org/10.1016/j.jcp.2004.05.015}
  {\path{doi:10.1016/j.jcp.2004.05.015}}.

\bibitem{Johnsen2006}
E.~Johnsen, T.~Colonius, {Implementation of WENO schemes in compressible
  multicomponent flow problems}, J. Comput. Phys. 219~(2) (2006) 715--732.
\newblock \href {https://doi.org/10.1016/j.jcp.2006.04.018}
  {\path{doi:10.1016/j.jcp.2006.04.018}}.

\bibitem{Berger1989}
M.~J. Berger, P.~Colella, {Local Adaptive Mesh Refinement for Shock
  Hydrodynamics}, J. Comput. Phys. 82 (1989) 64--84.

\bibitem{Deiterding2020}
R.~Deiterding, M.~O. Domingues, K.~Schneider, Multiresolution analysis as a
  criterion for effective dynamic mesh adaptation--a case study for euler
  equations in the samr framework amroc, Computers \& Fluids (2020) 104583.

\bibitem{Schneider2010}
K.~Schneider, O.~V. Vasilyev, {Wavelet Methods in Computational Fluid
  Dynamics}, Annual Review of Fluid Mechanics 42~(1) (2010) 473--503.
\newblock \href {https://doi.org/10.1146/annurev-fluid-121108-145637}
  {\path{doi:10.1146/annurev-fluid-121108-145637}}.

\bibitem{Vasilyev2000}
O.~V. Vasilyev, C.~Bowman, {Second-Generation Wavelet Collocation Method for
  the Solution of Partial Differential Equations}, Journal of Computational
  Physics 165~(2) (2000) 660--693.
\newblock \href {https://doi.org/10.1006/JCPH.2000.6638}
  {\path{doi:10.1006/JCPH.2000.6638}}.

\bibitem{Regele2009}
J.~D. Regele, O.~V. Vasilyev, {An adaptive wavelet-collocation method for shock
  computations}, International Journal of Computational Fluid Dynamics 23~(7)
  (2009) 503--518.
\newblock \href {https://doi.org/10.1080/10618560903117105}
  {\path{doi:10.1080/10618560903117105}}.

\bibitem{Vasilyev2003}
O.~V. Vasilyev, {Solving multi-dimensional evolution problems with localized
  structures using second generation wavelets}, International Journal of
  Computational Fluid Dynamics 17~(2) (2003) 151--168.
\newblock \href {https://doi.org/10.1080/1061856021000011152}
  {\path{doi:10.1080/1061856021000011152}}.

\bibitem{Alam2006}
J.~M. Alam, N.~K.-R. Kevlahan, O.~V. Vasilyev, {Simultaneous space–time
  adaptive wavelet solution of nonlinear parabolic differential equations},
  Journal of Computational Physics 214~(2) (2006) 829--857.
\newblock \href {https://doi.org/10.1016/J.JCP.2005.10.009}
  {\path{doi:10.1016/J.JCP.2005.10.009}}.

\bibitem{Nejadmalayeri2015}
A.~Nejadmalayeri, A.~Vezolainen, E.~Brown-Dymkoski, O.~V. Vasilyev, {Parallel
  adaptive wavelet collocation method for PDEs}, Journal of Computational
  Physics 298 (2015) 237--253.
\newblock \href {https://doi.org/10.1016/j.jcp.2015.05.028}
  {\path{doi:10.1016/j.jcp.2015.05.028}}.

\bibitem{Harten1995}
A.~Harten, {Multiresolution algorithms for the numerical solution of hyperbolic
  conservation laws}, Communications on Pure and Applied Mathematics 48~(12)
  (1995) 1305--1342.
\newblock \href {https://doi.org/10.1002/cpa.3160481201}
  {\path{doi:10.1002/cpa.3160481201}}.

\bibitem{Bihari1996}
B.~L. Bihari, {Multiresolution schemes for conservation laws with viscosity},
  Journal of Computational Physics 123~(1) (1996) 207--225.
\newblock \href {https://doi.org/10.1006/jcph.1996.0017}
  {\path{doi:10.1006/jcph.1996.0017}}.

\bibitem{Bihari1997}
B.~L. Bihari, A.~Harten, {Multiresolution schemes for the numerical solution of
  2-D conservation laws I}, SIAM Journal on Scientific Computing 18~(2) (1997)
  315--354.

\bibitem{Kaibara2001}
M.~K. Kaibara, S.~M. Gomes, {A Fully Adaptive Multiresolution Scheme for Shock
  Computations}, in: Godunov Methods, Springer US, Boston, MA, 2001, pp.
  497--503.
\newblock \href {https://doi.org/10.1007/978-1-4615-0663-8{\_}49}
  {\path{doi:10.1007/978-1-4615-0663-8{\_}49}}.

\bibitem{Cohen2003}
A.~Cohen, S.~Kaber, S.~M{\"{u}}ller, M.~Postel, {Fully adaptive multiresolution
  finite volume schemes for conservation laws}, Mathematics of Computation
  72~(241) (2003) 183--225.

\bibitem{Roussel2003}
O.~Roussel, K.~Schneider, A.~Tsigulin, H.~Bockhorn, {A conservative fully
  adaptive multiresolution algorithm for parabolic PDEs}, Journal of
  Computational Physics 188~(2) (2003) 493--523.
\newblock \href {https://doi.org/10.1016/S0021-9991(03)00189-X}
  {\path{doi:10.1016/S0021-9991(03)00189-X}}.

\bibitem{Castro2016}
D.~A. Castro, S.~M. Gomes, J.~Stolfi, {High-order adaptive finite-volume
  schemes in the context of multiresolution analysis for dyadic grids},
  Computational and Applied Mathematics 35~(1) (2016) 1--16.
\newblock \href {https://doi.org/10.1007/s40314-014-0159-2}
  {\path{doi:10.1007/s40314-014-0159-2}}.

\bibitem{Maulik2018}
R.~Maulik, O.~San, R.~Behera, {An adaptive multilevel wavelet framework for
  scale-selective WENO reconstruction schemes}, International Journal for
  Numerical Methods in Fluids~(January) (2018) 1--31.
\newblock \href {https://doi.org/10.1002/fld.4489}
  {\path{doi:10.1002/fld.4489}}.

\bibitem{Muller2007}
S.~M{\"{u}}ller, Y.~Stiriba, {Fully Adaptive Multiscale Schemes for
  Conservation Laws Employing Locally Varying Time Stepping}, Journal of
  Scientific Computing 30~(3) (2007) 493--531.
\newblock \href {https://doi.org/10.1007/s10915-006-9102-z}
  {\path{doi:10.1007/s10915-006-9102-z}}.

\bibitem{Domingues2008}
M.~O. Domingues, S.~M. Gomes, O.~Roussel, K.~Schneider, {An adaptive
  multiresolution scheme with local time stepping for evolutionary PDEs},
  Journal of Computational Physics 227~(8) (2008) 3758--3780.
\newblock \href {https://doi.org/10.1016/j.jcp.2007.11.046}
  {\path{doi:10.1016/j.jcp.2007.11.046}}.

\bibitem{Domingues2009}
M.~O. Domingues, S.~M. Gomes, O.~Roussel, K.~Schneider, {Space–time adaptive
  multiresolution methods for hyperbolic conservation laws: Applications to
  compressible Euler equations}, Applied Numerical Mathematics 59~(9) (2009)
  2303--2321.
\newblock \href {https://doi.org/10.1016/j.apnum.2008.12.018}
  {\path{doi:10.1016/j.apnum.2008.12.018}}.

\bibitem{Kaiser2019}
J.~W.~J. Kaiser, N.~Hoppe, S.~Adami, N.~A. Adams, An adaptive local
  time-stepping scheme for multiresolution simulations of hyperbolic
  conservation laws, Journal of Computational Physics: X 4 (2019) 100038.

\bibitem{Descombes2017}
S.~Descombes, M.~Duarte, T.~Dumont, T.~Guillet, V.~Louvet, M.~Massot,
  Task-based adaptive multiresolution for time-space multi-scale
  reaction-diffusion systems on multi-core architectures, The SMAI journal of
  computational mathematics 3 (2017) 29--51.
\newblock \href {https://doi.org/10.5802/smai-jcm.19}
  {\path{doi:10.5802/smai-jcm.19}}.

\bibitem{Brix2011}
K.~Brix, S.~Melian, S.~M{\"{u}}ller, M.~Bachmann, {Adaptive Multiresolution
  Methods: Practical issues on Data Structures, Implementation and
  Parallelization}, ESAIM: Proceedings 34 (2011) 151--183.
\newblock \href {https://doi.org/10.1051/proc/201134003}
  {\path{doi:10.1051/proc/201134003}}.

\bibitem{Sutter2005}
H.~Sutter, {The free lunch is over: A fundamental turn toward concurrency in
  software}, Dr. Dobb's Journal (2005) 1--9\href
  {https://doi.org/10.1002/minf.201100042} {\path{doi:10.1002/minf.201100042}}.

\bibitem{Hager2011a}
G.~Hager, G.~Wellein, {Introduction to high performance computing for
  scientists and engineers}, CRC Press, 2011.

\bibitem{Ferreira2017}
C.~R. Ferreira, M.~Bader, {Load Balancing and Patch-Based Parallel Adaptive
  Mesh Refinement for Tsunami Simulation on Heterogeneous Platforms Using Xeon
  Phi Coprocessors}, in: Proceedings of the Platform for Advanced Scientific
  Computing Conference on - PASC '17, ACM Press, New York, New York, USA, 2017,
  pp. 1--12.
\newblock \href {https://doi.org/10.1145/3093172.3093237}
  {\path{doi:10.1145/3093172.3093237}}.

\bibitem{Hejazialhosseini2010}
B.~Hejazialhosseini, D.~Rossinelli, M.~Bergdorf, P.~Koumoutsakos, {High order
  finite volume methods on wavelet-adapted grids with local time-stepping on
  multicore architectures for the simulation of shock-bubble interactions},
  Journal of Computational Physics 229~(22) (2010) 8364--8383.
\newblock \href {https://doi.org/10.1016/j.jcp.2010.07.021}
  {\path{doi:10.1016/j.jcp.2010.07.021}}.

\bibitem{Han2011}
L.~H. Han, T.~Indinger, X.~Y. Hu, N.~A. Adams, {Wavelet-based adaptive
  multi-resolution solver on heterogeneous parallel architecture for
  computational fluid dynamics}, in: Computer Science - Research and
  Development, Vol.~26, 2011, pp. 197--203.
\newblock \href {https://doi.org/10.1007/s00450-011-0167-z}
  {\path{doi:10.1007/s00450-011-0167-z}}.

\bibitem{Sroka2019}
M.~Sroka, T.~Engels, P.~Krah, S.~Mutzel, K.~Schneider, J.~Reiss, {An Open and
  Parallel Multiresolution Framework Using Block-Based Adaptive Grids},
  Springer, Cham, 2019, pp. 305--319.
\newblock \href {https://doi.org/10.1007/978-3-319-98177-2{\_}19}
  {\path{doi:10.1007/978-3-319-98177-2{\_}19}}.

\bibitem{Roe1981}
P.~L. Roe, {Approximate Riemann solvers, parameter vectors, and difference
  schemes}, Journal of computational physics 43~(2) (1981) 357--372.

\bibitem{Rusanov1962}
V.~V. Rusanov, The calculation of the interaction of non-stationary shock waves
  and obstacles, USSR Computational Mathematics and Mathematical Physics 1~(2)
  (1962) 304--320.

\bibitem{Toro1994}
E.~F. Toro, M.~Spruce, W.~Speares, {Restoration of the contact surface in the
  HLL-Riemann solver}, Shock waves 4~(1) (1994) 25--34.

\bibitem{Fu2016}
L.~Fu, X.~Y. Hu, N.~A. Adams, {A family of high-order targeted ENO schemes for
  compressible-fluid simulations}, Journal of Computational Physics 305 (2016)
  333--359.
\newblock \href {https://doi.org/10.1016/j.jcp.2015.10.037}
  {\path{doi:10.1016/j.jcp.2015.10.037}}.

\bibitem{Balsara2016}
D.~S. Balsara, S.~Garain, C.-W. Shu, {An efficient class of WENO schemes with
  adaptive order}, Journal of Computational Physics 326 (2016) 780--804.
\newblock \href {https://doi.org/10.1016/j.jcp.2016.09.009}
  {\path{doi:10.1016/j.jcp.2016.09.009}}.

\bibitem{Morton1966}
G.~Morton, {A computer oriented geodetic data base and a new technique in the
  file sequencing} (1966) 20.

\bibitem{Harlow1971}
F.~Harlow, A.~Amsden, {Fluid dynamics}, Tech. rep., Los Alamos National Labs
  (1971).

\bibitem{Liu1994}
{Weighted Essentially Non-oscillatory Schemes}, Journal of Computational
  Physics 115 (1994) 200--212.
\newblock \href {https://doi.org/10.1006/jcph.1994.1187}
  {\path{doi:10.1006/jcph.1994.1187}}.

\bibitem{Rossinelli2011}
D.~Rossinelli, B.~Hejazialhosseini, D.~G. Spampinato, P.~Koumoutsakos,
  {Multicore/multi-gpu accelerated simulations of multiphase compressible flows
  using wavelet adapted grids}, SIAM Journal on Scientific Computing 33~(2)
  (2011) 512--540.

\bibitem{Hoppe2019}
N.~{Hoppe}, I.~{Pasichnyk}, M.~{Allalen}, S.~{Adami}, N.~A. {Adams}, Node-level
  optimization of a 3{D} block-based multiresolution compressible flow solver
  with emphasis on performance portability, in: 2019 International Conference
  on High Performance Computing Simulation (HPCS), 2019, pp. 732--740.
\newblock \href {https://doi.org/10.1109/HPCS48598.2019.9188088}
  {\path{doi:10.1109/HPCS48598.2019.9188088}}.

\bibitem{Bader2013}
M.~Bader, {Space-Filling Curves}, Vol.~9 of Texts in Computational Science and
  Engineering, Springer Berlin Heidelberg, Berlin, Heidelberg, 2013.
\newblock \href {https://doi.org/10.1007/978-3-642-31046-1}
  {\path{doi:10.1007/978-3-642-31046-1}}.

\bibitem{Hilbert1891}
D.~Hilbert, {{\"U}ber die stetige Abbildung einer Linie auf ein
  Fl{\"a}chenst{\"u}ck}, Math. Ann. 38 (1891) 459--460.

\bibitem{Sod1978}
G.~A. Sod, A survey of several finite difference methods for systems of
  nonlinear hyperbolic conservation laws, Journal of Computational Physics
  27~(1) (1978) 1--31.
\newblock \href {https://doi.org/10.1016/0021-9991(78)90023-2}
  {\path{doi:10.1016/0021-9991(78)90023-2}}.

\bibitem{Harrison1908}
W.~J. Harrison, {The influence of viscosity on the oscillations of superposed
  fluids}, Proceedings of the London Mathematical Society s2-6~(1) (1908)
  396--405.
\newblock \href {https://doi.org/10.1112/plms/s2-6.1.396}
  {\path{doi:10.1112/plms/s2-6.1.396}}.

\bibitem{Chandrasekhar1961}
S.~Chandrasekhar, {Hydrodynamic and hydromagnetic stability}, 1961.

\bibitem{Shi2003}
J.~Shi, Y.~T. Zhang, C.~W. Shu, {Resolution of high order WENO schemes for
  complicated flow structures}, Journal of Computational Physics 186~(2) (2003)
  690--696.
\newblock \href {https://doi.org/10.1016/S0021-9991(03)00094-9}
  {\path{doi:10.1016/S0021-9991(03)00094-9}}.

\bibitem{Fleischmann2019}
N.~Fleischmann, S.~Adami, N.~A. Adams, {Numerical Symmetry-Preserving
  Techniques for Low-Dissipation Shock-Capturing Schemes}, Computers {\&}
  Fluids (may 2019).
\newblock \href {https://doi.org/10.1016/j.compfluid.2019.04.004}
  {\path{doi:10.1016/j.compfluid.2019.04.004}}.

\bibitem{Woodward1984}
P.~Woodward, P.~Colella, {The numerical simulation of two-dimensional fluid
  flow with strong shocks} (apr 1984).
\newblock \href {https://doi.org/10.1016/0021-9991(84)90142-6}
  {\path{doi:10.1016/0021-9991(84)90142-6}}.

\bibitem{Liska2003}
R.~Liska, B.~Wendroff, {Comparison of Several Difference Schemes on 1D and 2D
  Test Problems for the Euler Equations}, SIAM Journal on Scientific Computing
  25~(3) (2003) 995--1017.
\newblock \href {http://arxiv.org/abs/0911.1613} {\path{arXiv:0911.1613}},
  \href {https://doi.org/10.1137/S1064827502402120}
  {\path{doi:10.1137/S1064827502402120}}.

\bibitem{Toro2009}
E.~F. Toro, {Riemann solvers and numerical methods for fluid dynamics: a
  practical introduction}, 3rd Edition, Springer, Dordrecht ; New York, 2009.

\bibitem{Winter2019}
J.~Winter, J.~Kaiser, S.~Adami, N.~Adams, Numerical investigation of 3d
  drop-breakup mechanisms using a sharp interface level-set method, in: 11th
  International Symposium on Turbulence and Shear Flow Phenomena, TSFP 2019,
  2019.

\bibitem{Fedkiw1999}
R.~P. Fedkiw, T.~D. Aslam, B.~Merriman, S.~Osher, {A Non-oscillatory Eulerian
  Approach to Interfaces in Multimaterial Flows (The Ghost Fluid Method)},
  Journal of Computational Physics 152 (1999) 457--492.

\end{thebibliography}
